\documentclass[traditabstract]{aa}
\usepackage[utf8]{inputenc}
\usepackage{txfonts}
\usepackage{graphicx}
\usepackage{array}
\usepackage{natbib}
\usepackage{multirow}
\usepackage{relsize}
\usepackage{amsbsy}
\usepackage{epstopdf}
\usepackage{subfig}
\usepackage[english]{babel}

\begin{document}

\title {The VIMOS Ultra Deep Survey} 
\subtitle{Luminosity and stellar mass dependence of galaxy clustering at $z\sim3$ \thanks{Based on data obtained with the European Southern Observatory Very Large
Telescope, Paranal, Chile, under Large Program 185.A-0791.}} 
\titlerunning{Luminosity and stellar mass dependence of galaxy clustering at z$\sim3$ in VUDS}

\author{A. Durkalec \inst{1}
\and O. Le F\`evre\inst{2}
\and A. Pollo\inst{1,3}
\and G.~Zamorani \inst{4}
\and B.~C.~Lemaux \inst{5}
\and B.~Garilli\inst{6}
\and S.~Bardelli\inst{4}
\and N.~Hathi\inst{2,8}
\and A.~Koekemoer\inst{8}
\and J.~Pforr\inst{2,7}
\and E.~Zucca\inst{4}
}
\institute{National Centre for Nuclear Research, ul. Hoza 69, 00-681, Warszawa, Poland, \email{anna.durkalec@ncbj.gov.pl}
\and
Aix Marseille Universit\'e, CNRS, LAM (Laboratoire d'Astrophysique de Marseille) UMR 7326, 13388, Marseille, France
\and
Astronomical Observatory of the Jagiellonian University, Orla 171, 30-001 Cracow, Poland
\and
INAF - Osservatorio Astronomico di Bologna, Via Gobetti 93/3, 40129 Bologna - Italy 
\and
Department of Physics, University of California, Davis, One Shields Ave., Davis, CA 95616, USA
\and
INAF--IASF Milano, via Bassini 15, I--20133, Milano, Italy
\and
ESA/ESTEC SCI-S, Keplerlaan 1, 2201 AZ, Noordwijk
\and
Space Telescope Science Institute, 3700 San Martin Drive, Baltimore, MD 21218, USA
}


\abstract{
We present the study of the dependence of galaxy clustering on luminosity and stellar mass in the redshift range $2<z<3.5$ using 3236 galaxies with robust spectroscopic redshifts from the VIMOS Ultra Deep Survey (VUDS), covering a total area of $0.92$ deg$^2$.
We measure the two-point real-space correlation function $w_p(r_p)$ for four volume-limited sub-samples selected by stellar mass and four volume-limited sub-samples selected by $M_{UV}$ absolute magnitude. 
We find that the scale dependent clustering amplitude $r_0$ significantly increases with increasing luminosity and stellar mass.
For the least luminous galaxies ($M_{UV}<-19.0$) we measure a correlation length $r_0 = 2.87\pm0.22$ $h^{-1}$ Mpc and slope $\gamma = 1.59\pm0.07$, while for the most luminous ($M_{UV}<-20.2$) $r_0 = 5.35\pm0.50$ $h^{-1}$ Mpc and $\gamma = 1.92\pm0.25$.
This corresponds to a strong relative bias between these two sub-samples of $\Delta b/b^* = 0.43$.
Fitting a 5-parameter HOD model we find that the most luminous ($M_{UV}<-20.2$) and massive ($M_{\star}>10^{10}$ $h^{-1} M_{\sun}$) galaxies occupy the most massive dark matter haloes with $\langle M_h \rangle = 10^{12.30}$ $h^{-1} M_{\sun}$. 
Similar to the trends observed at lower redshift, the minimum halo mass $M_{min}$ depends on the luminosity and stellar mass of galaxies and grows from $M_{min} =10^{9.73}$ $h^{-1} M_{\sun}$ to $M_{min} = 10^{11.58}$ $h^{-1} M_{\sun}$ from the faintest to the brightest among our galaxy sample, respectively.
We find the difference between these halo masses to be much more pronounced than is observed for local galaxies of similar properties.
Moreover, at $z\sim3$, we observe that the masses at which a halo hosts, on average, one satellite and one central galaxy is $M_1\approx4M_{min}$ over all luminosity ranges, significantly lower than observed at $z\sim0$ indicating that the halo satellite occupation increases with redshift. 
The luminosity and stellar mass dependence is also reflected in the measurements of the large scale galaxy bias, which we model as $b_{g,HOD}(>L) = 1.92+25.36(L/L^*)^{7.01}$. 
We conclude our study with measurements of the stellar-to-halo mass ratio (SHMR).
We observe a significant model-observation discrepancy for low-mass galaxies, suggesting a higher than expected star formation efficiency of these galaxies. 
}

\keywords{Cosmology: observations -- large-scale structure of Universe -- Galaxies: high-redshift -- Galaxies: clustering}

\maketitle

\section{Introduction}

The large structure of the Universe consists of two main elements: the luminous, baryonic matter (e.g., in the form of stars, gas and dust) and the dominant underlying dark matter (DM).
The properties and evolution of the former components can, and have been, directly mapped with the use of large sky surveys, both at local and high redshifts using a variety of observations at different wavelengths.
As for the second, dark matter component, the situation is less clear.
Direct observations are currently difficult, but in the paradigm of the $\Lambda$CDM cosmology the visible baryonic matter indirectly traces the dark matter structure.
If we assume that all galaxies are hosted by dark matter haloes \citep{White1978}, the information about the underlying dark matter distribution can be extracted, e.g., using the mean occupation of galaxies in dark matter haloes. 
However, the relation between these two components is not straightforward.
In particular, the spatial distribution of baryonic matter is biased with respect to that of dark matter, which is a result of additional physics of the baryonic component, like star formation, supernova feedback and galaxy merging, that regulate formation and evolution of galaxies \citep[see, e.g.,][]{Kaiser1984, Bardeen1986, Mo1996, Kauffmann1997}.
It has been shown that the difference between the luminous and dark matter distributions depends both on the epoch of galaxy formation and the physical properties of galaxies \citep[e.g.,][]{Fry1996, Tegmark1998}.
Therefore, studies of the evolution of the luminous-dark matter relation (called \textit{bias}), and its dependence on various galaxy properties (like luminosity, stellar mass or colour) are crucial, because they can provide us with valuable information for investigating the nature of the underlying dark matter distribution and, in the wider perspective, understanding the evolution of the accelerating universe.   

There are various methods used to infer the properties of the dark matter through the observations of the luminous component. 
The most direct ones involve gravitational lensing \citep{Zwicky1937}, which is a unique observational technique that allows to probe both the nature and distribution of dark matter \citep[e.g.,][]{VanWaerbeke2000, Metcalf2001, Moustakas2003, Hoekstra2004, Massey2007, Fu2008,  Rines2013}. 
The gravitational lensing observations, however, are usually possible only for a special set of circumstances, as the objects available for exploration are limited by the geometry of lens and sources \citep[see, e.g., ][]{Blandford1992, Meylan2006}.
Other methods for studying dark-luminous matter relations are applied on the scales of individual galaxies, where e.g., studies of rotation curves \citep{Rubin1978} of stars or gas clouds within individual galaxies are used to explore the hosting dark matter halo masses and density profiles, improving the understanding of the role of dark matter haloes in galaxy formation and evolution \citep[e.g.,][]{Genzel2017, Dekel2017, Katz2017}.
On the large scales considered in this work the most effective methods make use of statistical tools.
Among them the most extensively used one is galaxy clustering based on galaxy correlation function measurements, which allows to understand the time evolution of luminous-dark matter relation and its dependence on galaxy properties.

The galaxy correlation function is a simple, yet powerful statistical tool \citep{Peebles1980} and it can be modelled using, among others, the two parameter power-law $\xi(r) = (r/r_0)^{-\gamma}$ \citep{Davis1983} model or can be modelled from Halo Occupation Distribution models \citep[HOD, ][]{Seljak2000, Peacock2000, Magliocchetti2003, Zehavi2004, Zheng2005}.
In the HOD framework, the theoretical description of the correlation function differs for different scales $r$, accounting for the fact that the clustering of galaxies residing in the same halo differs from clustering between galaxies residing in the separate haloes. 
For small scales ($r \leqslant 1.5 h^{-1}$ Mpc) the \textit{one-halo} term is dominant, as it describes exclusively the clustering of galaxies that reside within a single dark matter halo.
On the opposite side, on large scales ($r \geqslant 3 h^{-1}$ Mpc), the \textit{two-halo} term is dominant, which describes the clustering of galaxies residing in separate dark matter haloes.

Using these two prescriptions of the galaxy correlation function it has been shown that galaxy clustering, and by extension the galaxy-dark matter relation, strongly depends on various galaxy properties.
In general, at local ($z\sim0$) and intermediate ($z<2$) redshifts, luminous and massive galaxies tend to be more strongly clustered than their less luminous and less massive counterparts.
Additionally, it has been found that the clustering strength varies as a function of morphology, colour and spectral type. 
Galaxies with bulge dominated morphologies, red colours, or spectral types indicating old stellar populations also exhibit stronger clustering and a preference for dense environments 
\citep[e.g., ][]{Norberg2002, Pollo2006, delaTorre2007, Coil2008, Meneux2006, Meneux2008, Meneux2009, Abbas2010, Hartley2010, Zehavi2011, Coupon2012, Mostek2013, Marulli2013, Beutler2013, Guo2015, Skibba2015}.
These studies are in good agreement with the hierarchical theory of galaxy formation and evolution \citep{White1987, Kauffmann1997, Benson2001}.

A lot of effort has been put into testing whether or not similar clustering dependencies can be observed at high redshift ($z>2$).
Some evidence for a difference between the clustering of massive, luminous and faint galaxies has been found \citep[e.g., ][]{Daddi2003, Adelberger2005, OLF2005, Ouchi2005, Lee2006, Hildebrandt2009, Wake2011, Lin2012, Bielby2014}.   
However, most of these observational constraints suffer from a combination of many types of selection biases, due to the limited sample size and volume explored of galaxy surveys performed at $z>2$.
Until now, high redshift samples have been either too small to allow a subdivision into galaxy classes or they targeted special types of galaxies (like extremely massive red objects or sources selected using a Lyman-break or B$z$K technique) that cannot be easily related to galaxy populations at lower redshifts. 
Therefore, the overall picture of the possible dependence of galaxy clustering on luminosity and stellar mass at these high redshifts is still difficult to establish.

In this paper we attempt to overcome some of these difficulties and provide improved constraints on the dependence of galaxy clustering with luminosity and stellar mass at high redshifts.
We compute the projected two-point correlation function $w_p(r_p)$ for galaxy samples limited in luminosity and stellar mass in the redshift range $2<z<3.5$ using data sample from VIMOS Ultra Deep Survey \citep[VUDS, ][]{OLF2015}.
There are two main features of VUDS that are advantageous for our studies.
First, VUDS, being a spectroscopic survey, provides a very reliable redshift measurement of a large number of galaxies in a relatively large field.
Second, since its target selection is based mainly on photometric redshifts, the VUDS survey targets a representatively sampled population of star forming galaxies, with luminosities close to the characteristic ($\sim L^*$) luminosity, that are relatively easy to compare to low-redshift objects. 
Consequently, we are able to present reliable correlation function measurements, with power-law and HOD fitting, as well as measurements of the galaxy bias, and satellite fraction at $z\sim3$, and discuss all these results in terms of the current scenario of the density field evolution.
Additionally, the comparison between VUDS clustering measurements with similar studies performed at lower redshifts allows us to put constraints on the cosmic evolution of the relationship between DM and galaxy properties, hence between gravity and cosmology on one side and processes associated with baryonic physics on the other side.

The paper is organized as follows. 
In Section \ref{sec:data} we briefly describe the properties of the VUDS survey and our selected samples. 
The methods used to measure the correlation function and derive power-law and HOD fits are presented in Section \ref{sec:method}. 
Results and comparison of our findings to other works are described in Section \ref{sec:results}. 
We discuss the luminosity and stellar mass dependence, as well as the redshift evolution of galaxy clustering, galaxy bias, halo mass, satellite fraction and stellar-to-halo mass ratio in Section \ref{sec:discussion}, before concluding in Section \ref{sec:summary}.

Throughout all this paper, we adopt a flat $\Lambda$CDM cosmological model with $\Omega_m = 0.3175$ and $\Omega_{\Lambda} = 0.6825$ \citep{Planck2014}.
The Hubble constant is parametrized via $h = H_0/100$ to ease the comparison with previous works. 
We report correlation length measurements in comoving coordinates and express magnitudes in the AB system.


\section{Data}
\label{sec:data}
\subsection{VUDS Survey summary}
\renewcommand{\arraystretch}{1.2} 
\begin{table}
    \begin{center}
     \caption{Properties of the galaxy sample in the range $2<z<3.5$, as used in this study.}
      \scalebox{1.0} {
	\begin{tabular}{p{2.5cm} p{1.5cm} p{1.5cm} p{1.5cm} } \hline \hline
	VUDS field	&	$N_g$	&	$z_{median}$	&	$S_{eff}$ [deg$^2$] \\ \hline
	COSMOS		& 	1605	&	2.79	&	0.50	\\ 
	VVDS-02h	& 	1237	&	2.63	&	0.31	\\ 
	ECDFS		& 	{\b394}	&	2.57	&	0.11	\\ \hline
	\textbf{Total}		&	{\b 3236}	&	2.7	&	0.92	\\ \hline
	\end{tabular}
     \label{tab:sample_numbers}
      }
      \end{center}
\end{table}
\setlength{\tabcolsep}{-2pt} 
\begin{figure*}[t!]
 \centering
  \begin{tabular}{ccc}
  \includegraphics[angle=270]{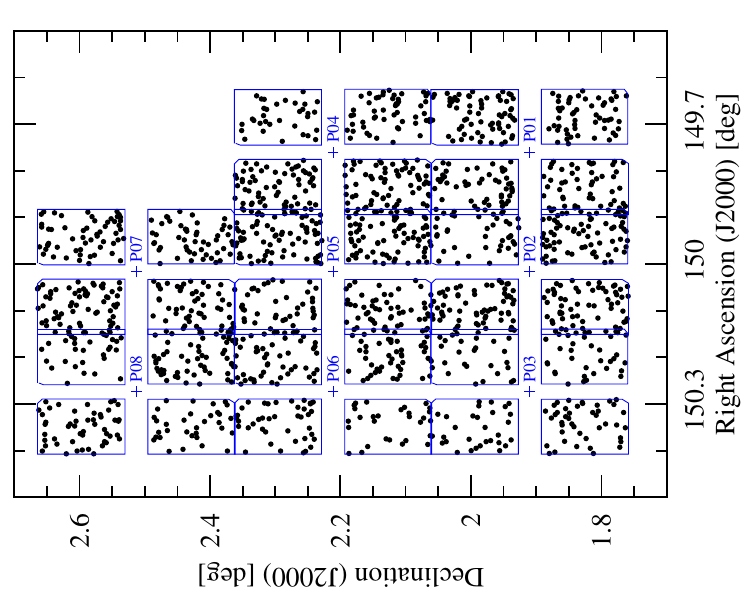} 
  & \includegraphics[angle=270]{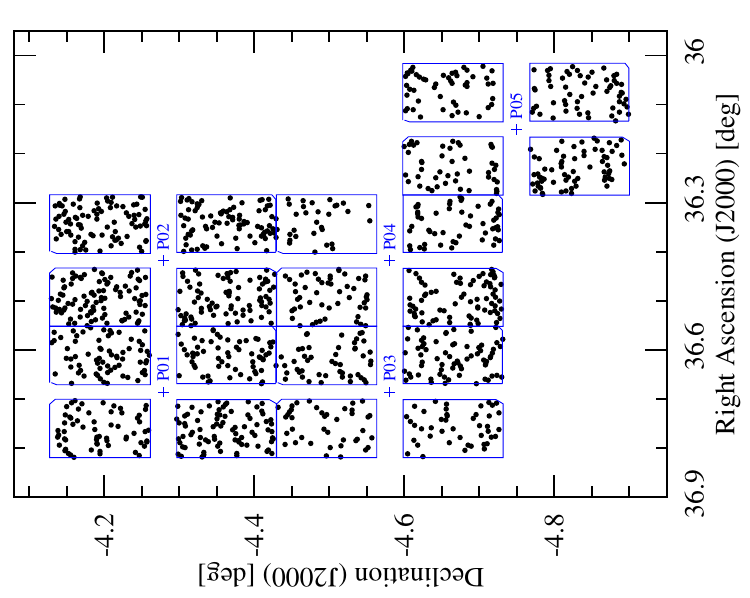} 
  & \includegraphics[angle=270]{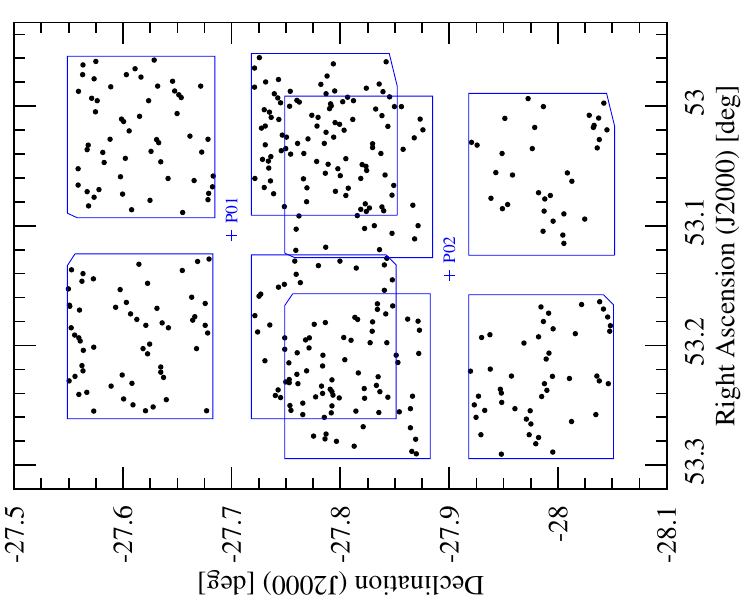} 
  \end{tabular} 
  \caption{Spatial distribution of galaxies with spectroscopic redshifts $2<z<3.5$ in three independent VUDS fields: COSMOS (\textit{left} panel), VVDS-02h (\textit{central} panel) and ECDFS (\textit{right} panel).
	  The blue crosses indicate VIMOS pointing centres.}
 \label{fig:fields_spatial}
\end{figure*}
\begin{figure}
 \centering
    \includegraphics[angle=270]{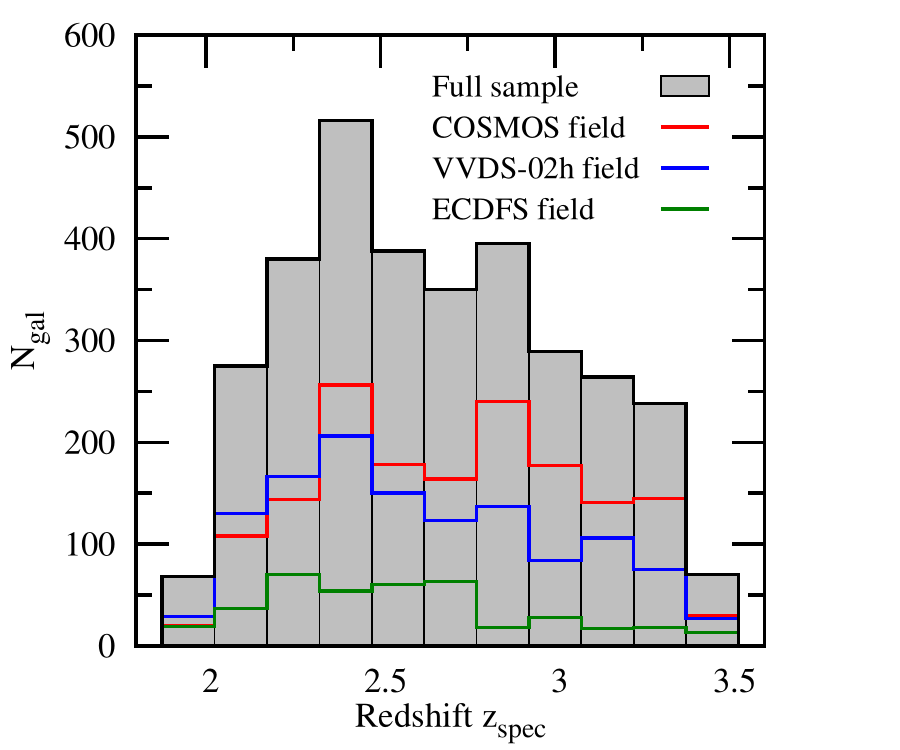} 
    \caption{Redshift distribution of the VUDS galaxy sample in the redshift range $2.0 < z < 3.5$ used in this study. 
	    The filled grey histogram represents the total sample of galaxies, while the red, blue and green histograms represent the contribution from COSMOS, VVDS-02h and ECDFS fields, respectively.}
 \label{fig:redshift_distribution}
\end{figure}
Our galaxy sample is drawn from the VIMOS Ultra Deep Survey (VUDS).
Details about the survey strategy, target selection, as well as data processing and redshift measurements are presented in \cite{OLF2015}.
Below we provide only a brief summary of these survey features, that are relevant to the study of the galaxy clustering presented in this paper.

VUDS is a spectroscopic survey targeting $\sim$ 10 000 galaxies in the redshift range $2 < z < 6+$.
The survey covers a total area of $\sim$1 deg$^2$ across three independent fields (see Fig. \ref{fig:fields_spatial}), thus reducing the effect of cosmic variance, which is important for galaxy clustering measurements. 
The majority ($\sim 88\%$) of targets are selected based on photometric redshifts ($z_{phot} + 1\sigma$ $\geqslant$ 2.4) derived from deep multi-band photometry available for the VUDS fields, and supplemented with the targets selected by various magnitude and colour-colour criteria (mainly Lyman Break Galaxies, LBGs). 
As the result, VUDS sample at redshift $z>2$, exceeds greatly the number of spectroscopically confirmed galaxies from all previous surveys, allowing, e.g., for selection of various volume-limited sub-samples characterised by different galaxy properties.
Moreover, due to its target selection method, VUDS can be considered as a largely representative sample of star-forming galaxies with luminosities close to the characteristic luminosity, i.e., $\sim 0.3L^* < L_{UV} < 3L^*$ \citep{Cassata2013} observed at redshift $z>2$.
However, the dusty galaxy population at high redshift in VUDS sample is almost certainly underrepresented.

The core engine for redshift measurement in VUDS is the cross-correlation of the observed spectrum with the reference templates using the EZ redshift measurement code \citep{Garilli2010}. 
At the end of the process, each redshift is assigned a flag, that expresses the reliability of the measurement \citep[for details see][]{OLF2015}.
In our study we are using only the most reliable objects, with the high $75-100\%$ probability of the redshift measurement being correct ($z_{flag} = 2,3,4,$ and 9).
It is worth to mention, that the galaxies assigned with lower flags, e.g., $z_{flag}=1$, do not appear to occupy a distinct region of stellar mass/$M_{UV}$ phase space relative to the $z_{flag}\geq2$ sample, so due to this flag selection we do not exclude any specific type of galaxies from the sample.
The influence of this selection on the clustering measurements and the correction methods used, are fully described in \cite{Durkalec2015a}.

The selected VUDS sample also benefits from an extended multi-wavelength data set \citep[see][]{OLF2015}. 
The multi-wavelength photometry is used to compute absolute magnitudes from the SED fitting using the 'Le Phare' code \citep{Arnouts1999, Ilbert2006}, as described in detail by \cite{Ilbert2005} and references therein \citep[see also ][]{Tasca2015}.

Stellar masses are measured using GOSSIP+ (Galaxy  Observed-Simulated SED Interactive Program) software, which performs a joint fitting of both spectroscopy and multi-wavelengths photometry data with stellar population models, as described in detail by \cite{Thomas2016}.
We note that this stellar mass measurement method differs from the commonly used ones based on, e.g., SED fitting on the multi-wavelength photometry.
We decided to use the GOSSIP+ stellar masses, due to the larger number of reliable $M_{\star}$ measurements available for the VUDS sample.
Based on the tests performed prior to this study, we observe no noticeable difference in the correlation function shape and/or correlation amplitude when GOSSIP+ derived or Le Phare derived stellar masses are used.  

\subsection{Luminosity and stellar mass sub-samples selection}
\label{sec:lum_mass_subsamples}
Our full sample consists of 3236 objects with reliable spectroscopic redshifts in the range $2<z<3.5$ observed in three independent fields, COSMOS, VVDS-02h and ECDFS, that cover a total area of 0.92 deg$^2$ (the sum of VIMOS slitmask outline after accounting for overlaps, see Fig. \ref{fig:fields_spatial}), which corresponds to a volume of $\sim 1.75 \times 10^7$ Mpc$^3$. 
The spatial distribution of galaxies in each field is presented in Fig. \ref{fig:fields_spatial}, while Fig. \ref{fig:redshift_distribution} shows their redshift distribution.
The general properties of the whole sample, including the number of galaxies, median redshift and the effective area, are listed in Tab. \ref{tab:sample_numbers}.

For the following analysis we selected four volume-limited luminosity sub-samples, with the selection cuts made in the UV-rest frame absolute magnitudes, computed at a rest wavelength of 1500$\AA{}$ ($M_{UV,1500}$, also denoted as $M_{FUV}$ and further in this work simplified to $M_{UV}$), and four stellar mass sub-samples, in order to study respectively the luminosity and stellar mass dependence of the galaxy clustering within the redshift range $2<z<3.5$.
We chose this specific redshift range to be able to study galaxy clustering of as faint galaxies as possible, and at the same time, to maintain volume completeness and large number of galaxies in the various sub-samples.   
On the other side, the choice of the UV wavelength for the luminosity selection is driven by the fact that VUDS is an optically selected survey. 
The full wavelength coverage of VUDS is 3650 - 9350 $\AA{}$, which corresponds to the UV rest frame wavelength coverage at redshift $\sim 3$.

\begin{figure}
   \subfloat{\includegraphics[angle=270]{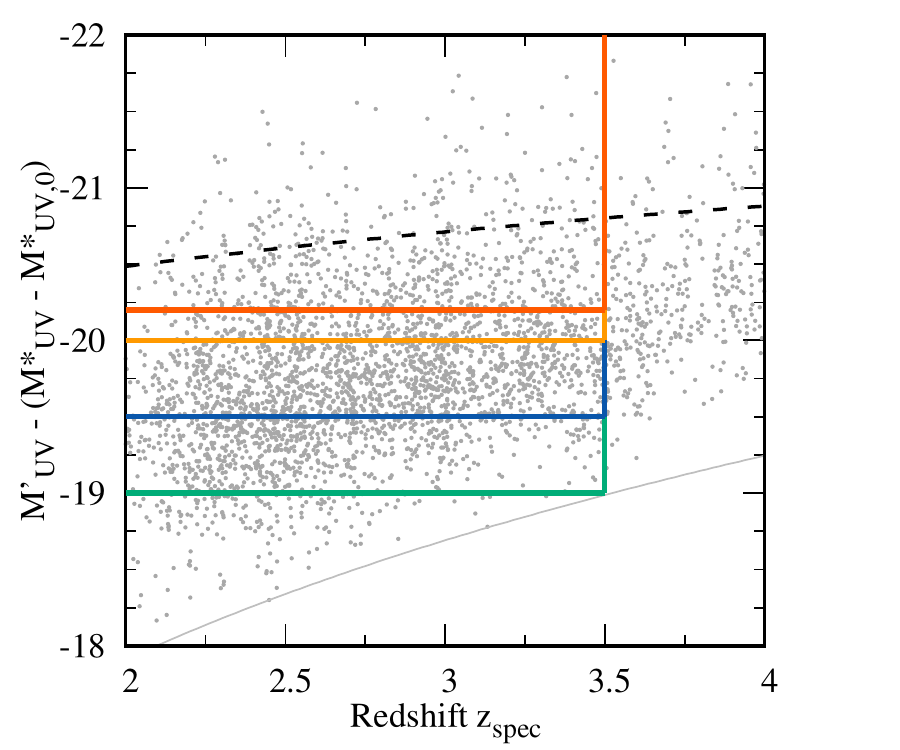}} \\ [-3ex]
   \subfloat{\includegraphics[angle=270]{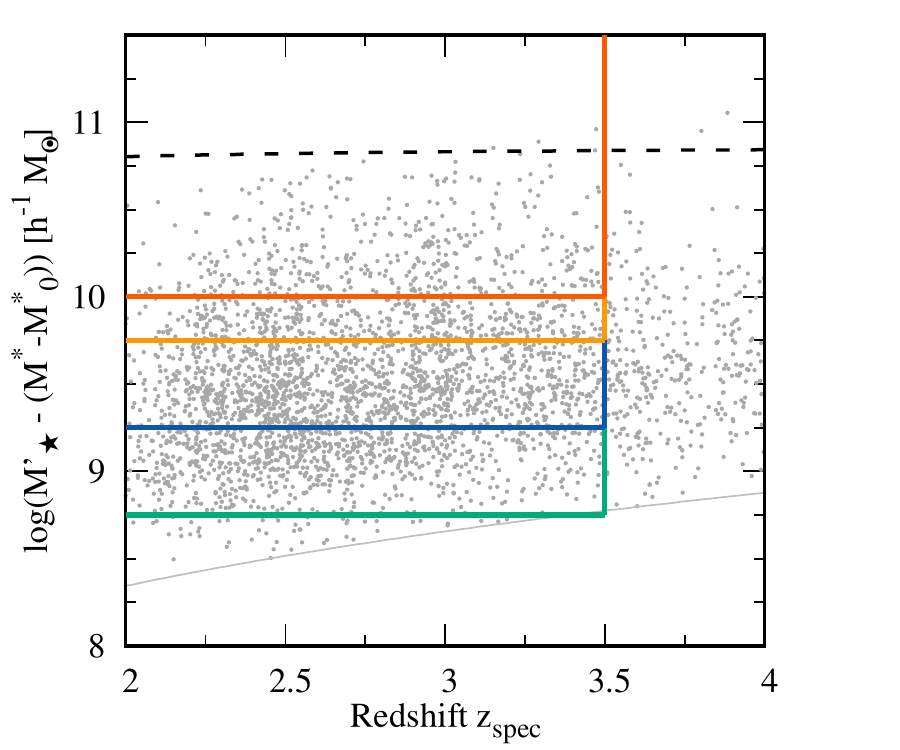}}
   \caption{Construction of the volume-limited galaxy sub-samples with different luminosity (\textit{upper} panel) and stellar mass (\textit{lower} panel). 
	     In both figures grey dots represent the distribution of VUDS galaxies as a function of spectroscopic redshift $z$. 
	     At each redshift UV-band absolute magnitudes and stellar masses are normalized to the characteristic absolute magnitudes, or to the characteristic stellar mass, respectively (see Sec. \ref{sec:lum_mass_subsamples}).
	     The different colour lines delineate the selection cuts for selected UV absolute magnitude and stellar mass sub-samples as defined in Tab. \ref{tab:lum_subsamples} and Tab. \ref{tab:mass_subsamples}.
	     The dashed black line represents the evolution of the not corrected characteristic UV absolute magnitude $M^*_{UV}$ (upper panel), or characteristic stellar mass $M^*$ (lower panel).
	     The grey line indicates the volume limit of the VUDS sample.}
 \label{fig:subsamples}
\end{figure}

All of selected sub-samples have been chosen to contain a number of galaxies sufficient for a reliable measurement of the correlation function (based on tests performed on VUDS data prior to this research, see \citealt{Durkalec2015a}).
Selection cuts for different sub-samples are shown in Fig. \ref{fig:subsamples}.
Additionally, general properties of these sub-samples including number of galaxies, median redshifts, UV median absolute magnitudes $M^{med}_{UV}$ and median stellar masses $\log M_{\star}^{med}$ of each sub-sample are listed in Tab. \ref{tab:lum_subsamples} and Tab. \ref{tab:mass_subsamples}.

To account for the mean brightening of galaxies due to their evolution and to ease the comparison between measurements based on samples from various epochs, we normalized the absolute magnitudes and stellar masses, at each redshift, to the corresponding value of the characteristic absolute magnitude $M^*_{UV}$ of the Schechter luminosity function in the UV band or to the characteristic stellar mass $\log M^*$ respectively.
Therefore, for the absolute magnitudes we compute $M_{UV} = M'_{UV} - (M_{UV}^* - M_{UV,0}^*)$, where $M'_{UV}$ is the original (not corrected) absolute magnitude in the UV filter, $M_{UV}^*$ is the characteristic absolute magnitude and $M_{UV,0}^*$ is the characteristic luminosity for galaxies at $z=0$.  
Similar correction is applied for the stellar masses.
The values of the characteristic absolute magnitudes have been estimated based on the work of \cite{Bouwens2015, Mason2015, Hagen2015, Finkelstein2015, Sawicki2006}, while the characteristic stellar masses have been taken from \cite{Ilbert2013} and \cite{Perez2008}.
The details of the methods used to determine the values of characteristic absolute magnitudes and stellar masses at a given redshift are presented in Appendix \ref{app:ev_correction}.

\setlength{\tabcolsep}{6pt}
\begin{table}
  \begin{center}
     \caption{Properties of the galaxy luminosity sub-samples, as used in this study.}
     \label{tab:lum_subsamples}
	\begin{tabular}{p{1cm}p{1cm}p{1cm}p{1cm}p{1cm}} \hline \hline
	Sample	&	$M_{UV}^{max}$	&	$N_g$	&	$z_{median}$	&	$M_{UV}^{med}$ \\ \hline
	1	&	-19.0		& 	2987	&	2.72		&	-19.84	\\ 
	2	&	-19.5		& 	2241	&	2.77		&	-20.03	\\ 
	3	&	-20.0		& 	986	&	2.83		&	-20.39	\\
	4	&	-20.2		&	616	&	2.84		&	-20.56	\\ \hline
	\end{tabular}
\bigskip
     \caption{Properties of the galaxy stellar mass sub-samples, as used in this study.}
     \label{tab:mass_subsamples}
	\begin{tabular}{p{1cm}p{1cm}p{1cm}p{1cm}p{1cm}} \hline \hline
	Sample	&	$\log M_{\star}^{min}$	&	$N_g$	&	$z_{median}$	&	$\log M_{\star}^{med}$  \\ \hline
	1	&	8.75		& 	3089	&	2.70		&	9.48	\\ 
	2	&	9.25		& 	2304	&	2.75		&	9.64	\\ 
	3	&	9.75		& 	989	&	2.82		&	10.10	\\
	4	&	10.0		&	522	&	2.83		&	10.24	\\ \hline
	\end{tabular}
   \end{center}
\end{table}

\section{Measurement methods}
\label{sec:method}
This work is an extension of our previous studies presented in \cite{Durkalec2015a}, and all the methods used in this study to quantify the galaxy clustering are similar to those presented therein.
This includes computation techniques, error estimations, analysis of systematics in the correlation function measurements and correction methods.
For a detailed description we refer the reader to Sec. 3 of \cite{Durkalec2015a}, while below we provide a short summary of the procedures.  

We measure the real-space correlation function $\xi(r_p,\pi)$ of the combined data from three independent VUDS fields through the Landy-Szalay estimator \citep{Landy1993}.
The differences in size and galaxy numbers between the fields have been accounted for by an appropriate weighting scheme. 
In particular, each pair was multiplied by the number of galaxies per unit volume.
\begin{equation}
 \xi(r_p,\pi) = \sum_{i=1}^{n_{field}}w_i\left(GG_i-2GR_i + RR_i\right)/\sum_{i=1}^{n_{field}}w_iRR_i,
\end{equation}
where $w_i=\left(N_{g,i}/V_i\right)^2$ and $GG$, $GR$, $RR$ are the number of distinct galaxy-galaxy, galaxy-random and random-random pairs with given separations lying in the intervals of $(r_p,r_p+dr_p)$ and $(\pi,\pi+d\pi)$, respectively.   
Integrating the measured $\xi(r_p,\pi)$ along the line of sight gives us the two-point projected correlation function $w_p(r_p)$, which is the two-dimensional counterpart of the real-space correlation function, free from the redshift-space distortions \citep{Davis1983}.
\begin{equation}
 w_p(r_p) = 2\int^{\infty}_0\xi\left(r_p,\pi\right)d\pi.
 \label{eq:wprp}
\end{equation}
In practice, a finite upper integral limit $\pi_{max}$ has to be used in order to avoid adding uncertainties to the result.
A value that is too small results in missing small-scale signal of correlation function $w_p(r_p)$, while a value that is too large has the effect of inducing an unjustified increase in the $w_p(r_p)$ amplitude \citep[see, e.g., ][]{Guzzo1997, Pollo2005}.
After performing a number of tests for different $\pi_{max}$, we find that $w_p(r_p)$ is insensitive to the choice of $\pi_{max}$ in the range $15 < \pi_{max} < 20 h^{-1}$ Mpc. 
Therefore, we choose $\pi_{max} = 20 h^{-1}$ Mpc, which is the maximum value for which the correlation function measurement was not appreciably affected by the mentioned uncertainties.

\begin{figure*}
 \centering
 \begin{tabular}{cc}
 \includegraphics[angle=270]{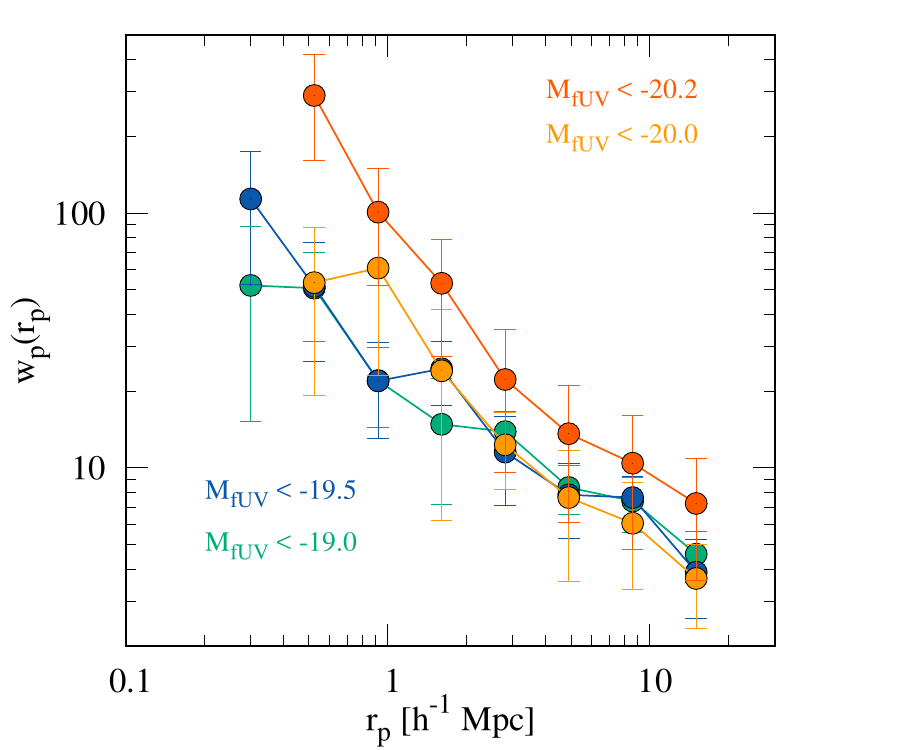} & \includegraphics[angle=270]{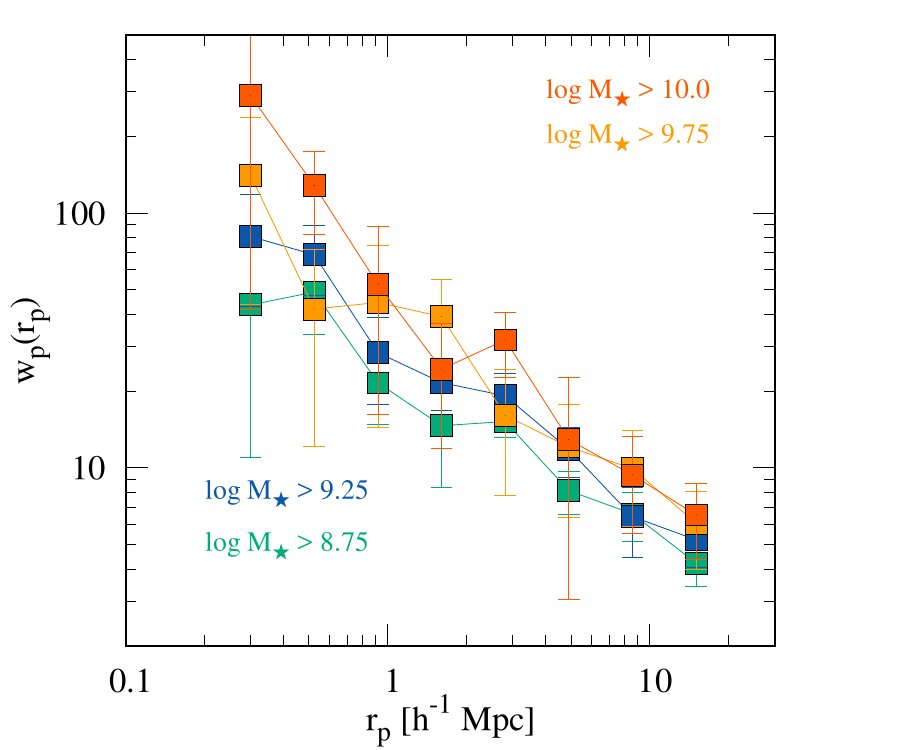} \\
 \end{tabular}
 \caption{Projected correlation functions for volume-limited samples corresponding to different luminosity (left panel, circles) and stellar mass (right panel, squares) bins, as labelled.}
  \label{fig:cf}
\end{figure*}

All correlation function measurements presented in this paper have been corrected for the influence of various systematics originating in the VUDS survey properties, by introducing the correction scheme developed in \cite{Durkalec2015a}.
In particular, we accounted for the galaxies excised from the observations due to the VIMOS layout and other geometrical constraints introduced by the target selection (see Fig. \ref{fig:fields_spatial}). 
Also, the correcting scheme addresses the possible underestimation of the correlation function related to the small fraction of incorrect redshifts present in the sample, as well as small scale underestimations observed in tests based on the VUDS mock catalogues.

To estimate the two-point correlation function errors we apply a combined method \citep[see ][]{Durkalec2015a}, which makes use of the so-called blockwise bootstrap re-sampling method with $N_{boot}=100$ \citep{Barrow1984} coupled to $N_{mock}=66$ independent VUDS mock catalogues \citep[see ][ for details about mocks]{Durkalec2015a, delaTorre2013}, similar to the method proposed by \cite{Pollo2005}.
The associated covariance matrix $\textrm{\textbf{C}}_{ik}$ between the values $w_p$ on \textit{i}th and \textit{k}th scale has been computed using:
\begin{equation}
 \textrm{\textbf{C}}_{ik} = \left\langle\left(w_p^j(r_i)-\langle w_p^j(r_i)\rangle_j\right)\left(w_p^j(r_k)-\langle w_p^j(r_k)\rangle_j\right)\right\rangle_j
\end{equation}
where "$\langle\rangle$" indicates an average over all bootstrap or mock realizations, the $w_p^j(r_k)$ is the value of $w_p$ computed at $r_p = r_i$ in the cone $j$, where $1 < j < N_{mock}$ for the VUDS mocks and $1 < j < N_{boot}$ for the bootstrap data.

Throughout this study we use two approximations of the shape of the real-space correlation function.
The first one is a power-law function $\xi(r) = (r/r_0)^{-\gamma}$ , where $r_0$ and $\gamma$ are the correlation length and slope, respectively. 
With this parametrization, the integral in Eq. \ref{eq:wprp} can be computed analytically and $w_p(r_p)$ can be expressed as
\begin{equation}
 w_p(r_p) = r_p\left(\frac{r_0}{r_p}\right)^{\gamma}\frac{\Gamma\left(\frac{1}{2}\right)\Gamma\left(\frac{\gamma-1}{2}\right)}{\Gamma\left(\frac{\gamma}{2}\right)},
\end{equation}
where $\Gamma$ is the Euler's Gamma Function.
Despite of its simplicity, a power-law model remains an efficient and simple approximation of galaxy clustering properties.

A second, more detailed description of the real-space correlation function, used here, has been done in the framework of the Halo Occupation Distribution (HOD) models.
Following a commonly used, analytical prescription, we parametrized the halo occupation model in the way used, e.g., by \cite{Zehavi2011} and motivated by \cite{Zheng2007}.
The mean halo occupation function $\langle N_g(M_h)\rangle$, i.e., the number of galaxies that occupy the dark matter halo of a given mass is the sum of the mean occupation functions for the central and satellite galaxies,
\begin{equation}
 \langle N_{g}(M_h)\rangle = \langle N_{cen}(M_h)\rangle + \langle N_{sat}(M_h)\rangle,
\end{equation}
where,
\begin{eqnarray}
 \langle N_{cen}(M_h)\rangle & = & \frac{1}{2}\left[1+\textrm{erf}\left(\frac{\log M_h-\log M_{min}}{\sigma_{\log M}}\right) \right] \\
 \langle N_{sat}(M_h)\rangle & = & \langle N_{cen}(M_h)\rangle \times \left(\frac{M_h-M_0}{M_1'}\right)^{\alpha}.
 \label{eq.halo_occupation_sat}
\end{eqnarray}
This model includes five free parameters, two of which represent characteristic halo masses, that describe the mass scales of halos hosting central galaxies and their satellites. 
The characteristic mass $M_{min}$ is the minimum mass needed for half of the haloes to host one central galaxy above the assumed luminosity (or mass) threshold, i.e., $\langle N_{cen}(M_{min})\rangle = 0.5$, whereas the second characteristic mass $M_1$ is the mass of haloes that on average have one additional satellite galaxy above the luminosity (or mass) threshold, i.e., $\langle N_{sat}(M_{1})\rangle = 1$. 
Note that $M_1$ is different from $M_1'$ from Eq. \ref{eq.halo_occupation_sat}.
However, both quantities are related to each other and in most cases $M_1 \sim M_1'$ (see Tab. \ref{tab:parameters}).
The remaining three free parameters are:  $\sigma_{\log M}$ - related to the scatter between the galaxy luminosity (or stellar mass) and halo mass $M_h$, the cutoff mass scale $M_0$, and the high-mass power-law slope $\alpha$ of the satellite galaxy mean occupation function.

The HOD parameter space for each galaxy sample has been explored by using the Population Monte Carlo (PMC) technique \citep{Wraith2009, Kilbinger2011}, using the full covariance error matrix, as described in \cite{Durkalec2015a}.
From the best-fitting HOD parameters we derived quantities describing the halo and galaxy properties, like the average host halo mass $\langle M_h \rangle$,
\begin{equation}
 \langle M_h \rangle(z) = \int dM_h\ M_h\ n(M_h,z)\ \frac{\langle N_g(M_h)\rangle}{n_g(z)},
\end{equation}
the large-scale galaxy bias $b_g$,
\begin{equation}
 b_g(z) = \int dM_h\ b_h(M_h)\ n(M_h,z)\ \frac{\langle N_g(M_h)\rangle}{n_g(z)},
 \label{eq:bias}
\end{equation}
and the fraction of satellite galaxies per halo $f_s$
\begin{equation}
 f_s = 1 - \int dM_h n(M_h,z) \frac{N_c(M_h)}{n_g(z)},
 \label{eq:satellite_fraction}
\end{equation}
where $n(M_h,z)$ is the dark matter mass function, $b_h(M_h,z)$ is the large-scale halo bias, and $n_g(z)$ represents the number density of galaxies,
\begin{equation}
 n_g(z) = \int dM_h\ n(M_h,z)\ \langle N_g(M_h)\rangle.
\end{equation}


\setlength{\tabcolsep}{-2pt} 
\section{Results}
\label{sec:results}
\begin{figure*}
   \centering
   \subfloat{
   \begin{tabular}{cc}
     \includegraphics[angle=270]{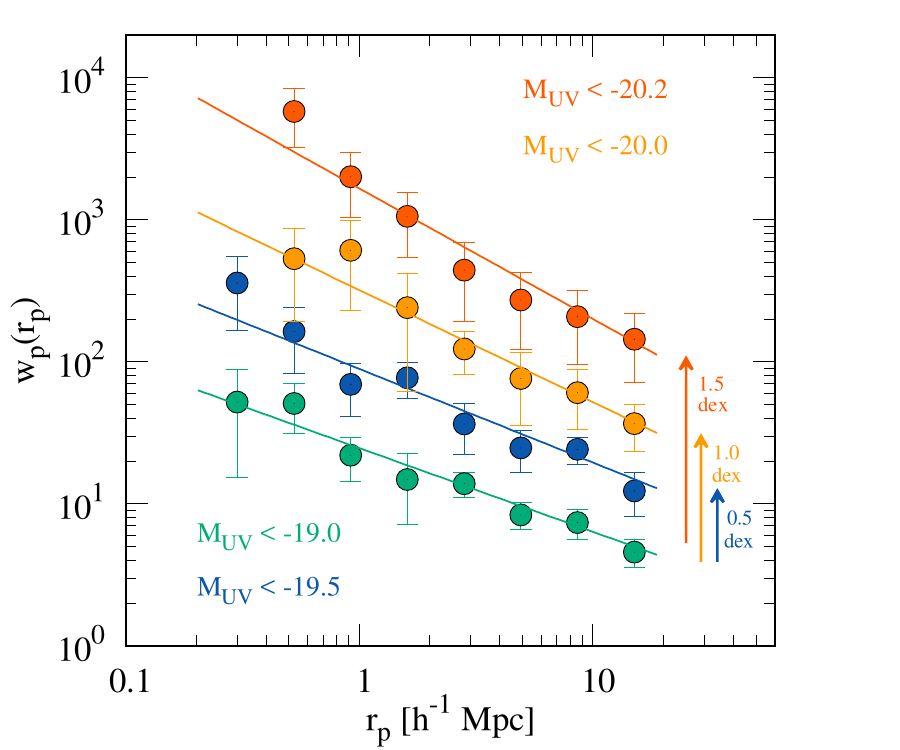} & \includegraphics[angle=270]{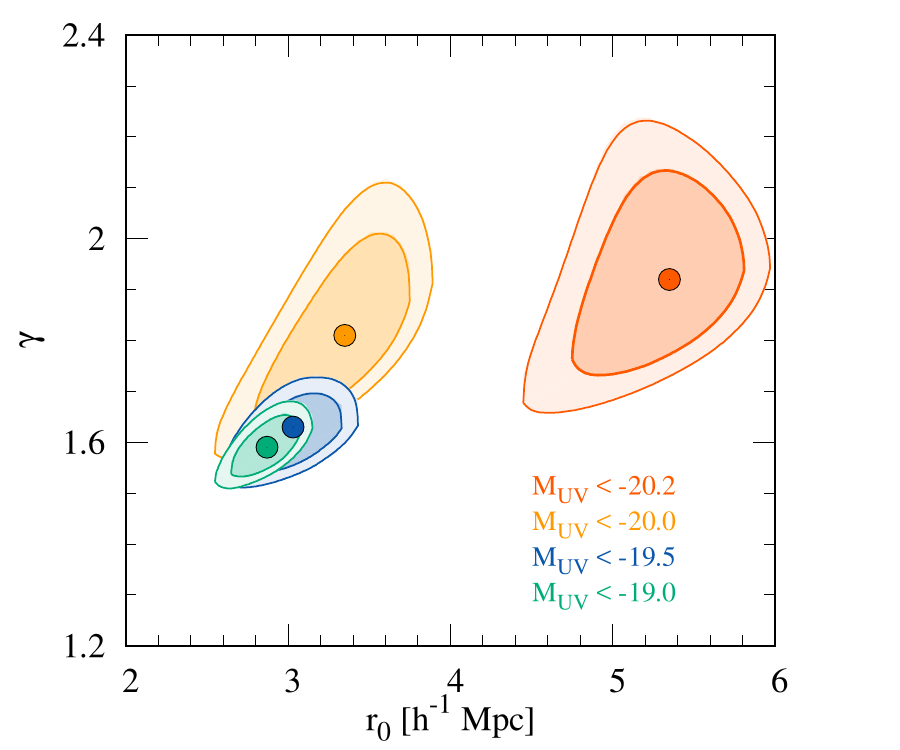} 
   \end{tabular}
   } \\ [-4ex]
   \subfloat{
   \begin{tabular}{cc}
      \includegraphics[angle=270]{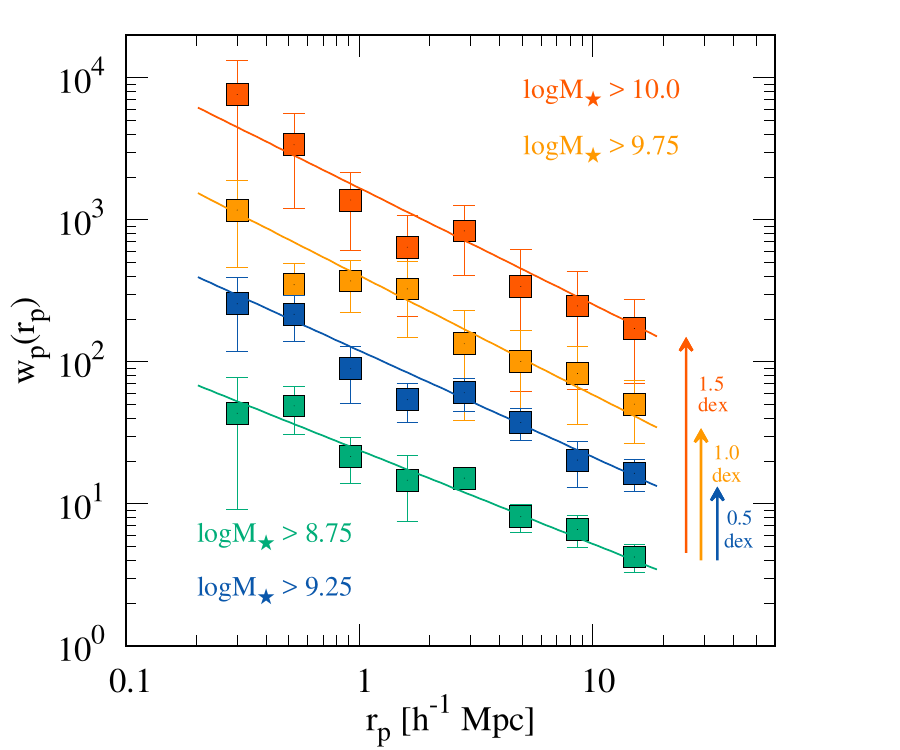} & \includegraphics[angle=270]{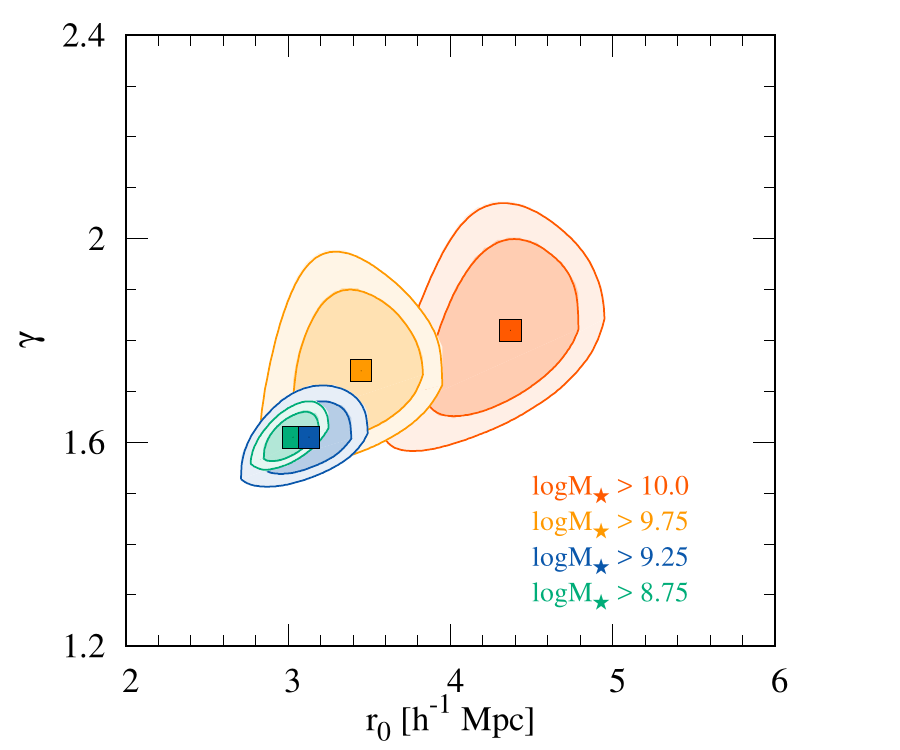}
   \end{tabular}
   }
   \caption{Projected two-point correlation function $w_p(r_p)$ associated with the best-fitting power-law function (\textit{left} side) and best-fit power-law parameters $r_0$ and $\gamma$ along with 68.3$\%$ and 95.4$\%$ joint confidence levels (\textit{right} side) in four UV absolute magnitude sub-samples (\textit{upper} panel) and four stellar mass sub-samples (\textit{lower} panel).
	 The symbols and error bars (see Sec. \ref{sec:method} for the error estimation method) denote measurements of the composite correlation function for different luminosity (circles) and stellar mass (squares) sub-samples selected from VUDS survey in the redshift range $2<z<3.5$.
	 For clarity, offsets are applied both to the data points and best-fitting curves of the $w_p(r_p)$, i.e., the values of $w_p(r_p)$ and associated best-fits for galaxy sub-samples with increasing luminosity and stellar masses have been staggered by 0.5 dex each.
	 Error contours on the fit parameters are obtained taking into account the full covariance matrix. The 68.3$\%$ and 95.4$\%$ joint confidence levels are defined in terms of the corresponding likelihood intervals that we obtain from our fitting procedure.}
 \label{fig:cf_results}
\end{figure*}

\begin{figure*}
   \centering
   \subfloat{
   \begin{tabular}{cc}
     \includegraphics[angle=270]{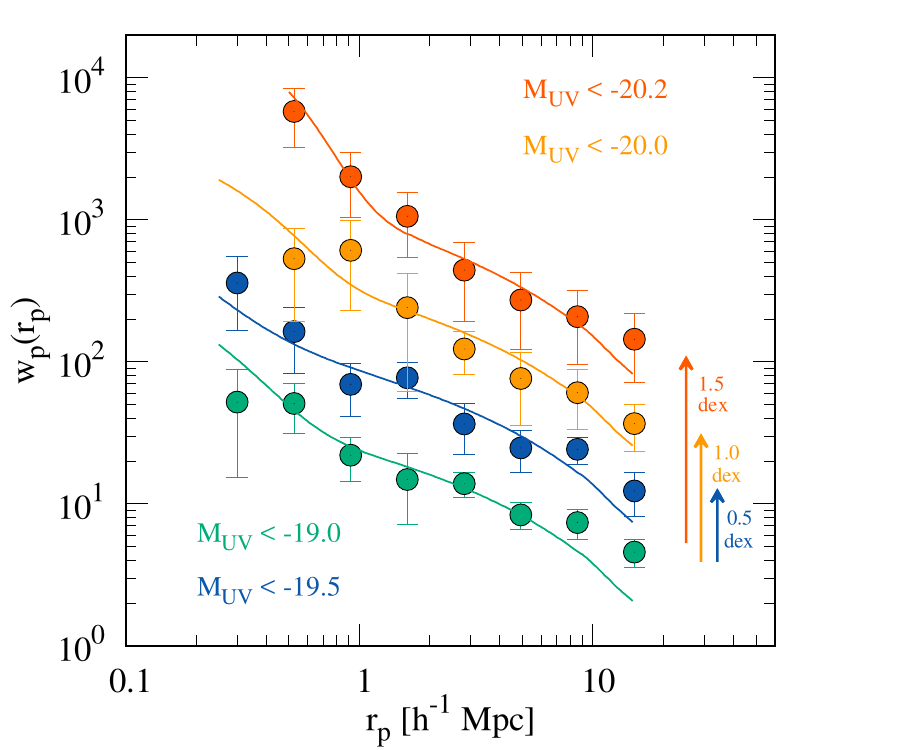} & \includegraphics[angle=270]{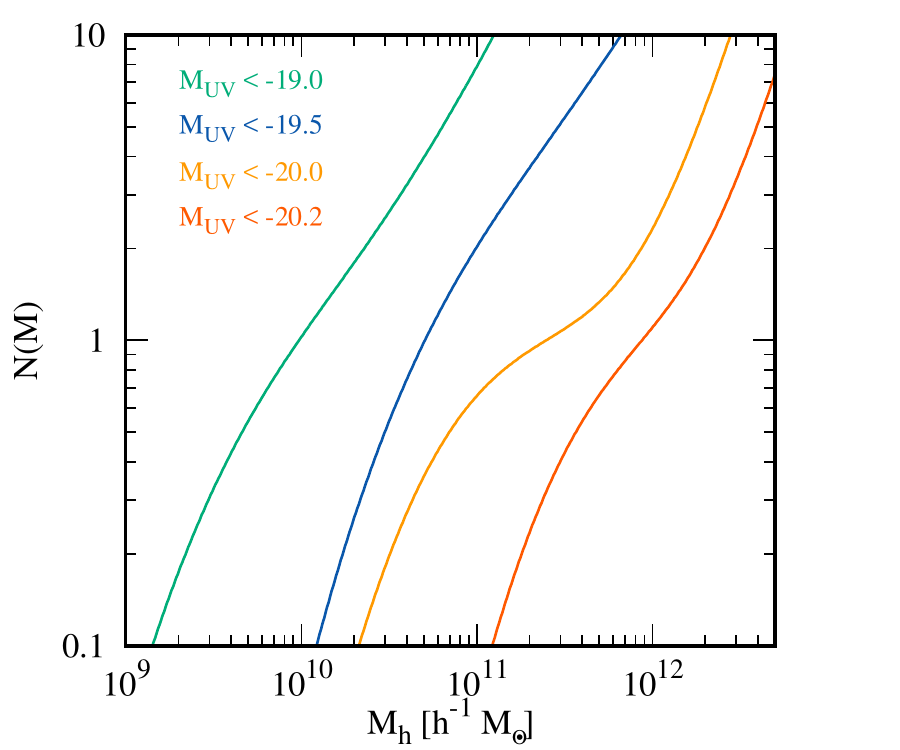} 
   \end{tabular}
   } \\ [-4ex]
   \subfloat{
   \begin{tabular}{cc}
      \includegraphics[angle=270]{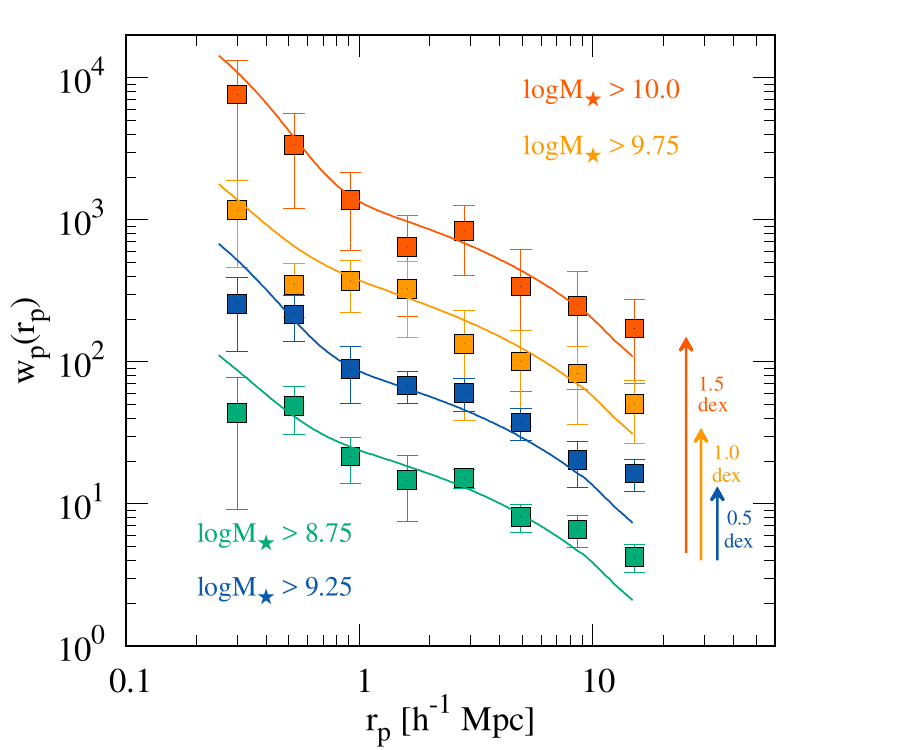} & \includegraphics[angle=270]{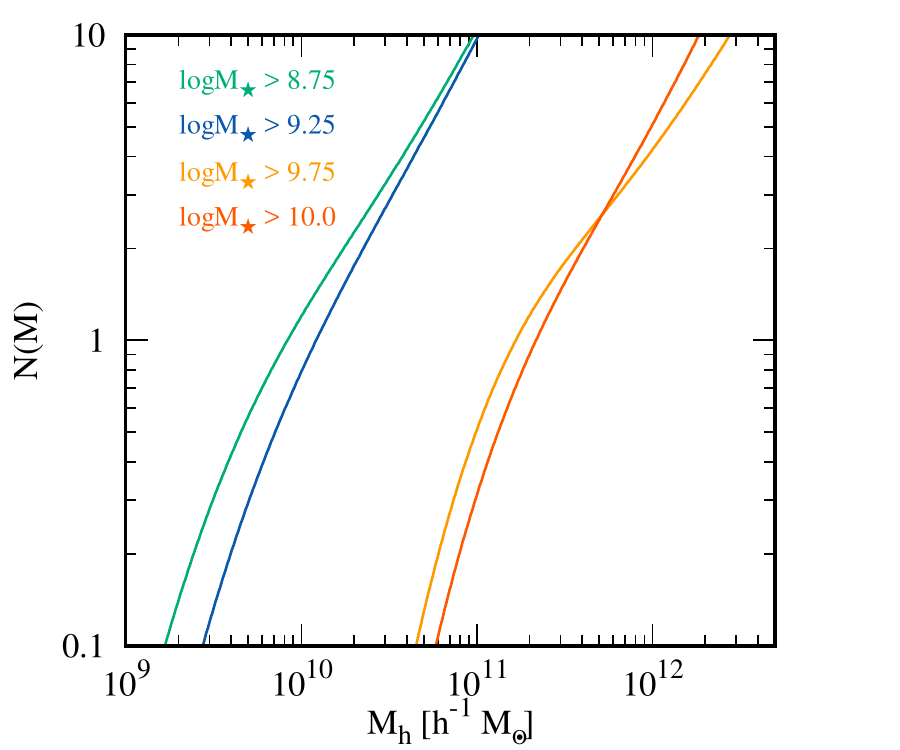}
   \end{tabular}
   }
   \caption{Projected two-point correlation function $w_p(r_p)$ associated with the best-fitting HOD models (\textit{left} side) and evolution of the halo occupation function of the best-fit HOD model (\textit{right} side) in four UV absolute magnitude sub-samples (\textit{upper} panel) and four stellar mass sub samples (\textit{lower} panel).
	 The symbols and error bars (see Sec. \ref{sec:method} for the error estimation method) denote measurements of the composite correlation function for different luminosity (circles) and stellar mass (squares) sub-samples selected from the VUDS survey in the redshift range $2<z<3.5$.
	 For clarity, offsets are applied to both the data points and best-fitting curves of the $w_p(r_p)$, i.e., the values of $w_p(r_p)$ and associated best-fits for the galaxy sub-samples with increasing luminosity and stellar masses have been staggered by 0.5 dex each.}  
	 
 \label{fig:cf_results_hod}
\end{figure*}
\setlength{\tabcolsep}{6pt} 
The two-point projected correlation function $w_p(r_p)$ has been measured in four volume-limited luminosity sub-samples and four stellar mass sub-samples selected from a total number of 3236 spectroscopically confirmed VUDS galaxies observed in the redshift range $2<z<3.5$.
The composite correlation functions (from three VUDS fields, see Sec. \ref{sec:method}) measured for each of these luminosity and stellar mass sub-samples are presented in Fig. \ref{fig:cf}, while the associated best power-law and HOD fits are shown in Fig. \ref{fig:cf_results}.

In the case of luminosity limited sub-samples the minimum scale $r_p$ that can be reliably measured varies slightly for different galaxy sub-samples. 
For the two faintest sub-samples we measure a correlation signal on scales $0.3 < r_p < 15$ $h^{-1}$ Mpc, while for the more luminous sub-samples it can be measured only on scales $0.5 < r_p < 15$ $h^{-1}$ Mpc.
We set these particular limits after performing a range of tests on correlation function measured for each of VUDS luminosity sub-samples (see Sec. \ref{sec:lum_mass_subsamples}).
The lower $r_p$ limit is set at the lowest scale for which (1) we are able to measure a correlation function signal, i.e., $w_p(r_p)$ has a positive value, and/or (2) we are able to reliably correct \citep[see ][for details about the used correction methods]{Durkalec2015a} the underestimation of the correlation function that occurs due to missing close galaxy pairs (result of the low number of galaxies in the sample and/or VIMOS limitations and positions of spectral slits).
The maximum scale limit of $r_p$ has been chosen as a result of similar tests, and under the same conditions.
This time, however, the distant galaxy pairs at large $r_p$ are missing due to the finite size of VUDS fields.

In practice, we therefore limit our measurement to scales for which the number of galaxy-galaxy pairs in VUDS data is sufficient to measure correlation function with uncertainties that do not exceed the value of $w_p(r_p)$, and are not affected by volume effects.

\subsection{Luminosity and stellar mass dependence - power-law fitting of the CF}
\label{subsec:powerlaw_fitting}
The best power-law fits of $w_p(r_p)$, parametrized with two free parameters $r_0$ and $\gamma$ (see Sec. \ref{sec:method}), are presented in the left panel of Fig. \ref{fig:cf_results}.
The best-fitting parameters for all luminosity and stellar mass sub-samples are listed in Tab. \ref{tab:parameters} and their 68.3$\%$ and 95.4$\%$ joint confidence levels are shown in the right panel of Fig. \ref{fig:cf_results}.

At redshift $z\sim3$ we observe a pronounced dependence of galaxy clustering on both luminosity and stellar mass, with the brightest and most massive galaxies more strongly clustered than their fainter and less massive counterparts (see Fig. \ref{fig:cf}).
This dependence is reflected in the increase of the correlation length $r_0$.
We find that $r_0$ rises from $r_0=2.87\pm0.22$ $h^{-1}$ Mpc for the least luminous galaxy sub-sample (with $M_{UV}^{med} = -19.84$) to $r_0=5.35\pm0.50$ $h^{-1}$ Mpc for the most luminous galaxies (with $M_{UV}^{med} = -20.56$).
This observed luminosity dependence is systematic, but it becomes more significant for the most luminous galaxies.
The correlation functions of the galaxies with increasing luminosities moving from $M_{UV}<-19.0$ to $M_{UV}<-20.0$ are very similar at scales $r_p > 2$ $h^{-1}$ Mpc, which results in a subtle increase in $r_0$ between these sub-samples (see Tab. \ref{tab:parameters}).
The rapid growth in the correlation length, by $\Delta r_0 \sim 2$ $h^{-1}$ Mpc, can be observed afterwards for the brightest galaxies ($M_{UV}<-20.2$).

A similar behaviour occurs for the galaxies selected according to their stellar masses, with the correlation length increasing from $r_0 = 3.03\pm0.18$ $h^{-1}$ Mpc for the least massive sub-sample ($\log M_{\star}^{med} = 9.48$ $h^{-1}$ Mpc) to $r_0 = 4.37\pm0.48$ $h^{-1}$ Mpc measured for the most massive ones ($\log M_{\star}^{med} = 10.24$ $h^{-1}$ Mpc).
However, in this case the change in the correlation function between sub-samples of increasing stellar mass appears to be smoother.

The second of the two free parameters, the slope $\gamma$, has also a tendency to grow with increasing luminosity and stellar mass. 
We find that for the luminosity selected sub-samples the value of $\gamma$ rises from $\gamma = 1.59\pm0.07$  for the faint galaxies to $\gamma = 1.92\pm0.25$ for the brightest ones.
Similarly, the slope of the power-law fit changes from $\gamma = 1.61\pm0.06$ to $\gamma = 1.82\pm0.20$ for the stellar mass selected sub-samples.
This increase in the value of $\gamma$ is likely related to the continuously stronger one-halo term measured for sub-samples with increasing luminosities and stellar masses, as discussed below.

\subsection{Luminosity and stellar mass dependence - HOD modelling}
\label{sec:hod_results}
In the left panel of Fig. \ref{fig:cf_results_hod} we present the measurements of the projected real-space correlation function $w_p(r_p)$ and the best-fitting HOD models for the four volume limited UV absolute magnitude (upper panel) and stellar mass (lower panel) sub-samples at redshift $z\sim3$.
As shown, for all selected galaxy samples the best-fitting HOD models reproduce the measurements of the projected correlation function well.
However, it is noticeable that in all cases there are some deviations with respect to the model, which predicts correlation function values at large scales ($r_p > 10 h^{-1}$ Mpc) lower than measured. 
Given the measurement errors, these deviations are more significant for the two least massive and least luminous sub-samples.
We verified that these deviations are mostly driven by the behaviour of the correlation function measured in the COSMOS field, the field with the most galaxies distributed over the largest area in our sample (it comprises of $\sim50\%$ of our galaxy sample, see Tab. \ref{tab:sample_numbers}), hence with a significant influence on the combined correlation function.
The flattening of $w_p(r_p)$ measured for the COSMOS field at large separations $r_p>5$ $h^{-1}$ Mpc can be explained by the presence of an extremely large structure in the COSMOS field which spans a size comparable to that covered by VUDS-COSMOS (see Appendix \ref{app:cosmic_varaince} and Cucciati et al. 2017, in prep).

In Tab. \ref{tab:parameters} we list the values of the best-fitting HOD parameters (inferred using the full error covariance matrix), with their 1$\sigma$ errors.
Similarly to what is seen at lower redshifts \citep[e.g., ][]{Zehavi2011, Abbas2010, Zheng2007} we observe a mass growth of the dark matter haloes hosting galaxies with rising luminosity and stellar mass. 
The minimum halo mass $M_{min}$, for which at least $50\%$ of haloes host one central galaxy, increases from $M_{min} = 10^{9.73\pm0.51}$ $h^{-1} M_{\sun}$ to $M_{min} = 10^{11.58\pm0.62}$ $h^{-1} M_{\sun}$ for galaxies with the median UV absolute magnitude  $M_{UV}^{med} = -19.84$ and $M_{UV}^{med} = -20.56$, respectively.
At the same time, for galaxy sub-samples selected according to stellar mass, $M_{min}$ grows from $M_{min} = 10^{9.75 \pm 0.48}$ $h^{-1} M_{\sun}$ to $M_{min} = 10^{11.23\pm0.56}$ $h^{-1} M_{\sun}$ for galaxies with $\log M_{\star}^{med} = 9.48$ $h^{-1} M_{\sun}$ to $\log M_{\star}^{med} = 10.24$ $h^{-1} M_{\sun}$.

We also observe a growth of another characteristic halo mass, $M_1$, with the luminosity and stellar mass of galaxies.
The limiting mass of dark matter halo hosting on average one additional satellite galaxy above the luminosity (or mass) threshold increases from $M_1 = 10^{10.33\pm0,74}$ $h^{-1} M_{\sun}$ for the faintest galaxy sub-sample to $M_1=10^{12.29\pm0.48}$ $h^{-1} M_{\sun}$ for the most luminous galaxies. 
Similarly, for the stellar mass selected sub-samples $M_1$ rises from $M_1 = 10^{10.21\pm0.69}$ $h^{-1} M_{\sun}$ to $M_1=10^{11.57\pm0.65}$ $h^{-1} M_{\sun}$ from the less to the most massive galaxy sub-samples, respectively.

These changes, both, of the minimum $M_{min}$ and 'satellite' $M_1$ masses of dark matter haloes hosting galaxies with different properties are in agreement with the predictions of the hierarchical scenario of structure formation as discussed in Sec. \ref{sec:halo_mass_discussion}.

Additionally, we observe an increase with luminosity of the high-mass slope $\alpha$ of the satellite occupation in the UV absolute magnitude selected galaxy sub-samples.
For the two brightest sub-samples ($M_{UV} < -20.0$ and $M_{UV}<-20.2$) $\alpha$ is noticeably higher $\alpha = 1.95\pm0.23$, than observed for the fainter galaxy populations, where $\alpha$ takes values around unity. 
This observed difference is likely related to the more pronounced one-halo term for the most luminous galaxy sample.
It indicates that satellite galaxies are more likely to occupy most massive dark matter haloes.
The situation is less clear for the stellar mass selected sub-samples, where, given the measurement uncertainties, we do not observe any significant change in the slope $\alpha$ for the four different stellar mass sub-samples. 

All these differences in the HOD parameter values measured for galaxy populations with different luminosities and stellar masses are reflected in the evolution of the halo occupation function presented in the right panels of Fig. \ref{fig:cf_results_hod}.
The halo occupation function shifts towards higher halo masses when going towards brighter and more massive galaxy sub-samples showing that more luminous and more massive galaxies occupy, respectively, more massive haloes.
For the luminosity selected sub-samples this shift of the halo occupation function is rather continuous, while for the stellar mass selected galaxies there is a rapid 1 dex increase in halo masses moving from the two least massive to the two most massive galaxy populations.

Such a rapid shift in the halo mass related to a relatively small change in the stellar mass has not been reported in the literature.
At $z\sim2$ \cite{McCracken2015}, based on the angular correlation function measurements, finds a continuous growth of both minimum (from $M_{min}\sim10^{12.4}$ $M_{\sun}$ to $M_{min}\sim10^{12.6}$ $M_{\sun}$) and 'satellite' halo masses (from $M_1\sim10^{13.45}$ $M_{\sun}$ to $M_1=10^{14.0}$ $M_{\sun}$) for galaxies with stellar masses ranging from $M_{\star}^{thresh}=10^{10.2}$ $M_{\sun}$ to $M_{\star}^{thresh}=10^{10.8}$ $M_{\sun}$.
Similarly, at $z\sim1.5$ \cite{Hatfield2016} measure a steady increase in the minimum halo mass by $\Delta \log M_{min} = 0.5$ $M_{\sun}$ for sub-samples of galaxies with stellar masses from $M_{\star}^{thresh}=10^{10.1}$ $M_{\sun}$ to $M_{\star}^{thresh}=10^{10.6}$ $M_{\sun}$.
These studies, however, do not cover stellar masses smaller than $M_{\star}^{thresh}\sim10^{10}$ $M_{\sun}$, which is the threshold limit of the most massive galaxy sub-sample used in this work.
    
The presence of the halo mass discontinuity with respect to the increasing stellar mass of galaxies and lack of such discontinuity observed for luminosity selected sub-samples suggests that the relationship between the luminosity of a galaxy and the corresponding halo mass significantly differs from the relationship between its stellar mass and the mass of the dark matter halo.
This in turn implies that the processes determining the galaxy luminosity, even if related to the evolution of the hosting halo, could be more complex than the relation between the halo and galaxy stellar mass.
    
The observed discontinuity in halo mass, with respect to small difference in stellar mass, directly influence the observed stellar-to-halo mass relation.
In particular we observe that, at $z\sim3$, low mass end of this relation deviates from the theoretical predictions by, e.g., \cite{Behroozi2013} and \cite{Moster2013}.
We discuss the possible implications of this result in more details in Sec. \ref{sec:shmr}.

\section{Discussion}
\label{sec:discussion}
\subsection{Dependency of galaxy clustering on their luminosity and stellar-mass}
\label{sec:discussion_r0}
Our most important conclusion is that at redshift  $z\sim3$ galaxy clustering depends on luminosity and stellar mass. 
As presented in Fig. \ref{fig:cf} and described in section \ref{subsec:powerlaw_fitting}, we observe a constant increase of $r_0$ from faint and low massive samples to the most luminous and the most massive ones.
This implies that at high redshift the most luminous and most massive galaxies are more strongly clustered than their fainter and less massive counterparts, with a higher clustering observed on both small and large spatial scales.

This luminosity and stellar mass dependence of galaxy clustering can be explained in the framework of the hierarchical mass growth paradigm.
In this scenario, the mass overdensities of the density field collapsed overcoming the cosmological expansion.
The initially stronger overdensities grew faster, hence their stronger clustering pattern imprinted in the dark matter density field.
With time, the resulting dark matter haloes merged together, forming larger haloes, which served as the environment where galaxies formed and evolved \citep{Press1974, White1976}.
The strongest and most clustered overdensities produced the largest haloes, containing the corresponding amount of baryons, which - in turn - agglomerated to produce the largest and the most massive (consequently also the most luminous) galaxies.
This behaviour is reflected in the N-body simulations complemented by the semi-analytical models which show that the galaxy luminosity and stellar mass are tightly correlated with the mass of their haloes.
In consequence, the clustering of a particular galaxy sample is expected to be largely determined by the clustering of haloes that host these galaxies \citep{Conroy2006, Wang2007}.

This simple picture, however, complicates when we need to take the evolution of galaxies, driven by baryonic physics, into account.
This makes more difficult to predict how exactly luminosity and stellar mass dependence of galaxy clustering changes with time.
In particular, the star formation occurs only after baryonic matter reaches a certain critical density and proceeds in a different way depending, e.g., on the initial galaxy mass, halo mass, and interactions with other galaxies \citep[see, e.g.,][]{White1978, DeLucia2007, LopezSanjuan2011, Tasca2014}.
Therefore, the evolution of luminosity and stellar mass clustering dependence is not only related to the growth of dark matter halo masses, but also to the physics of baryons that make up the galaxies.
We expect that for the most massive galaxies occupying the most massive dark matter haloes the build up of stellar mass is eventually limited by various feedback effects \citep[e.g., ][]{Blanton1999}, while the less massive galaxies occupying less massive dark matter haloes continue to form stars (downsizing, see e.g., \citealt{DeLucia2006}).
In consequence, we expect to observe a strong luminosity and stellar mass dependence of galaxy clustering at $z\sim3$ and its weakening with time.

However, the question of whether there is a differential evolution between low and high luminosity galaxies or low and high stellar mass galaxies from $z\sim3$ to $z\sim0$ remains open.
The comparisons of the strength of galaxy clustering at different redshifts are difficult. 
The clustering amplitudes observed at different epochs cannot be easily related due to the differences in the selection methods used to sample galaxies in different surveys, which in turn results in sampling different galaxy populations at different redshifts. 
Still, we find that our results - a higher clustering amplitude observed for more luminous galaxies on both small and large spatial scales $r_p$ -  are consistent with the results based on the data from low (e.g., the SDSS survey - \citealt{Zehavi2011}, \citealt{Guo2015}, the 2dF survey \citealt{Norberg2002}) and intermediate (e.g., the DEEP2 survey - \citealt{Coil2006}, the VVDS survey \citealt{Pollo2006}, \citealt{Abbas2010}, the zCOSMOS - \citealt{Meneux2009}, the VIPERS survey \citealt{Marulli2013}) redshift ranges.
For example, based on the large SDSS $z\sim0$ galaxy sample \cite{Zehavi2011} found that the correlation length increases by $\Delta r_0 \sim 6.5$ $h^{-1}$ Mpc between galaxies with $M_r < -18.0 $ and $M_r < -22.0$. 
Moreover, similarly to our work, the luminosity dependence is more pronounced for bright samples, and less significant for the fainter ones (see Sec. \ref{subsec:powerlaw_fitting}).
At intermediate redshift ranges, e.g., \cite{Marulli2013} analysing data from the VIPERS survey, found that at $z\sim1$ the correlation length increases from $r_0=4.29\pm0.19$ $h^{-1}$ Mpc to $r_0 = 5.87\pm0.43$ $h^{-1}$ Mpc for galaxies with $M_B<-20.5$ and $M_B<-21.5$, respectively.   
Consistently with these findings at lower redshifts, also at $z\sim3$, we find a $\Delta r_0 \sim 2.5$ increase between the faintest ($M_{UV}<-19.0$) and the brightest ($M_{UV}<-20.2$) galaxies and a $\Delta r_0\sim1.5$ increase between stellar mass selected sub-samples.  
As mentioned at the beginning of this paragraph, due to the fact that all these measurements possibly consider different galaxy populations, we are not able to draw a detailed conclusion whether or not luminosity and stellar mass clustering dependence is stronger (or weaker) at high redshift in comparison to the local universe.
What can be safely said, however, is that dependence of clustering with luminosity and stellar mass is present and strong at $z\sim3$, as it is observed at $z\sim0$ ($\Delta r_0$ of the same order of magnitude at both redshifts), and therefore much of the processes which produced luminosity and stellar mass clustering dependence must have been at work at significantly higher redshift than $z\sim3$.

\subsection{The relative and large scale galaxy bias of different luminosity and stellar mass sub-samples}
\label{sec:bias}
\begin{figure}
\centering
 \includegraphics[angle=270]{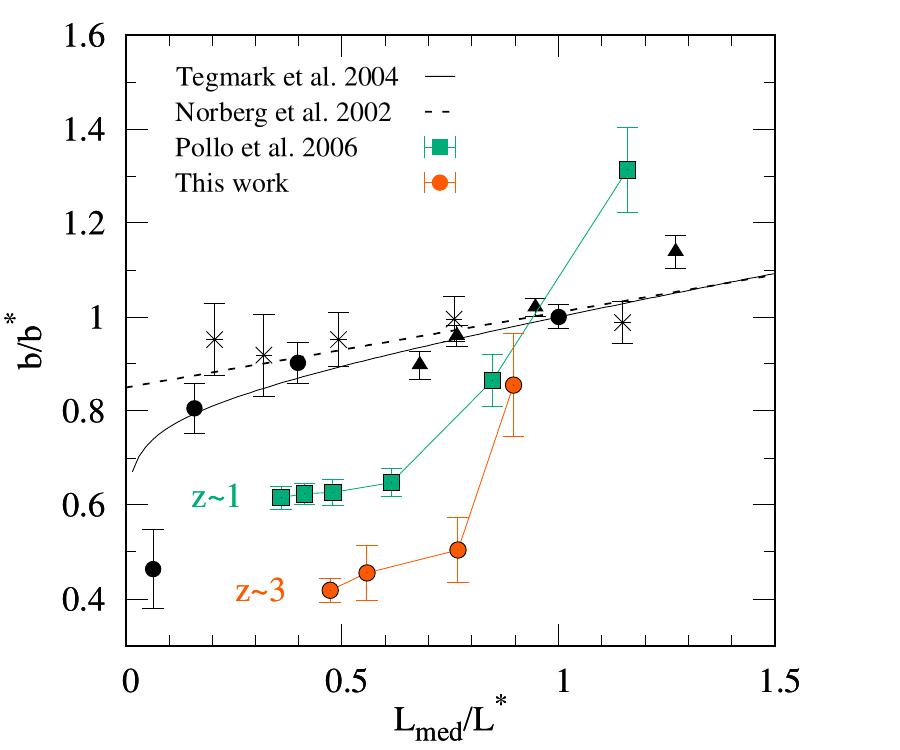}
 \caption{The relative bias $b/b^*$ (see Eq. \ref{eq:rel_bias}) for the selected VUDS luminosity sub-samples at $z\sim 3$ (orange circles) as a function of luminosity, with $L^*$ as a reference point.
	  The results from this work are compared to similar studies at lower redshift ranges: at $z\sim0.1$ from \cite{Zehavi2011} (filled black circles) and \cite{Norberg2002} (black crosses), and at $z\sim0.9$ from \cite{Pollo2006} (green circles).
	  The lines indicate the analytic fit of the 2dFGRS data from \cite{Norberg2002} (black solid line) and SDSS data from \citep{Tegmark2004} (black dashed line), as described in the text.} 
	  
  \label{fig:rel_bias}
\end{figure}
\setlength{\tabcolsep}{-2pt}
\begin{figure*}
 \centering
 \begin{tabular}{cc}
     \includegraphics[angle=270]{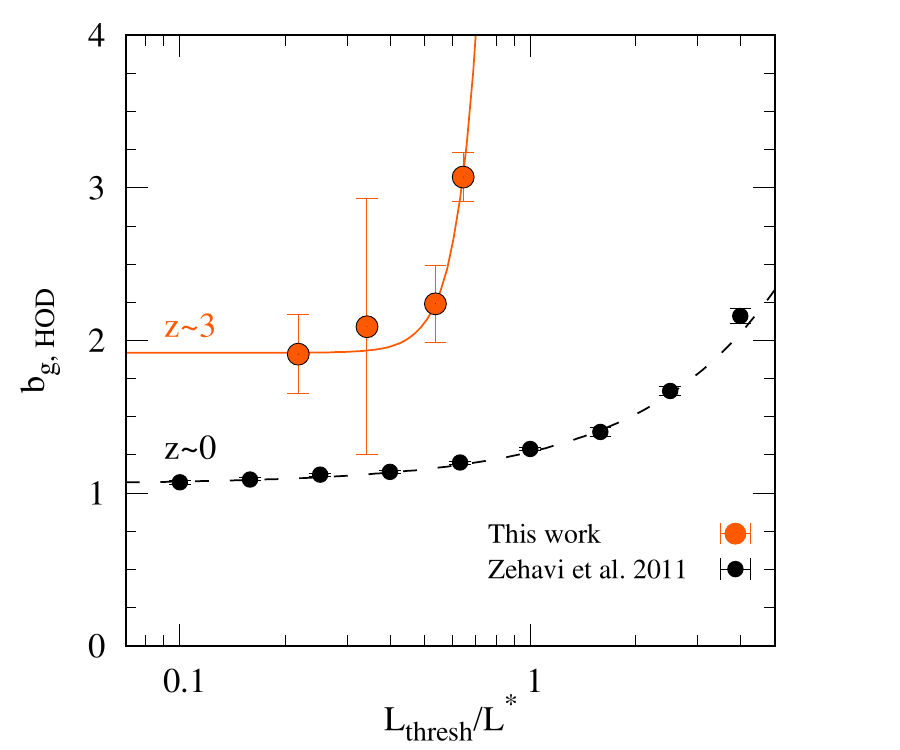} & \includegraphics[angle=270]{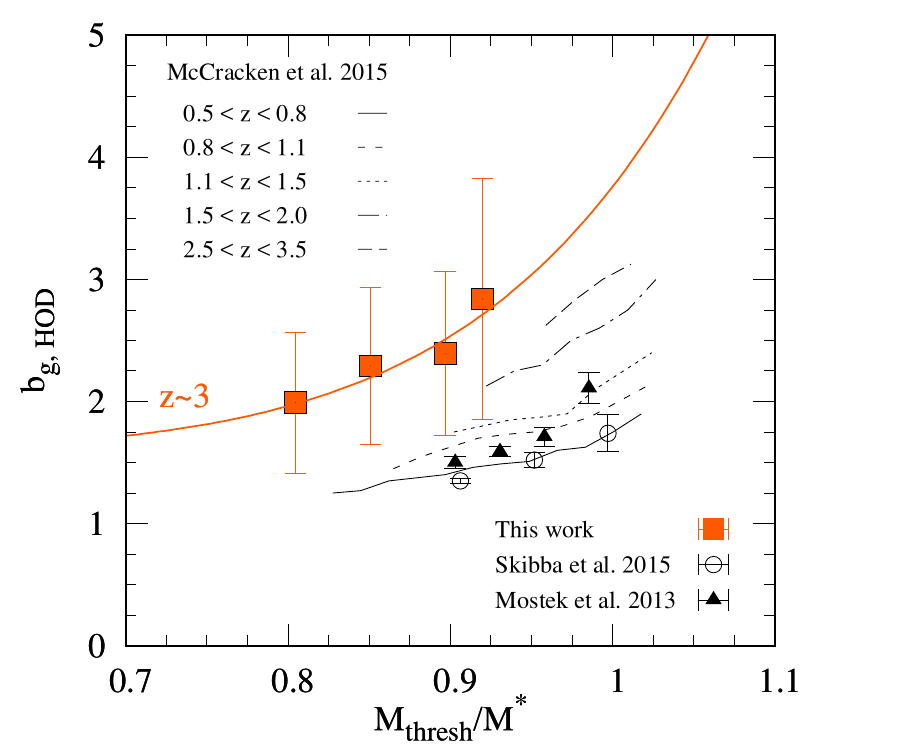}  
   \end{tabular}
   \caption{Large scale galaxy bias $b_{g,HOD}$ as a function of luminosity, with $L^*$ as a reference point ({\it left} panel) and as a function of stellar mass, with $M^*$ as a reference point ({\it right} panel).
	    In both plots the coloured points indicate measurements at $z\sim3$ from this work and the solid lines are the best-fit functions of the bias-luminosity dependence $b_{g,HOD}(>L)$ (Eq. \ref{eq:bias_lum}) and bias-stellar mass dependence $b_{g,HOD}(>M)$ (Eq. \ref{eq:bias_mass}) as described in Sec. \ref{sec:bias}. 
	    In the left panel the black circles show the results from \cite{Zehavi2011} at $z\sim0$ and the dashed line represents the empirical fit of the bias-luminosity dependence in the functional form given therein.
	    In the right panel we show for comparison the $b_g$ measurements at $z\sim0.5$ from \cite{Skibba2015}, represented by open circles, at $z\sim1$ by \cite{Mostek2013}, indicated by black triangles and from different samples in the redshift range $0.5<z<2.5$ measured by \cite{McCracken2015} shown by black lines as labelled.  
	    }
   \label{fig:bias_hod}
\end{figure*}
Using the best-fitting power-law parameters $r_0$ and $\gamma$ we interpret our results in terms of the relation between the distribution of galaxies and the underlying dark matter density field for galaxy populations with different luminosities. 
We compare the values of the relative galaxy bias $b/b^*$ measured from the VUDS survey to the bias of galaxy populations with different luminosities measured at lower redshift ranges, taken from the literature.

The relative bias parameter, $b/b^*$, is based on the amplitude of the correlation function relative to that of $L^*$ galaxies and can be defined as the relative bias of the generic \textit{i}th sample with a given median luminosity $L_{med}$, with respect to that corresponding to $L^*$, as
\begin{equation}
 \frac{b_i}{b^*} = \sqrt{\frac{\left(r_0^i\right)^{\gamma_i}}{\left(r_0^*\right)^{\gamma^*}}r^{\gamma^*-\gamma_i}}.
 \label{eq:rel_bias}
\end{equation}
In our study we use a fixed scale $r = 1 h^{-1}$ Mpc \citep[see also][ for a slightly different definition]{Meneux2006}.
To apply this formula, first we need to estimate the values of $r_0^*$ and $\gamma^*$ for $M^*_{UV}$ galaxies. 
We obtain them through a linear fit to the relation between correlation length and absolute magnitude of the sample normalised to the characteristic absolute magnitude at median redshift $r_0(M_{UV}-M^*_{UV})$ and $\gamma(M_{UV}-M^*_{UV})$ measured in this work.

Fig. \ref{fig:rel_bias} shows the relative bias measured for the VUDS galaxies with the luminosities sampled at $z\sim3$ compared to various results at lower redshifts, along with the analytic fit of the 2dFGRS data $b/b^* = 0.85+0.16L/L^*$ from \cite{Norberg2002} and  $b/b^* = 0.85+0.15L/L^* - 0.04(M-M^*)$ based on the SDSS sample \citep{Tegmark2004}.    

In each luminosity sub-sample the relative bias at $z\sim3$ of galaxies with $L_{med}/L^* < 1$ is significantly lower than the one observed at lower redshifts for galaxies with similar $L_{med}/L^*$ ratios.
However, none of our sub-samples has $L_{med}>L^*$, thus we cannot exclude the possibility that for galaxies with $L_{med}>L^*$ the relative bias would be higher, which is very likely, taking into account the trend visible in Fig. \ref{fig:rel_bias}.
Additionally, we observe that the value of $b/b^*$ rises more steeply with $L_{med}/L^*$ for high redshift galaxies than observed locally.
At $z\sim3$ the relative bias increases from low values $b/b^* = 0.42\pm0.03$ at low luminosities to $b/b^*=0.85\pm0.11$ for the high luminosity sub-sample. 
\cite{Pollo2006} found a similar steep growth of the relative bias for galaxies observed at $z\sim1$.
At $z\sim0$, instead, $b/b^*$ increases only by $\sim0.1$ in the same $L_{med}/L^*$ interval, following the model from \cite{Norberg2002}. 
This appears to be an indication that going back in time the bias contrast of the most luminous galaxies with respect to the rest of the population becomes stronger and is consistent with the fact that fainter galaxies are found to be significantly less biased tracers of the mass than brighter galaxies even at high redshifts.
However, we need to take into account the possibility that the observed strengthening of $b/b^*$ relation with luminosity at higher redshifts can also be partially attributed to a more pronounced one-halo term at higher $z$ making the power-law fit of the clustering measurement less reliable. 

In order to break this ambiguity, we use the best-fitting parameters of the HOD model to estimate the large scale galaxy bias $b_{g,HOD}$, using Eq. \ref{eq:bias}.
The results obtained for the luminosity and stellar mass sub-samples at $z\sim3$ are given in Tab. \ref{tab:parameters} and presented in Fig. \ref{fig:bias_hod}, where for comparison we plot also the results obtained at lower redshifts.
For both, the UV absolute magnitude and stellar mass selected sub-samples, the values of $b_{g,HOD}$ measured at $z\sim3$ are significantly higher than locally, indicating that in the early stages of evolution galaxies are highly biased tracers of the underlying dark matter density field.
As shown in the right panel of Fig. \ref{fig:bias_hod}, the galaxy bias decreases systematically with cosmic time for all stellar masses extending to $z>3$ the trend found at lower redshifts \citep[e.g.,][]{McCracken2015}. 
The observed decrease in the galaxy bias with cosmic time can be explained in terms of the hierarchical scenario of structure formation. 
At early epochs the first galaxies are expected to form in the most dense regions, resulting in a high bias with respect to the underlying average mass density field.
As the mass density field evolves with time, these regions grow in size and mass, the gas trapped inside becomes too hot to collapse, effectively preventing the formation of new stars \citep[e.g., ][]{Blanton1999} and resulting in galaxy formation systematically moving to less dense, hence less biased, regions.

In addition to the redshift dependence of galaxy bias and in agreement with previous studies at lower redshifts \citep[e.g., ][]{Norberg2002, Tegmark2004, Meneux2008, Zehavi2011, Mostek2013} we observe a clear luminosity and stellar mass bias dependence, with the brightest and most massive galaxies being the most biased ones.

\setlength{\tabcolsep}{-2pt}
\begin{figure*}[t!]
 \centering
 \begin{tabular}{cc}
     \includegraphics[angle=270]{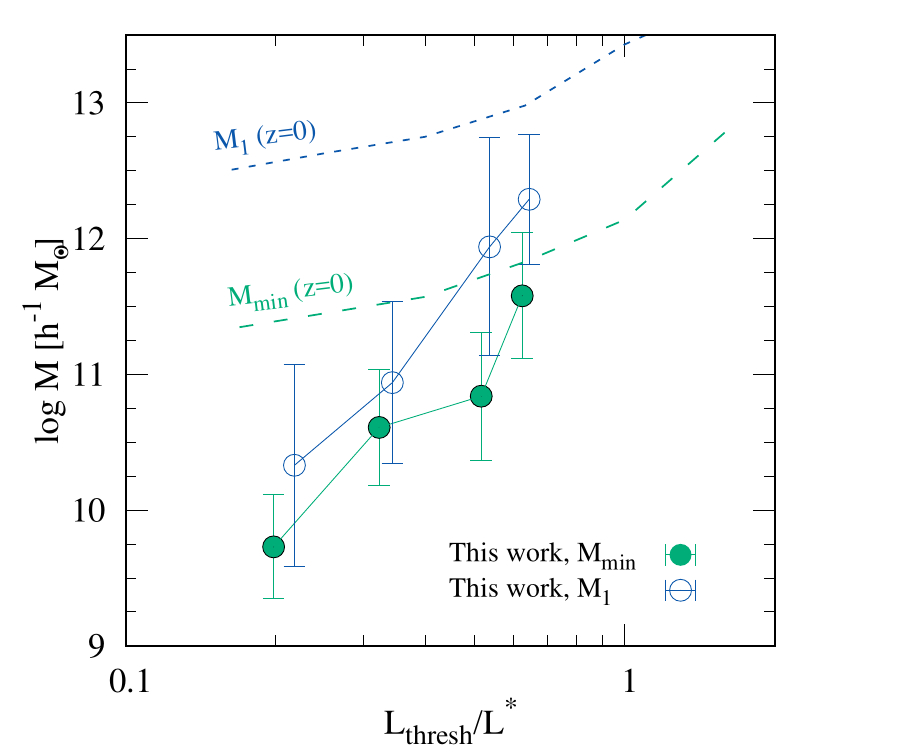} & \includegraphics[angle=270]{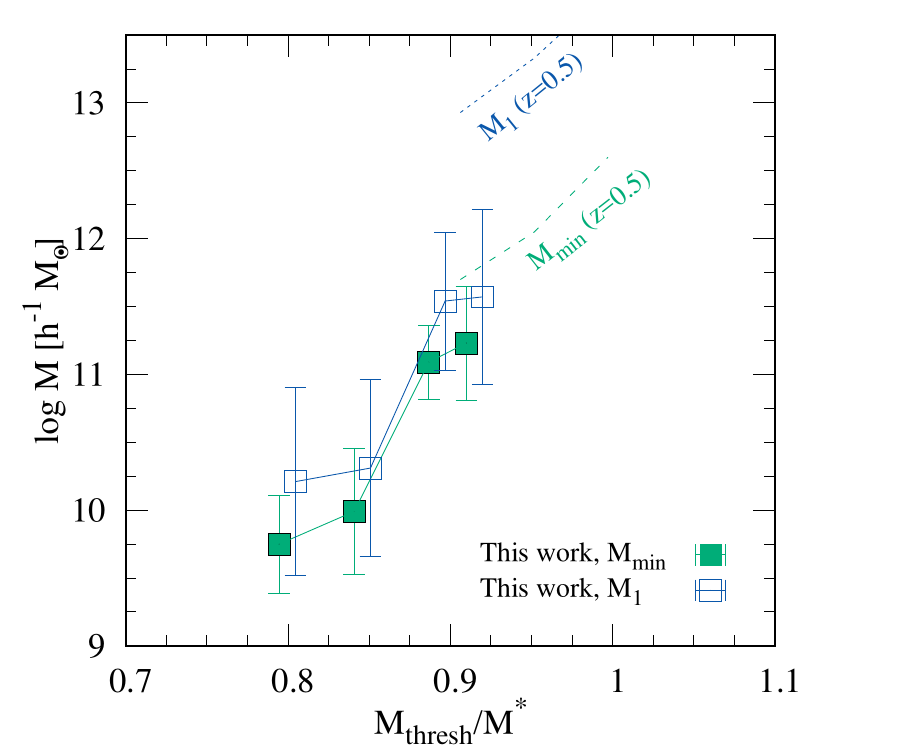} 
   \end{tabular}
   \caption{Characteristic halo masses from the best-fitting HOD models of the correlation function selected in luminosity versus $L_{thresh}/L^*$ (\textit{left} panel) and selected in stellar mass versus $M_{thresh}/M^*$ (\textit{right} panel). 
	  Minimum halo masses $M_{min}$ for which $50\%$ of haloes host one central galaxy above the threshold limit (filled symbols) and masses of haloes which on average host one additional satellite galaxy $M_1$ (open symbols) observed at $z\sim3$ are compared with similar results found by \cite{Zehavi2011} at $z\sim0$, for the luminosity selected galaxies, and by \cite{Skibba2015} at $z\sim0.5$, for the stellar mass selected galaxies (dotted and dashed lines).}
 \label{fig:mass_comparison}
\end{figure*}

In the left panel of Fig. \ref{fig:bias_hod} we show the large scale galaxy bias $b_{g,HOD}$ as a function of luminosity and compare it with the similar results from \cite{Zehavi2011}.
As presented, at $z\sim0$ the luminosity dependence of bias is nearly flat for galaxies with luminosities $L\leqslant L^*$ and then rises at brighter luminosities.
According to \cite{Zehavi2011} this relation is best fitted by the functional form $b_g(>L)\times(\sigma_8/0.8) = 1.06+0.21(L/L^*)^{1.12}$.
We adopt a similar formula to model the galaxy bias-luminosity relation, and at $z\sim3$ we find that for the luminosity threshold samples $b_{g,HOD}(>L)$ is best fitted by
\begin{equation}
 b_{g,HOD}(>L) = 1.92+25.36(L/L^*)^{7.01}
 \label{eq:bias_lum}
\end{equation}
represented by a solid line in the left panel of Fig. \ref{fig:bias_hod}. 
Here $L$ is the UV luminosity and $L^*$ corresponds to the characteristic absolute magnitude $M_{UV}^*(z=3)$ obtained as described in Appendix \ref{app:ev_correction}.
Our estimate of the dependence of the large scale bias on galaxy luminosity is nearly flat for galaxies with luminosities $L\leqslant 0.5L^*$ and rises very sharply for brighter ones. 
Therefore, in agreement with the analysis of the relative bias discussed above, this suggests that the bias contrast between bright and faint galaxies becomes stronger when going back in time.

In the right panel of Fig. \ref{fig:bias_hod} we also present the large scale galaxy bias measurements for the stellar mass selected sub-samples. 
We compare our results with the similar measurements at $z\sim0.5$ from \cite{Skibba2015} (open circles), at $z\sim1$ from \cite{Mostek2013} (filled triangles), and from \cite{McCracken2015} over the redshift $0.5<z<3.5$ (black lines), based on the large PRIMUS, DEEP2 and UltraVISTA galaxy samples respectively. 
In addition to the $b_g$ values at $z\sim3$ being higher than observed at lower redshifts (discussed earlier), we find that the galaxy bias rises toward more massive galaxies from $b_{g,HOD}=1.99\pm0.58$ measured for galaxies with $M_{med} = 10^{9.48}$ $h^{-1} M_{\sun}$ to $b_{g,HOD}=2.84\pm0.99$ for the most massive galaxy sub-sample with $M_{med} = 10^{10.24}$ $h^{-1} M_{\sun}$.
These galaxy bias values are also in excellent agreement with measurements based on N-body simulations performed by \cite{Chiang2013}, who at $z=3$ find $b_g=2.24$ and $b_g=2.71$ for galaxies with stellar masses $M>10^9$ $M_{\sun}$ and  $M>10^{10}$ $M_{\sun}$, respectively.
Like for the luminosity selected galaxies, we made an attempt to model  this bias-stellar mass relation at $z\sim3$.
We find that the best fitting function, represented in Fig. \ref{fig:bias_hod} by a solid line, is given by

\begin{equation}
 b_{g,HOD}(>M) = 1.59+2.17(M/M^*)^{7.88},
 \label{eq:bias_mass}
\end{equation}
where $M$ is the galaxy stellar mass and $M^*$ is the characteristic stellar mass at $z\sim3$.

\begin{figure*}
 \centering
 \begin{tabular}{cc}
     \includegraphics[angle=270]{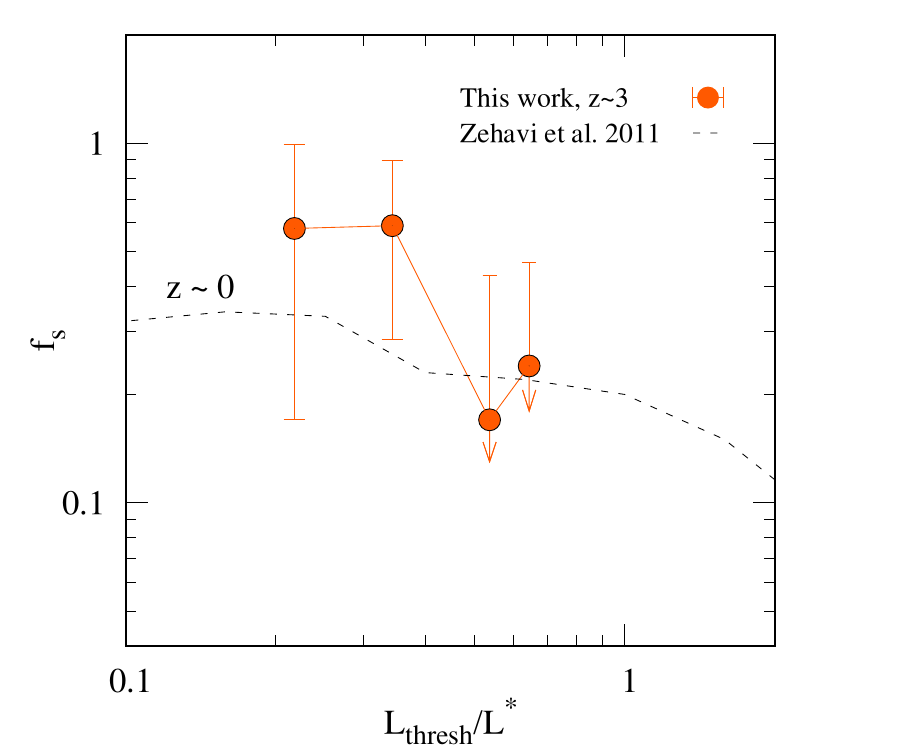} & \includegraphics[angle=270]{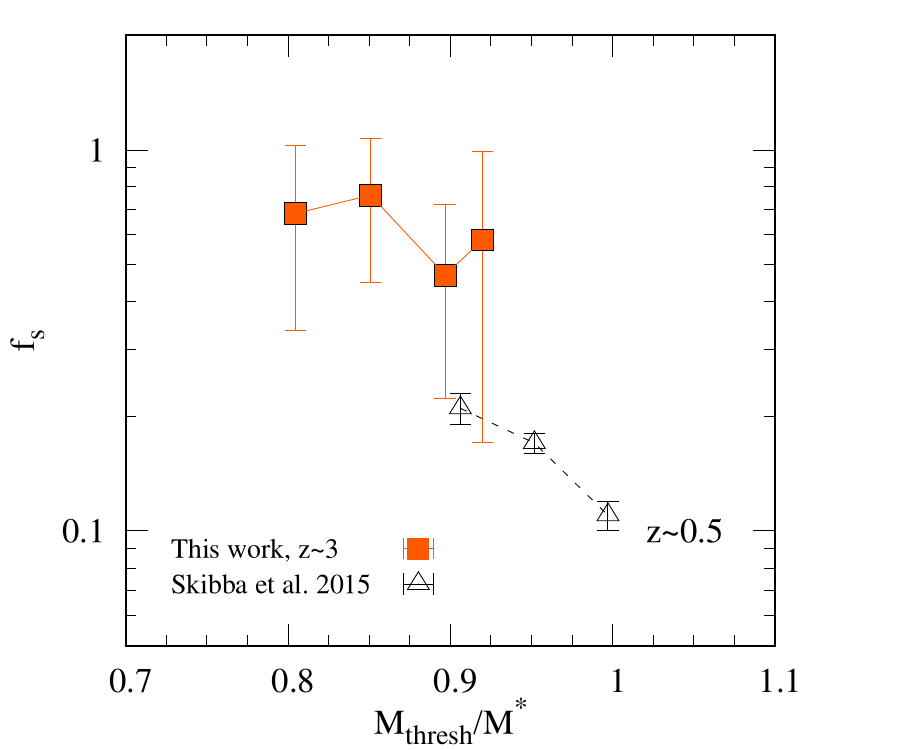} 
   \end{tabular}
   \caption{Satellite fraction $f_s$ as a function of threshold luminosity, with $L^*$ as a reference point (\textit{left} panel) and as a function of threshold stellar mass, with $M^*$ as a reference point (\textit{right} panel).
	    Results obtained in this work at $z\sim3$ (filled symbols) are compared with similar measurements from lower redshift ranges. 
	    In the left panel the dashed line marks satellite fraction as measured at $z\sim0$ by \cite{Zehavi2011}, while in the right panel results found by \cite{Skibba2015} at $z\sim0.5$ are shown with open triangles.}
 \label{fig:satellite_fraction}
\end{figure*}

\subsection{Halo masses of different galaxy populations}
\label{sec:halo_mass_discussion}
\setlength{\tabcolsep}{6pt}
In Fig. \ref{fig:mass_comparison} we show the values of two characteristic halo masses, $M_{min}$ and $M_1$, in terms of the sample threshold luminosity (left panel) and stellar mass (right panel) relative to the characteristic luminosity and stellar mass, $L_{thresh}/L^*$ and $M_{thresh}/M^*$ respectively, at different redshifts.
The minimum halo mass needed for half of the haloes to host one central galaxy above the luminosity or stellar mass threshold $M_{min}$ (filled symbols) and the mass of haloes with on average one additional satellite galaxy above the luminosity or stellar mass threshold $M_1$ (open symbols), measured at $z\sim3$ are compared with similar results at $z\sim0$ from \cite{Zehavi2011}, represented by dashed and dotted lines, respectively.
As shown, the values of both $M_{min}$ and $M_1$ at $z\sim3$ for all galaxy luminosities are lower than measured in the local universe.  
This observation suggests that in order to host at least one central galaxy, above the luminosity or stellar mass threshold, the dark matter haloes at low redshift need to accumulate a larger amount of mass than is seen at higher redshifts.

In Sec. \ref{sec:hod_results} we noted that the minimum halo masses grow with increasing luminosity and stellar mass of the galaxy sample.
Similar growth is reported at lower redshifts \citep[e.g., ][]{Zheng2007, Abbas2010, Zehavi2011, Coupon2012, delaTorre2013}; however, as presented in Fig. \ref{fig:mass_comparison}, at $z\sim3$ the contrast between halo masses of faint and bright galaxies is much larger than observed in the local universe for galaxies with similar $L_{thresh}/L^*$.
This implies that at high redshift the bright and most massive galaxies are much more likely to occupy the most massive dark matter haloes.
Combining this with the earlier observation that a lower mass dark matter halo is needed to host a galaxy of higher luminosity/stellar mass at higher redshift, suggests that the processes responsible for the following increase of the mass of the halo and the stellar mass of the galaxies operate on different timescales and are both stellar mass and epoch dependent.

With the increasing $M_{min}$ we observe a proportional growth of $M_1$. 
At all luminosities the values of $M_{min}$ and $M_1$ present an approximately constant ratio $M_1/M_{min}\approx4$ .
This indicates that at $z\sim3$ the halo hosting one central and one satellite galaxy (above a luminosity threshold) needs to be only 4 times more massive than the halo which hosts only one central galaxy.
For the stellar mass selected sub-samples this factor is even smaller $M_1/M_{min}\approx2.5$.
For comparison, in the local universe the ratio between $M_1$ and $M_{min}$ is higher. 
At $z\sim0$ \cite{Zehavi2011} find the scale factor $M_1\approx17M_{min}$ for the SDSS galaxies selected by their r-band absolute magnitude, while at intermediate redshift $z\sim1$ \cite{Zheng2007} and \cite{Skibba2015} observe a slightly lower factor of $\approx15$ and \cite{McCracken2015} report values of $\approx 10$ for galaxies at $1.5<z<2.0$.
These results, combined with our observations at $z\sim3$, can be interpret as evidence that (1) at higher redshifts, dark matter haloes consist of many recently accreted satellites, and (2) the $M_1/M_{min}$ ratio evolves with redshift, with smaller values observed at higher redshifts, which is in agreement with other studies \citep[e.g., ][]{delaTorre2013, Skibba2015, McCracken2015} and can be explained by the relation between halo versus galaxy merging \citep[]{Conroy2006, Wetzel2009}.
The dark matter halo mergers create an infall of satellite galaxies onto a halo, while the galaxy major mergers destroy them.
If the halo mergers occur more often than the galaxy mergers, we can expect a large satellite population, resulting in a small $M_1/M_{min}$ ratio.  

According to the high-resolution N-body simulations performed by \cite{Wetzel2009}, at $z>2.5$ the merger rate of subhaloes (effectively galaxies) is significantly lower than that of haloes. 
For example, at $z\sim3$ haloes of mass $M_h \approx 10^{12}$ h$^{-1}$ $M_{\sun}$ are expected to experience $\sim0.9$ mergers/Gyr, compared to only $\sim0.4$ mergers/Gyr expected for sub-haloes \citep[based on a preliminary VUDS sample][find an even lower value of major galaxy mergers, $0.17$ mergers/Gyr]{Tasca2014}. 
This implies that at $z>2.5$ the satellite galaxies are created faster than they are destroyed.
Moreover, the halo versus galaxy merger ratio decreases with time and at $z<1.6$ the two merger rates are approximately the same.
Therefore, there is an expected rapid rise in the satellite halo occupation at redshifts higher than $z\sim2$ and its slow levelled evolution afterwards.
Simulation predictions from \cite{Wetzel2009} are indirectly confirmed by our measurements.
The high number of satellites per halo at high redshift is reflected in the small ratio of $M_1/M_{min}$, while a smaller halo occupation at lower redshift corresponds to its increase with time.

From the observational side the galaxy major merger rate has been shown to rapidly rise from $z\sim0$ to $z\sim1.5$ \citep[e.g., ][]{deRavel2009,LopezSanjuan2011,LopezSanjuan2013} and to decrease for higher redshifts $z>2$ \citep{Tasca2014}.
This indicates that the peak of galaxy merging activity occurred around $z\sim1.5-2$ \citep[see also][]{Conselice2008}, hence later than the lower redshift limit of our galaxy sample.
These observational results combined with large scale N-body simulations predictions, mentioned earlier, might explain (1) the observed low value of $M_1/M_{min}$ at $z\sim3$ and (2) its increase with cosmic time.

\subsection{Satellite fraction}
We compute the fraction of satellite galaxies per halo $f_s$ for all luminosity and stellar mass sub-samples using the HOD best-fitting parameters (Eq. \ref{eq:satellite_fraction}).
The results, as a function of threshold luminosity, with $L^*$ as a reference point (left panel) and threshold stellar mass, with $M^*$ as a reference point (right panel), are shown in Fig. \ref{fig:satellite_fraction}. 
We compare our measurements at $z\sim3$, represented by filled symbols, with similar results obtained at $z\sim0$ by \cite{Zehavi2011} (left panel, dashed line) and at $z\sim0.5$ by \cite{Skibba2015} (right panel, open triangles).

These results have implications for satellite abundances as a function of luminosity and stellar mass, as well as a function of redshift.
At $z\sim3$ we observe a luminosity dependence of satellite abundance.
The satellite fraction drops from $\sim60\%$ for the faintest galaxy population to $\sim20\%$ for the brightest ones. 
A smaller value of $f_s$ for the brightest galaxies does not necessarily mean that there are no other satellite galaxies occupying a dark matter halo, but rather that there are no bright satellite galaxies.
Therefore, our results would suggest that, at high redshift it is more probable that a dark matter haloes host faint satellite galaxies, rather than very bright ones. 
A similar, however less steep, trend is present in the local universe \citep{Zehavi2011}.
For galaxies selected according to their stellar mass the situation is less clear. 
Taking into account the uncertainties of our measurement we are not able to determine if $f_s$ changes with the stellar mass of galaxies, as observed at lower redshift ranges \citep{Skibba2015}. 
At face value our data suggest the possible presence of a small drop, by $\Delta f_s \sim0.1$, from the least massive to the most massive galaxies, but it is not a significant change (at the level of $0.5\sigma$).

From the perspective of the redshift evolution, we observe that the satellite fraction of the two faintest galaxy sub-samples and of all the stellar mass selected galaxy sub-samples is higher at $z\sim3$ than it is observed at lower redshift.
This means that at high redshift it is more likely that a halo hosts a satellite galaxy above a given threshold limit, than locally.
This high satellite abundance observed for star-forming galaxies with $L\sim L^*$ at high redshift can be explained using the same reasoning as presented in Sec. \ref{sec:halo_mass_discussion}.
It suggests that the infall of the satellite galaxies, as a result of halo mergers, onto a dark matter halo is faster than their destruction via galaxy major mergers \citep{Wetzel2009}.
Therefore, the subhaloes that form after halo mergers are likely to remain intact and this leads to a large number of satellite galaxies at high redshift, resulting in the measured high satellite fraction.
It is necessary to mention, however, that this conclusion applies to star forming galaxies, with $L\sim L^*$, as the used data sample does not include a population of faint galaxies at $z\sim3$.

\subsection{The stellar to halo mass relation for low mass galaxies}
\label{sec:shmr}

\begin{figure}
\centering
 \includegraphics[angle=270]{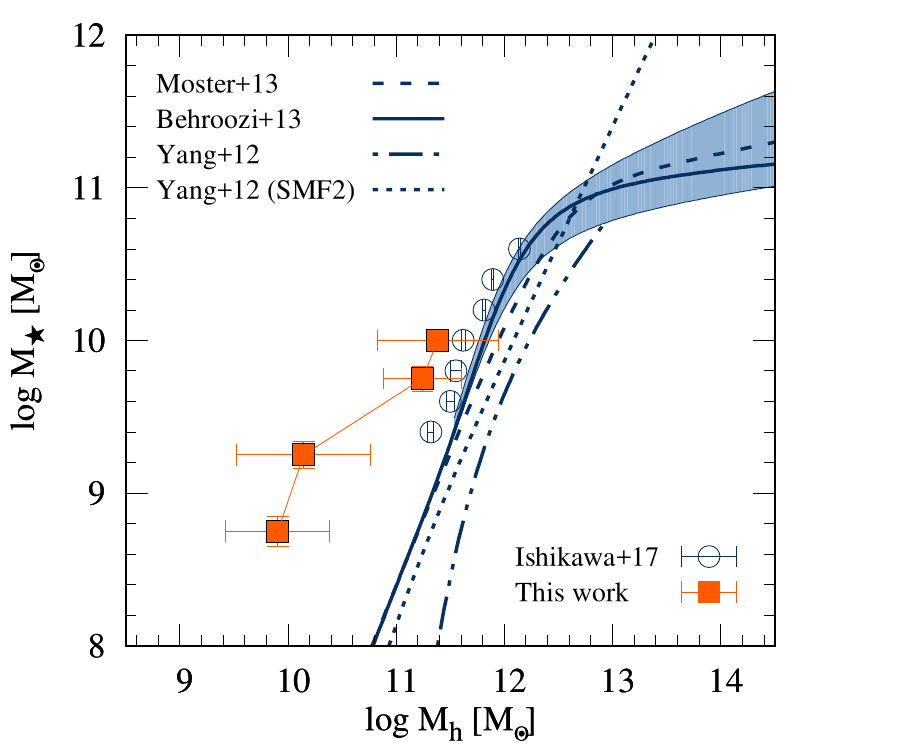}
 \caption{Stellar mass - halo mass relation (SHMR) of central galaxies obtained for different stellar mass selected sub-samples at $z\sim3$ (orange symbols).
	  The halo masses are represented by the best-fit parameter $M_{min}$, while the associated stellar masses of the galaxies are represented by the threshold limits $M_{\star}^{thresh}$ of each sub-sample.
	  The measurements from this work are compared with the results based on the $z=3$ LBGs sample from \cite{Ishikawa2017}.
	  We also plot the $z = 3$ model predictions by \cite{Behroozi2013}, \cite{Moster2013} (abundance matching), and \cite{Yang2012} (correlation function HOD modelling) represented by different lines, as labelled.
	  \cite{Yang2012} paper includes best-fit SHMR models for two different stellar mass functions and we plot both of them.
	  The blue shaded area corresponds to the 68$\%$ confidence limits of \cite{Behroozi2013}.
	  } 
  \label{fig:shmr}
\end{figure}

In this section we focus on the relationship between halo mass and stellar mass of each galaxy sample, in the literature simply referred to as the stellar-to-halo mass relation \citep[SHMR, see e.g.,][]{Behroozi2010, Behroozi2013, Moster2013, Leauthaud2012, Yang2012, Durkalec2015b}.

In Fig. \ref{fig:shmr} we present the SHMR at $z\sim3$ for all stellar mass sub-samples used in this paper (filled squares).
Due to the construction of the sub-samples (threshold limited) and the halo occupation model used, we plot the parameter $M_{min}$ as the one that represents the halo mass associated with the threshold stellar masses $M_{\star}^{thresh}$ of the galaxy sub-samples.
The errors associated with the stellar mass threshold limit are computed as the average of the errors on $M_{\star}$ for each stellar mass sub-sample separately.

We compare our results with the $z=3$ theoretical model predictions by \cite{Behroozi2013} and \cite{Moster2013}, which both use the abundance matching method to infer stellar-to-halo mass relation, and with models by \cite{Yang2012}, which are based on galaxy clustering and HOD modelling.
We find that, for the massive galaxies, with stellar masses $M_{\star} > 10^{9.75}$ $M_{\sun}$ our results are in agreement with these models.
However, for galaxies with low stellar masses ($M_{\star} < 10^{9.25}$ $M_{\sun}$), there is a striking difference between our $z\sim3$ measurements of SHMR and the theoretical model predictions.  
For these galaxies all models predict significantly more massive (by 1 dex) dark matter haloes than inferred from our measurements.
For instance, we estimate haloes of $M_h=10^{9.75}$ $M_{\sun}$ hosting galaxies with minimum stellar masses of $M_{\star}^{thresh}=10^{8.75}$ $M_{\sun}$, while model predictions by \cite{Behroozi2013} place the same galaxies in much more massive haloes of $M_h\sim10^{11}$ $M_{\sun}$.
In other words, we observe that the low-mass galaxies at $z\sim3$ have formed stars more efficiently than it is expected from these models, that all assume a much steeper decrease of the effective star formation with decreasing halo mass. 

Such discrepancies between model predictions and observational constraints at high redshift have not been reported before in the literature.
E.g., in our previous studies \citep{Durkalec2015b} based on the preliminary VUDS observations and for sub-samples covering a wider redshift range ($2.0 < z < 5.0$), and higher stellar masses, we found the SHMR in broad agreement with theoretical model predictions.
At $z\sim2$ for numerous stellar mass sub-samples \cite{McCracken2015} compared the HOD based SHMR measurements with the abundance matching based models and found them to be in broad agreement.
Similarly, at $z=3$, \cite{Ishikawa2017} reports an excellent agreement of their SHMR measurements with the model predictions by \cite{Behroozi2013} for a large Lyman break galaxy (LBGs) sample.
It is important to note, however, that galaxies used in these studies do not reach the stellar mass range below $M_{\star}=10^{9.1}$ $M_{\sun}$ \footnote{$M_{\star}=10^{9.1}$ $M_{\sun}$ in \cite{Durkalec2015b}, $M_{\star}=10^{9.4}$ $M_{\sun}$ in \cite{Ishikawa2017} and $M_{\star}=10^{10}$ $M_{\sun}$ in \cite{McCracken2015}}, while the stellar mass limit of our least massive sub-sample is significantly smaller ($10^{8.75}$ $M_{\sun}$). 
The same limitation applies to the theoretical models of SHMR at high redshift, which are not constrained by observations at the low stellar mass end \citep[e.g., the SHMR model by][at $z=3$ is constrained only down to $M_{\star}=10^{9.4}$ $M_{\sun}$]{Behroozi2013}. 

The SHMR is most commonly parametrized either with a double power-law function \citep{Behroozi2010, Yang2012, Moster2013}, or with a the five parameter function proposed by \cite{Behroozi2013}, which retains a power law form for halo masses $M_h << 10^{11.97}$ at $z=3$.
Our results suggest that, at high redshifts, this power-law shape is broken at the low mass end below $M_h = 10^{11}$ $M_{\sun}$ (see Fig. \ref{fig:shmr}).
In particular, according to our measurements, the stellar to halo mass ratio is higher than predicted for this halo mass range.
This is in agreement with the conclusion by \cite{Behroozi2013} who note that the low-mass end of the SHMR cannot be predicted by extrapolating results from massive galaxies and fit with the power-law function alone.

A similar higher-than-expected stellar mass to halo mass ratio is observed for dwarf galaxies \citep[e.g.,][]{Boylan2012, Ferrero2012, Miller2014, Brook2014, Read2017}.
While the low-mass galaxy sub-samples used in this paper are not as low mass as the dwarf galaxies observed in the local group (the minimum stellar mass of VUDS galaxies used in our sample is $M_{\star} = 10^{8.75}$ $M_{\sun}$ while the masses of local dwarf galaxies are $10^6-10^9$ $M_{\sun}$), the low mass observation-models discrepancy of SHMR we observe is consistent with these low-mass low redshift samples and it is possible that similar processes are behind it at high $z$ for the higher mass galaxies. 
 
A possible explanation of the discrepancy between the observed SHMR of low mass galaxies and models may lie in the flaws of the abundance matching technique (used in the presented theoretical models to infer SHMR), coupled with our still poor understanding of the feedback effects that influence not only the galaxy stellar mass assembly, but also on the mass distribution of the hosting dark matter haloes \citep[e.g.][]{Pontzen2012, DiCintio2014, Ogiya2014, Katz2017}. 
The abundance matching technique uses simulated dark matter distributions. It is well known, however, that N-body simulations predict a dark matter halo mass function much steeper than the galaxy stellar mass function derived from observations \citep{Press1974, Jenkins2001, Sheth2001, Springel2005}.
Moreover, this difference increases while moving toward low, both stellar and halo, mass regimes (our point of interest here). 
This is usually reconciled by assuming that galaxy formation is directly connected to the halo mass and galaxies do not form efficiently in low mass haloes, which leads to an overestimation of halo masses for the low mass galaxies, when the galaxies are matched with haloes under the assumption that dark matter-only simulations represent structure formation and that every halo hosts a galaxy (which is the case in the abundance matching method). 
This overestimation of the halo masses derived by models, with respect to the observations, is the one visible in Fig. \ref{fig:shmr} for the galaxies with $M_{\star}<10^{9.5}$. 

The relation between dark matter halo mass and galaxy stellar mass is therefore not direct. 
It can be additionally influenced by, e.g., the strong feedback effects, which affect the star formation in low-mass galaxies more strongly than in more massive ones. 
In particular the strong positive feedback (either SN or AGN originated) would result in higher than expected star formation efficiency of low-mass galaxies visible as the model-observation discrepancy for these galaxies in Fig. \ref{fig:shmr}.

At low redshifts the feedback effects have been proposed as the ones that have the major impact on the evolution of dwarf galaxies \citep[see, e.g.,][]{Ferrara2000, Fujita2004, Mashchenko2008, Sawala2011, Kawata2014,Onorbe2015,Chen2016, Papastergis2016}.
Our SHMR measurements, i.e., the higher than expected star formation efficiency, suggest that at $z\sim3$ a positive feedback effects have a significant influence on stellar mass assembly in not only dwarf galaxies ($M_{\star} < 10 ^9$), like it is observed locally, but also in more massive ones, which at $z=0$ are not observed to be strongly affected.
This conclusion can be supported by the fact that a strong feedback effects, both positive and negative, has been observed in abundance in nearly all star-forming galaxies at high $z$ \citep[e.g.,][]{Pettini2001, Shapley2003, Weiner2009, Steidel2010, Jones2012, Newman2012, Erb2015, Talia2017, OLF2017}.

However, we note that other processes might be at work, hence this interpretation may not be the only one and that only further observations of low-mass high redshift galaxies might help to resolve the problem.
For example, positive feedback might not be sufficient to alleviate model to observations at low-mass end, and we need to account also for the possible existence of 'dark haloes', i.e., haloes that are completely devoid of stars \citep[see, e.g.,][]{Sawala2013, Sawala2015}.
A high number of such haloes would strongly affect the accuracy of models based on the abundance matching techniques.
Also, regardless of the fact that introducing a strong positive feedback in low-mass galaxies at high redshifts is physically motivated, it might not produce the correct star formation histories, resulting in a more numerous population of passive galaxies than it is observed locally, as suggested by, e.g., \cite{Fontanot2009, Weinmann2012} and \cite{Moster2013}.
We, therefore, conclude that a mixture of both effects, i.e., strong positive feedback effects and high number of empty dark matter haloes is a possible explanation of the observed trends.

\section{Summary and conclusions}
\label{sec:summary}
\renewcommand{\arraystretch}{1.5} 

\begin{table*}
 \begin{center}
     \caption{Best-fitting power-law and HOD parameters, with other derived parameters (as described in Sec. \ref{sec:method}) for the luminosity and stellar mass sub-samples used in this work. 
	     For the power-law fit, the number of degrees of freedom (dof) is 6 (8 measured $w_p$ values minus the 2 fitted parameters), while for the HOD dof$=3$.
	     All masses are given in units of $h^{-1} M_{\sun}$ and correlation length $r_0$ is given in $h^{-1}$ Mpc. }
	\begin{tabular}{p{1.45cm}|p{1.63cm}p{1.63cm}p{1.63cm}p{1.63cm}|p{1.63cm}p{1.63cm}p{1.63cm}p{1.63cm}} \hline \hline
	\multirow{2}{*}{\begin{tabular}[x]{@{}c@{}}Sample/\\Parameter\end{tabular}}	& \multicolumn{4}{c|}{$M_{UV}^{max}$}												& \multicolumn{4}{c}{$\log M_{\star}^{min}$} \\ \cline{2-9}
						& \multicolumn{1}{c}{$-19.0$}	& \multicolumn{1}{c}{$-19.5$}	& \multicolumn{1}{c}{$-20.0$}	& \multicolumn{1}{c|}{$-20.2$}	& \multicolumn{1}{c}{$8.75$}	& \multicolumn{1}{c}{$9.25$}	& \multicolumn{1}{c}{$9.75$}	& \multicolumn{1}{c}{$10.0$} \\ \hline \hline
	
	$r_0$					& $2.87\pm0.22$ 			& $3.03\pm0.32$ 			& $3.35\pm0.42$  		& $5.35\pm0.50$			& $3.03\pm0.18$			& $3.13\pm0.30$ 		& $3.45\pm0.42$ 		& $4.37\pm0.48$ 	\\
	$\gamma$				& $1.59\pm0.07$ 	 		& $1.63\pm0.09$  			& $1.81\pm0.23$			& $1.92\pm0.25$ 		& $1.61\pm0.06$ 		& $1.61\pm0.09$			& $1.74\pm0.19$			& $1.82\pm0.20$ 	\\ \hline		
	$\log M_{min}$				& $9.73\pm0.51$				& $10.61\pm0.57$			& $10.84\pm0.63$		& $11.58\pm0.62$		& $9.75\pm0.48$			& $9.99\pm0.62$ 		& $11.09\pm0.36$ 		& $11.23\pm0.56$		\\
	$\log M_1'$				& $10.27\pm0.89$			& $10.80\pm0.88$			& $11.93\pm0.81$ 		& $12.28\pm0.50$		& $10.13\pm0.87$ 		& $10.21\pm0.88$		& $11.49\pm0.62$		& $11.51\pm0.83$		\\
	$\log M_1$				& $10.33\pm0.74$			& $10.94\pm0.60$			& $11.94\pm0.80$		& $12.29\pm0.48$		& $10.21\pm0.69$		& $10.31\pm0.65$		& $11.54\pm0.51$		& $11.57\pm0.65$		\\
	$\log M_0$				& $9.05\pm0.96$				& $9.83\pm1.19$				& $8.98\pm1.23$		 	& $9.62\pm1.12$			& $8.92\pm0.98$			& $8.83\pm0.97$  		& $9.54\pm1.19$  		& $9.31\pm1.22$			\\		
	$\sigma_{\log M}$			& $0.64\pm0.13$				& $0.56\pm0.21$				& $0.57\pm0.20$		 	& $0.54\pm0.16$			& $0.57\pm0.16$			& $0.58\pm0.18$  		& $0.46\pm0.16$  		& $0.48\pm0.17$			\\
	$\alpha$				& $1.16\pm0.25$				& $0.94\pm0.27$				& $1.86\pm0.35$		 	& $1.95\pm0.23$			& $1.12\pm0.23$		 	& $1.19\pm0.25$ 		& $1.01\pm0.27$  		& $1.27\pm0.27$			\\
	$\log \langle M_h \rangle$		& $11.79\pm0.58$			& $11.90\pm0.45$			& $12.09\pm0.46$	 	& $12.36\pm0.71$		& $11.91\pm0.45$ 		& $12.06\pm0.42$	 	& $11.95\pm0.34$	 	& $12.24\pm0.47$		\\
	$b_g$					& $1.91\pm0.26$				& $2.09\pm0.84$				& $2.24\pm0.25$		 	& $3.07\pm0.16$			& $1.99\pm0.58$			& $2.29\pm0.64$		 	& $2.39\pm0.67$		 	& $2.84\pm0.99$			\\	
	$f_s$					& $0.58\pm0.41$				& $0.59\pm0.31$				& $0.17\pm0.34$		 	& $0.24\pm0.30$			& $0.68\pm0.44$ 		& $0.76\pm0.31$		 	& $0.47\pm0.25$			& $0.58\pm0.41$			\\ \hline
	\end{tabular}
     \label{tab:parameters}
      \end{center}
\end{table*}
In this paper we study the luminosity and stellar mass dependence of galaxy clustering at redshift $z\sim3$ using a large spectroscopic sample of 3236 star-forming galaxies from the VUDS survey.
We measure the real-space correlation function $w_p(r_p)$ in four volume-limited luminosity sub-samples, with the cuts made in UV absolute magnitude, and four stellar mass sub-samples.
Our measurements are quantified in the framework of two approximations. 
The first one is the power-law model $\xi(r) = \left (r/r_0\right)^{-\gamma}$, with two free parameters.
The second one is based on the halo occupation distribution model (HOD), with five free parameters.

The main results and conclusions of our study can be summarised as follows:
\begin{itemize}
\setlength{\itemsep}{5pt}
 \item We observe an increase of the correlation length $r_0$ with the luminosity and stellar mass of the galaxy populations, indicating a luminosity and stellar mass dependence of galaxy clustering at $z\sim3$. 
       For UV luminosity selected sub-samples $r_0$ rises from $r_0=2.87\pm0.22$ $h^{-1}$ Mpc to $r_0=5.35\pm0.50$ $h^{-1}$ Mpc over a threshold UV absolute magnitude ranging from $M_{UV}=-19.0$ to $M_{UV}=-20.2$.
       A similar trend is found for stellar mass selected galaxy samples, where the correlation length grows from $r_0= 3.03\pm0.18$ $h^{-1}$ Mpc to $r_0=4.37\pm0.48$ $h^{-1}$ Mpc over a relatively small stellar mass range $\Delta \log M_{\star} = 1.25$ $h^{-1} M_{\sun}$.
       Based on these observations we conclude that at $z\sim3$ the luminous and most massive galaxies exist preferentially in denser regions of the universe than their less luminous and less massive counterparts. 
       This trend is consistent with similar trends reported at lower redshifts and is still strong at $z\sim3$.
       It indicates  that mechanisms which led to luminosity and stellar mass clustering dependence must have been at work at a significantly higher redshift than $z\sim3$.
       
 \item Based on the power-law approximation of the correlation function we interpret our results in terms of the relation between the distribution of galaxies and the underlying dark matter density field, called bias $(b)$, relative to the $b^*$ of the $L^*$ galaxies.
       We note that at $z\sim3$ the measured values of $b/b^*$, in each luminosity sub-sample, are significantly lower than observed for the local and intermediate redshift ranges for galaxies of similar properties.
       Additionally we observe that the relative galaxy bias grows with the increasing luminosity of the sample from low values $b/b^* = 0.41\pm0.03$ at low luminosities to $b/b^* = 0.86\pm0.1$ for the high luminosity sub-sample. 
       This growth of $b/b^*$ at $z\sim3$ with luminosity is much steeper than measured for local galaxies, indicating that going back in time the bias contrast of the most luminous galaxies to the rest of the population was stronger.
       This is consistent with the fact that fainter galaxies are found to be significantly less biased tracers of the mass than the brighter galaxies, now confirmed at high redshifts.   
 
 \item Taking advantage of the HOD best-fitting parameters we measure the large scale galaxy bias $b_{g,HOD}$.
       We interpret our results in terms of both redshift evolution, and as a function of luminosity and stellar mass.
       As expected in the framework of the hierarchical scenario of structure formation and evolution, we observe that the $b_{g,HOD}$ measured at $z \sim 3$ is significantly higher than locally, indicating that in the early stages of the evolution of the universe galaxies were more biased tracers of the underlying dark matter density field than it is observed nowadays.
       In addition to redshift evolution, we also note a clear luminosity and stellar mass $b_{g,HOD}$ dependence, with the brightest and most massive galaxies being the most biased ones.
       We find that the luminosity dependence is much steeper than observed in the local universe. 
       The large scale galaxy bias grow by $\Delta b_{g,HOD} = 1.16$, while at $z\sim0$ it increases only by $\Delta b_{g,HOD} = 0.09$ over the same luminosity range.
       A similar growth is observed for stellar mass selected galaxies, with the large scale galaxy bias rising from $b_{g,HOD} = 1.99\pm0.58$ to $b_{g,HOD}=2.84\pm0.99$ over the threshold stellar mass range of $\Delta \log M_{\star} = 1.25$.
       Following \cite{Zehavi2011}, we made an attempt to model the galaxy bias-luminosity and galaxy bias-stellar mass relation, and at $z\sim3$ we find that for the luminosity threshold samples $b_{g,HOD} (> L)$ is best fitted by $b_{g,HOD} (> L) = 1.92 + 25.36(L/L^*)^{7.01}$, while for the stellar mass threshold samples the best fit is $b_{g,HOD}(>M) = 1.59+2.17(M/M^*)^{7.88}$.
 
 \item We report values of the best-fitting HOD parameters for all volume limited UV absolute magnitude and stellar mass sub-samples at redshift $z\sim3$.
       Similarly to what is seen at lower redshift we observe a growth of the dark matter halo characteristic masses $M_{min}$ and $M_1$ with rising luminosity and stellar mass of the galaxy population, indicating that bright and most massive galaxies are likely to occupy the most massive dark matter haloes.  
       Both quantities grow proportionally with a scaling relation of $M_1/M_{min}\approx4$ for the luminosity selected sub-samples, and $M_1/M_{min}\approx2.5$ for the stellar mass selected galaxies.
       These values are much lower than observed at $z\sim0$, where this ratio is reported to have values of $M_1/M_{min}\approx15-20$ \citep{Zehavi2011,McCracken2015,Skibba2015}, which suggests that at high redshift dark matter haloes  contain mainly recently accreted satellite galaxies.  
       We discuss (1) the observed low value of $M_1/M_{min}$ at $z\sim3$ and (2) its increase with cosmic time in terms of the halo versus galaxy merging relation. 
       We infer that DM halo mergers are more frequent than galaxy mergers at $z\sim3$. 
       Our results are consistent with high resolution N-body simulations (see Sec. \ref{sec:halo_mass_discussion}).

 \item We discuss further the satellite galaxies that occupy dark matter haloes at $z\sim3$ by measuring the satellite fraction $f_s$.
       Again our results have implications for the satellite abundances as a function of luminosity and stellar mass, but also as a function of redshift.
       At $z\sim3$ we find that the satellite fraction of the faintest galaxies reaches $f_s\sim 60\%$, while for the brightest galaxies it drops to $\sim20\%$. 
       Therefore our results suggest that it is more probable that dark matter haloes host more faint satellite galaxies than very bright ones.
       For stellar mass selected sub-samples, the satellite fraction remains constant over the sampled stellar mass range, with $f_s\sim50-60\%$.
 
 \item Finally we focus on the stellar to halo mass relation (SHMR) obtained 	for different stellar mass sub-samples.
	   We find that our $z\sim3$ stellar to halo mass ratio is higher than expected in models, e.g., \cite{Behroozi2013} for the low-mass galaxies ($M_{\star}<10^{9.25}$ $M_{\sun}$, Fig. \ref{fig:shmr}). 
       This suggests that the low-mass galaxies are producing stars more effectively than expected.
       We discuss the possibility that strong SNe or AGN feedback effects are at work, that would at least partly explain the observed discrepancy between observations and models for low-mass galaxies at z$\sim$3.

 \end{itemize}

Measurements presented in this paper are the first of their kind performed at $z>2$ based on a large unbiased sample of spectroscopically confirmed redshifts.
As such they provide a valuable benchmark for the interpretation of the co-evolution of galaxies and large scale structures at early epochs of galaxy formation (from the times when the Universe was only 1.5 Gyr old) and put constraints on the efficiency of the processes which drive the star formation and mass assembly in galaxies at that time. 
Moreover, as shown in this paper, our results very well complement lower-z measurements regarding the galaxy clustering dependencies.
All of this information can be used, among others, as an input to improve galaxy formation models (like semi-analytical models) and simulations (like the latest hydro-dynamical simulations), which are still uncertain at high redshifts and need to be confronted by improved observational constraints.
\begin{acknowledgements}
This work is supported by funding from the European Research Council Advanced Grant ERC-2010-AdG-268107-EARLY and by INAF Grants PRIN 2010, PRIN 2012 and PICS 2013.
AD is supported by the Polish National Science Centre grant UMO-2015/17/D/ST9/02121.
This work is based on data products made available at the CESAM data centre, Laboratoire d’Astrophysique de Marseille. 
This work partly uses observations obtained with MegaPrime/MegaCam, a joint project of CFHT and CEA/DAPNIA, at the Canada-France-Hawaii Telescope (CFHT) which is operated by the National Research Council (NRC) of Canada, the Institut National des Sciences de l’Univers of the Centre National de la Recherche Scientifique (CNRS) of France, and the University of Hawaii.
\end{acknowledgements}


\bibliographystyle{aa}
\bibliography{lum_mass}

\begin{thebibliography}{145}
\expandafter\ifx\csname natexlab\endcsname\relax\def\natexlab#1{#1}\fi

\bibitem[{{Abbas} {et~al.}(2010){Abbas}, {de la Torre}, {Le F{\`e}vre},
  {Guzzo}, {Marinoni}, {Meneux}, {Pollo}, {Zamorani}, {Bottini}, {Garilli}, {Le
  Brun}, {Maccagni}, {Scaramella}, {Scodeggio}, {Tresse}, {Vettolani},
  {Zanichelli}, {Adami}, {Arnouts}, {Bardelli}, {Bolzonella}, {Cappi},
  {Charlot}, {Ciliegi}, {Contini}, {Foucaud}, {Franzetti}, {Gavignaud},
  {Ilbert}, {Iovino}, {Lamareille}, {McCracken}, {Marano}, {Mazure}, {Merighi},
  {Paltani}, {Pell{\`o}}, {Pozzetti}, {Radovich}, {Vergani}, {Zucca}, {Bondi},
  {Bongiorno}, {Brinchmann}, {Cucciati}, {de Ravel}, {Gregorini},
  {Perez-Montero}, {Mellier}, \& {Merluzzi}}]{Abbas2010}
{Abbas}, U., {de la Torre}, S., {Le F{\`e}vre}, O., {et~al.} 2010, \mnras, 406,
  1306

\bibitem[{{Adelberger} {et~al.}(2005){Adelberger}, {Steidel}, {Pettini},
  {Shapley}, {Reddy}, \& {Erb}}]{Adelberger2005}
{Adelberger}, K.~L., {Steidel}, C.~C., {Pettini}, M., {et~al.} 2005, \apj, 619,
  697

\bibitem[{{Arnouts} {et~al.}(1999){Arnouts}, {Cristiani}, {Moscardini},
  {Matarrese}, {Lucchin}, {Fontana}, \& {Giallongo}}]{Arnouts1999}
{Arnouts}, S., {Cristiani}, S., {Moscardini}, L., {et~al.} 1999, \mnras, 310,
  540

\bibitem[{{Bardeen} {et~al.}(1986){Bardeen}, {Bond}, {Kaiser}, \&
  {Szalay}}]{Bardeen1986}
{Bardeen}, J.~M., {Bond}, J.~R., {Kaiser}, N., \& {Szalay}, A.~S. 1986, \apj,
  304, 15

\bibitem[{{Barrow} {et~al.}(1984){Barrow}, {Bhavsar}, \& {Sonoda}}]{Barrow1984}
{Barrow}, J.~D., {Bhavsar}, S.~P., \& {Sonoda}, D.~H. 1984, \mnras, 210, 19P

\bibitem[{{Behroozi} {et~al.}(2010){Behroozi}, {Conroy}, \&
  {Wechsler}}]{Behroozi2010}
{Behroozi}, P.~S., {Conroy}, C., \& {Wechsler}, R.~H. 2010, \apj, 717, 379

\bibitem[{{Behroozi} {et~al.}(2013){Behroozi}, {Wechsler}, \&
  {Conroy}}]{Behroozi2013}
{Behroozi}, P.~S., {Wechsler}, R.~H., \& {Conroy}, C. 2013, \apj, 770, 57

\bibitem[{{Benson} {et~al.}(2001){Benson}, {Frenk}, {Baugh}, {Cole}, \&
  {Lacey}}]{Benson2001}
{Benson}, A.~J., {Frenk}, C.~S., {Baugh}, C.~M., {Cole}, S., \& {Lacey}, C.~G.
  2001, \mnras, 327, 1041

\bibitem[{{Beutler} {et~al.}(2013){Beutler}, {Blake}, {Colless}, {Jones},
  {Staveley-Smith}, {Campbell}, {Parker}, {Saunders}, \&
  {Watson}}]{Beutler2013}
{Beutler}, F., {Blake}, C., {Colless}, M., {et~al.} 2013, \mnras, 429, 3604

\bibitem[{{Bielby} {et~al.}(2014){Bielby}, {Gonzalez-Perez}, {McCracken},
  {Ilbert}, {Daddi}, {Le F{\`e}vre}, {Hudelot}, {Kneib}, {Mellier}, \&
  {Willott}}]{Bielby2014}
{Bielby}, R.~M., {Gonzalez-Perez}, V., {McCracken}, H.~J., {et~al.} 2014, \aap,
  568, A24

\bibitem[{{Blandford} \& {Narayan}(1992)}]{Blandford1992}
{Blandford}, R.~D. \& {Narayan}, R. 1992, \araa, 30, 311

\bibitem[{{Blanton} {et~al.}(1999){Blanton}, {Cen}, {Ostriker}, \&
  {Strauss}}]{Blanton1999}
{Blanton}, M., {Cen}, R., {Ostriker}, J.~P., \& {Strauss}, M.~A. 1999, \apj,
  522, 590

\bibitem[{{Bouwens} \& {Illingworth}(2007)}]{Bouwens2007}
{Bouwens}, R.~J. \& {Illingworth}, G.~D. 2007, in Astronomical Society of the
  Pacific Conference Series, Vol. 380, Deepest Astronomical Surveys, ed.
  J.~{Afonso}, H.~C. {Ferguson}, B.~{Mobasher}, \& R.~{Norris}, 41

\bibitem[{{Bouwens} {et~al.}(2015){Bouwens}, {Illingworth}, {Oesch}, {Trenti},
  {Labb{\'e}}, {Bradley}, {Carollo}, {van Dokkum}, {Gonzalez}, {Holwerda},
  {Franx}, {Spitler}, {Smit}, \& {Magee}}]{Bouwens2015}
{Bouwens}, R.~J., {Illingworth}, G.~D., {Oesch}, P.~A., {et~al.} 2015, \apj,
  803, 34

\bibitem[{{Boylan-Kolchin} {et~al.}(2012){Boylan-Kolchin}, {Bullock}, \&
  {Kaplinghat}}]{Boylan2012}
{Boylan-Kolchin}, M., {Bullock}, J.~S., \& {Kaplinghat}, M. 2012, \mnras, 422,
  1203

\bibitem[{{Brook} {et~al.}(2014){Brook}, {Di Cintio}, {Knebe}, {Gottl{\"o}ber},
  {Hoffman}, {Yepes}, \& {Garrison-Kimmel}}]{Brook2014}
{Brook}, C.~B., {Di Cintio}, A., {Knebe}, A., {et~al.} 2014, \apjl, 784, L14

\bibitem[{{Cassata} {et~al.}(2013){Cassata}, {Le F{\`e}vre}, {Charlot},
  {Contini}, {Cucciati}, {Garilli}, {Zamorani}, {Adami}, {Bardelli}, {Le Brun},
  {Lemaux}, {Maccagni}, {Pollo}, {Pozzetti}, {Tresse}, {Vergani}, {Zanichelli},
  \& {Zucca}}]{Cassata2013}
{Cassata}, P., {Le F{\`e}vre}, O., {Charlot}, S., {et~al.} 2013, \aap, 556, A68

\bibitem[{{Chen} {et~al.}(2016){Chen}, {Bryan}, \& {Salem}}]{Chen2016}
{Chen}, J., {Bryan}, G.~L., \& {Salem}, M. 2016, \mnras, 460, 3335

\bibitem[{{Chiang} {et~al.}(2013){Chiang}, {Overzier}, \&
  {Gebhardt}}]{Chiang2013}
{Chiang}, Y.-K., {Overzier}, R., \& {Gebhardt}, K. 2013, \apj, 779, 127

\bibitem[{{Coil} {et~al.}(2006){Coil}, {Newman}, {Cooper}, {Davis}, {Faber},
  {Koo}, \& {Willmer}}]{Coil2006}
{Coil}, A.~L., {Newman}, J.~A., {Cooper}, M.~C., {et~al.} 2006, \apj, 644, 671

\bibitem[{{Coil} {et~al.}(2008){Coil}, {Newman}, {Croton}, {Cooper}, {Davis},
  {Faber}, {Gerke}, {Koo}, {Padmanabhan}, {Wechsler}, \& {Weiner}}]{Coil2008}
{Coil}, A.~L., {Newman}, J.~A., {Croton}, D., {et~al.} 2008, \apj, 672, 153

\bibitem[{{Conroy} {et~al.}(2006){Conroy}, {Wechsler}, \&
  {Kravtsov}}]{Conroy2006}
{Conroy}, C., {Wechsler}, R.~H., \& {Kravtsov}, A.~V. 2006, \apj, 647, 201

\bibitem[{{Conselice} {et~al.}(2008){Conselice}, {Rajgor}, \&
  {Myers}}]{Conselice2008}
{Conselice}, C.~J., {Rajgor}, S., \& {Myers}, R. 2008, \mnras, 386, 909

\bibitem[{{Coupon} {et~al.}(2012){Coupon}, {Kilbinger}, {McCracken}, {Ilbert},
  {Arnouts}, {Mellier}, {Abbas}, {de la Torre}, {Goranova}, {Hudelot}, {Kneib},
  \& {Le F{\`e}vre}}]{Coupon2012}
{Coupon}, J., {Kilbinger}, M., {McCracken}, H.~J., {et~al.} 2012, \aap, 542, A5

\bibitem[{{Daddi} {et~al.}(2003){Daddi}, {R{\"o}ttgering}, {Labb{\'e}},
  {Rudnick}, {Franx}, {Moorwood}, {Rix}, {van der Werf}, \& {van
  Dokkum}}]{Daddi2003}
{Daddi}, E., {R{\"o}ttgering}, H.~J.~A., {Labb{\'e}}, I., {et~al.} 2003, \apj,
  588, 50

\bibitem[{{Davis} \& {Peebles}(1983)}]{Davis1983}
{Davis}, M. \& {Peebles}, P.~J.~E. 1983, \apj, 267, 465

\bibitem[{{de la Torre} {et~al.}(2013){de la Torre}, {Guzzo}, {Peacock},
  {Branchini}, {Iovino}, {Granett}, {Abbas}, {Adami}, {Arnouts}, {Bel},
  {Bolzonella}, {Bottini}, {Cappi}, {Coupon}, {Cucciati}, {Davidzon}, {De
  Lucia}, {Fritz}, {Franzetti}, {Fumana}, {Garilli}, {Ilbert}, {Krywult}, {Le
  Brun}, {Le F{\`e}vre}, {Maccagni}, {Ma{\l}ek}, {Marulli}, {McCracken},
  {Moscardini}, {Paioro}, {Percival}, {Polletta}, {Pollo}, {Schlagenhaufer},
  {Scodeggio}, {Tasca}, {Tojeiro}, {Vergani}, {Zanichelli}, {Burden}, {Di
  Porto}, {Marchetti}, {Marinoni}, {Mellier}, {Monaco}, {Nichol}, {Phleps},
  {Wolk}, \& {Zamorani}}]{delaTorre2013}
{de la Torre}, S., {Guzzo}, L., {Peacock}, J.~A., {et~al.} 2013, \aap, 557, A54

\bibitem[{{de la Torre} {et~al.}(2007){de la Torre}, {Le F{\`e}vre}, {Arnouts},
  {Guzzo}, {Farrah}, {Iovino}, {Lonsdale}, {Meneux}, {Oliver}, {Pollo},
  {Waddington}, {Bottini}, {Fang}, {Garilli}, {Le Brun}, {Maccagni}, {Picat},
  {Scaramella}, {Scodeggio}, {Shupe}, {Surace}, {Tresse}, {Vettolani},
  {Zanichelli}, {Adami}, {Bardelli}, {Bolzonella}, {Cappi}, {Charlot},
  {Ciliegi}, {Contini}, {Foucaud}, {Franzetti}, {Gavignaud}, {Ilbert},
  {Lamareille}, {McCracken}, {Marano}, {Marinoni}, {Mazure}, {Merighi},
  {Paltani}, {Pell{\`o}}, {Pozzetti}, {Radovich}, {Zamorani}, {Zucca}, {Bondi},
  {Bongiorno}, {Brinchmann}, {Cucciati}, {Mellier}, {Merluzzi}, {Temporin},
  {Vergani}, \& {Walcher}}]{delaTorre2007}
{de la Torre}, S., {Le F{\`e}vre}, O., {Arnouts}, S., {et~al.} 2007, \aap, 475,
  443

\bibitem[{{De Lucia} {et~al.}(2007){De Lucia}, {Poggianti},
  {Arag{\'o}n-Salamanca}, {White}, {Zaritsky}, {Clowe}, {Halliday}, {Jablonka},
  {von der Linden}, {Milvang-Jensen}, {Pell{\'o}}, {Rudnick}, {Saglia}, \&
  {Simard}}]{DeLucia2007}
{De Lucia}, G., {Poggianti}, B.~M., {Arag{\'o}n-Salamanca}, A., {et~al.} 2007,
  \mnras, 374, 809

\bibitem[{{De Lucia} {et~al.}(2006){De Lucia}, {Springel}, {White}, {Croton},
  \& {Kauffmann}}]{DeLucia2006}
{De Lucia}, G., {Springel}, V., {White}, S.~D.~M., {Croton}, D., \&
  {Kauffmann}, G. 2006, \mnras, 366, 499

\bibitem[{{de Ravel} {et~al.}(2009){de Ravel}, {Le F{\`e}vre}, {Tresse},
  {Bottini}, {Garilli}, {Le Brun}, {Maccagni}, {Scaramella}, {Scodeggio},
  {Vettolani}, {Zanichelli}, {Adami}, {Arnouts}, {Bardelli}, {Bolzonella},
  {Cappi}, {Charlot}, {Ciliegi}, {Contini}, {Foucaud}, {Franzetti},
  {Gavignaud}, {Guzzo}, {Ilbert}, {Iovino}, {Lamareille}, {McCracken},
  {Marano}, {Marinoni}, {Mazure}, {Meneux}, {Merighi}, {Paltani}, {Pell{\`o}},
  {Pollo}, {Pozzetti}, {Radovich}, {Vergani}, {Zamorani}, {Zucca}, {Bondi},
  {Bongiorno}, {Brinchmann}, {Cucciati}, {de La Torre}, {Gregorini}, {Memeo},
  {Perez-Montero}, {Mellier}, {Merluzzi}, \& {Temporin}}]{deRavel2009}
{de Ravel}, L., {Le F{\`e}vre}, O., {Tresse}, L., {et~al.} 2009, \aap, 498, 379

\bibitem[{{Dekel} {et~al.}(2017){Dekel}, {Ishai}, {Dutton}, \&
  {Maccio}}]{Dekel2017}
{Dekel}, A., {Ishai}, G., {Dutton}, A.~A., \& {Maccio}, A.~V. 2017, \mnras,
  468, 1005

\bibitem[{{Di Cintio} {et~al.}(2014){Di Cintio}, {Brook}, {Macci{\`o}},
  {Stinson}, {Knebe}, {Dutton}, \& {Wadsley}}]{DiCintio2014}
{Di Cintio}, A., {Brook}, C.~B., {Macci{\`o}}, A.~V., {et~al.} 2014, \mnras,
  437, 415

\bibitem[{{Durkalec} {et~al.}(2015{\natexlab{a}}){Durkalec}, {Le F{\`e}vre},
  {de la Torre}, {Pollo}, {Cassata}, {Garilli}, {Le Brun}, {Lemaux},
  {Maccagni}, {Pentericci}, {Tasca}, {Thomas}, {Vanzella}, {Zamorani}, {Zucca},
  {Amor{\'{\i}}n}, {Bardelli}, {Cassar{\`a}}, {Castellano}, {Cimatti},
  {Cucciati}, {Fontana}, {Giavalisco}, {Grazian}, {Hathi}, {Ilbert}, {Paltani},
  {Ribeiro}, {Schaerer}, {Scodeggio}, {Sommariva}, {Talia}, {Tresse},
  {Vergani}, {Capak}, {Charlot}, {Contini}, {Cuby}, {Dunlop}, {Fotopoulou},
  {Koekemoer}, {L{\'o}pez-Sanjuan}, {Mellier}, {Pforr}, {Salvato}, {Scoville},
  {Taniguchi}, \& {Wang}}]{Durkalec2015b}
{Durkalec}, A., {Le F{\`e}vre}, O., {de la Torre}, S., {et~al.}
  2015{\natexlab{a}}, \aap, 576, L7

\bibitem[{{Durkalec} {et~al.}(2015{\natexlab{b}}){Durkalec}, {Le F{\`e}vre},
  {Pollo}, {de la Torre}, {Cassata}, {Garilli}, {Le Brun}, {Lemaux},
  {Maccagni}, {Pentericci}, {Tasca}, {Thomas}, {Vanzella}, {Zamorani}, {Zucca},
  {Amor{\'{\i}}n}, {Bardelli}, {Cassar{\`a}}, {Castellano}, {Cimatti},
  {Cucciati}, {Fontana}, {Giavalisco}, {Grazian}, {Hathi}, {Ilbert}, {Paltani},
  {Ribeiro}, {Schaerer}, {Scodeggio}, {Sommariva}, {Talia}, {Tresse},
  {Vergani}, {Capak}, {Charlot}, {Contini}, {Cuby}, {Dunlop}, {Fotopoulou},
  {Koekemoer}, {L{\'o}pez-Sanjuan}, {Mellier}, {Pforr}, {Salvato}, {Scoville},
  {Taniguchi}, \& {Wang}}]{Durkalec2015a}
{Durkalec}, A., {Le F{\`e}vre}, O., {Pollo}, A., {et~al.} 2015{\natexlab{b}},
  \aap, 583, A128

\bibitem[{{Erb}(2015)}]{Erb2015}
{Erb}, D.~K. 2015, \nat, 523, 169

\bibitem[{{Ferrara} \& {Tolstoy}(2000)}]{Ferrara2000}
{Ferrara}, A. \& {Tolstoy}, E. 2000, \mnras, 313, 291

\bibitem[{{Ferrero} {et~al.}(2012){Ferrero}, {Abadi}, {Navarro}, {Sales}, \&
  {Gurovich}}]{Ferrero2012}
{Ferrero}, I., {Abadi}, M.~G., {Navarro}, J.~F., {Sales}, L.~V., \& {Gurovich},
  S. 2012, \mnras, 425, 2817

\bibitem[{{Finkelstein} {et~al.}(2015){Finkelstein}, {Ryan}, {Papovich},
  {Dickinson}, {Song}, {Somerville}, {Ferguson}, {Salmon}, {Giavalisco},
  {Koekemoer}, {Ashby}, {Behroozi}, {Castellano}, {Dunlop}, {Faber}, {Fazio},
  {Fontana}, {Grogin}, {Hathi}, {Jaacks}, {Kocevski}, {Livermore}, {McLure},
  {Merlin}, {Mobasher}, {Newman}, {Rafelski}, {Tilvi}, \&
  {Willner}}]{Finkelstein2015}
{Finkelstein}, S.~L., {Ryan}, Jr., R.~E., {Papovich}, C., {et~al.} 2015, \apj,
  810, 71

\bibitem[{{Fontanot} {et~al.}(2009){Fontanot}, {De Lucia}, {Monaco},
  {Somerville}, \& {Santini}}]{Fontanot2009}
{Fontanot}, F., {De Lucia}, G., {Monaco}, P., {Somerville}, R.~S., \&
  {Santini}, P. 2009, \mnras, 397, 1776

\bibitem[{{Fry}(1996)}]{Fry1996}
{Fry}, J.~N. 1996, \apjl, 461, L65

\bibitem[{{Fu} {et~al.}(2008){Fu}, {Semboloni}, {Hoekstra}, {Kilbinger}, {van
  Waerbeke}, {Tereno}, {Mellier}, {Heymans}, {Coupon}, {Benabed}, {Benjamin},
  {Bertin}, {Dor{\'e}}, {Hudson}, {Ilbert}, {Maoli}, {Marmo}, {McCracken}, \&
  {M{\'e}nard}}]{Fu2008}
{Fu}, L., {Semboloni}, E., {Hoekstra}, H., {et~al.} 2008, \aap, 479, 9

\bibitem[{{Fujita} {et~al.}(2004){Fujita}, {Mac Low}, {Ferrara}, \&
  {Meiksin}}]{Fujita2004}
{Fujita}, A., {Mac Low}, M.-M., {Ferrara}, A., \& {Meiksin}, A. 2004, \apj,
  613, 159

\bibitem[{{Garilli} {et~al.}(2010){Garilli}, {Fumana}, {Franzetti}, {Paioro},
  {Scodeggio}, {Le F{\`e}vre}, {Paltani}, \& {Scaramella}}]{Garilli2010}
{Garilli}, B., {Fumana}, M., {Franzetti}, P., {et~al.} 2010, \pasp, 122, 827

\bibitem[{{Genzel} {et~al.}(2017){Genzel}, {Schreiber}, {{\"U}bler}, {Lang},
  {Naab}, {Bender}, {Tacconi}, {Wisnioski}, {Wuyts}, {Alexander}, {Beifiori},
  {Belli}, {Brammer}, {Burkert}, {Carollo}, {Chan}, {Davies}, {Fossati},
  {Galametz}, {Genel}, {Gerhard}, {Lutz}, {Mendel}, {Momcheva}, {Nelson},
  {Renzini}, {Saglia}, {Sternberg}, {Tacchella}, {Tadaki}, \&
  {Wilman}}]{Genzel2017}
{Genzel}, R., {Schreiber}, N.~M.~F., {{\"U}bler}, H., {et~al.} 2017, \nat, 543,
  397

\bibitem[{{Guo} {et~al.}(2015){Guo}, {Zheng}, {Zehavi}, {Behroozi}, {Chuang},
  {Comparat}, {Favole}, {Gottloeber}, {Klypin}, {Prada}, {Weinberg}, \&
  {Yepes}}]{Guo2015}
{Guo}, H., {Zheng}, Z., {Zehavi}, I., {et~al.} 2015, \mnras, 453, 4368

\bibitem[{{Guzzo} {et~al.}(1997){Guzzo}, {Strauss}, {Fisher}, {Giovanelli}, \&
  {Haynes}}]{Guzzo1997}
{Guzzo}, L., {Strauss}, M.~A., {Fisher}, K.~B., {Giovanelli}, R., \& {Haynes},
  M.~P. 1997, \apj, 489, 37

\bibitem[{{Hagen} {et~al.}(2015){Hagen}, {Hoversten}, {Gronwall}, {Wolf},
  {Siegel}, {Page}, \& {Hagen}}]{Hagen2015}
{Hagen}, L.~M.~Z., {Hoversten}, E.~A., {Gronwall}, C., {et~al.} 2015, \apj,
  808, 178

\bibitem[{{Hartley} {et~al.}(2010){Hartley}, {Almaini}, {Cirasuolo}, {Foucaud},
  {Simpson}, {Conselice}, {Smail}, {McLure}, {Dunlop}, {Chuter}, {Maddox},
  {Lane}, \& {Bradshaw}}]{Hartley2010}
{Hartley}, W.~G., {Almaini}, O., {Cirasuolo}, M., {et~al.} 2010, \mnras, 407,
  1212

\bibitem[{{Hatfield} {et~al.}(2016){Hatfield}, {Lindsay}, {Jarvis},
  {H{\"a}u{\ss}ler}, {Vaccari}, \& {Verma}}]{Hatfield2016}
{Hatfield}, P.~W., {Lindsay}, S.~N., {Jarvis}, M.~J., {et~al.} 2016, \mnras,
  459, 2618

\bibitem[{{Hathi} {et~al.}(2010){Hathi}, {Ryan}, {Cohen}, {Yan}, {Windhorst},
  {McCarthy}, {O'Connell}, {Koekemoer}, {Rutkowski}, {Balick}, {Bond},
  {Calzetti}, {Disney}, {Dopita}, {Frogel}, {Hall}, {Holtzman}, {Kimble},
  {Paresce}, {Saha}, {Silk}, {Trauger}, {Walker}, {Whitmore}, \&
  {Young}}]{Hathi2010}
{Hathi}, N.~P., {Ryan}, Jr., R.~E., {Cohen}, S.~H., {et~al.} 2010, \apj, 720,
  1708

\bibitem[{{Hildebrandt} {et~al.}(2009){Hildebrandt}, {Pielorz}, {Erben}, {van
  Waerbeke}, {Simon}, \& {Capak}}]{Hildebrandt2009}
{Hildebrandt}, H., {Pielorz}, J., {Erben}, T., {et~al.} 2009, \aap, 498, 725

\bibitem[{{Hoekstra} {et~al.}(2004){Hoekstra}, {Yee}, \&
  {Gladders}}]{Hoekstra2004}
{Hoekstra}, H., {Yee}, H.~K.~C., \& {Gladders}, M.~D. 2004, \apj, 606, 67

\bibitem[{{Ilbert} {et~al.}(2006){Ilbert}, {Arnouts}, {McCracken},
  {Bolzonella}, {Bertin}, {Le F{\`e}vre}, {Mellier}, {Zamorani}, {Pell{\`o}},
  {Iovino}, {Tresse}, {Le Brun}, {Bottini}, {Garilli}, {Maccagni}, {Picat},
  {Scaramella}, {Scodeggio}, {Vettolani}, {Zanichelli}, {Adami}, {Bardelli},
  {Cappi}, {Charlot}, {Ciliegi}, {Contini}, {Cucciati}, {Foucaud}, {Franzetti},
  {Gavignaud}, {Guzzo}, {Marano}, {Marinoni}, {Mazure}, {Meneux}, {Merighi},
  {Paltani}, {Pollo}, {Pozzetti}, {Radovich}, {Zucca}, {Bondi}, {Bongiorno},
  {Busarello}, {de La Torre}, {Gregorini}, {Lamareille}, {Mathez}, {Merluzzi},
  {Ripepi}, {Rizzo}, \& {Vergani}}]{Ilbert2006}
{Ilbert}, O., {Arnouts}, S., {McCracken}, H.~J., {et~al.} 2006, \aap, 457, 841

\bibitem[{{Ilbert} {et~al.}(2013){Ilbert}, {McCracken}, {Le F{\`e}vre},
  {Capak}, {Dunlop}, {Karim}, {Renzini}, {Caputi}, {Boissier}, {Arnouts},
  {Aussel}, {Comparat}, {Guo}, {Hudelot}, {Kartaltepe}, {Kneib}, {Krogager},
  {Le Floc'h}, {Lilly}, {Mellier}, {Milvang-Jensen}, {Moutard}, {Onodera},
  {Richard}, {Salvato}, {Sanders}, {Scoville}, {Silverman}, {Taniguchi},
  {Tasca}, {Thomas}, {Toft}, {Tresse}, {Vergani}, {Wolk}, \&
  {Zirm}}]{Ilbert2013}
{Ilbert}, O., {McCracken}, H.~J., {Le F{\`e}vre}, O., {et~al.} 2013, \aap, 556,
  A55

\bibitem[{{Ilbert} {et~al.}(2005){Ilbert}, {Tresse}, {Zucca}, {Bardelli},
  {Arnouts}, {Zamorani}, {Pozzetti}, {Bottini}, {Garilli}, {Le Brun}, {Le
  F{\`e}vre}, {Maccagni}, {Picat}, {Scaramella}, {Scodeggio}, {Vettolani},
  {Zanichelli}, {Adami}, {Arnaboldi}, {Bolzonella}, {Cappi}, {Charlot},
  {Contini}, {Foucaud}, {Franzetti}, {Gavignaud}, {Guzzo}, {Iovino},
  {McCracken}, {Marano}, {Marinoni}, {Mathez}, {Mazure}, {Meneux}, {Merighi},
  {Paltani}, {Pello}, {Pollo}, {Radovich}, {Bondi}, {Bongiorno}, {Busarello},
  {Ciliegi}, {Lamareille}, {Mellier}, {Merluzzi}, {Ripepi}, \&
  {Rizzo}}]{Ilbert2005}
{Ilbert}, O., {Tresse}, L., {Zucca}, E., {et~al.} 2005, \aap, 439, 863

\bibitem[{{Ishikawa} {et~al.}(2017){Ishikawa}, {Kashikawa}, {Toshikawa},
  {Tanaka}, {Hamana}, {Niino}, {Ichikawa}, \& {Uchiyama}}]{Ishikawa2017}
{Ishikawa}, S., {Kashikawa}, N., {Toshikawa}, J., {et~al.} 2017, \apj, 841, 8

\bibitem[{{Jenkins} {et~al.}(2001){Jenkins}, {Frenk}, {White}, {Colberg},
  {Cole}, {Evrard}, {Couchman}, \& {Yoshida}}]{Jenkins2001}
{Jenkins}, A., {Frenk}, C.~S., {White}, S.~D.~M., {et~al.} 2001, \mnras, 321,
  372

\bibitem[{{Jones} {et~al.}(2012){Jones}, {Kemper}, {Sargent}, {McDonald},
  {Gielen}, {Woods}, {Sloan}, {Boyer}, {Zijlstra}, {Clayton}, {Kraemer},
  {Srinivasan}, \& {Ruffle}}]{Jones2012}
{Jones}, O.~C., {Kemper}, F., {Sargent}, B.~A., {et~al.} 2012, \mnras, 427,
  3209

\bibitem[{{Kaiser}(1984)}]{Kaiser1984}
{Kaiser}, N. 1984, \apjl, 284, L9

\bibitem[{{Katz} {et~al.}(2017){Katz}, {Lelli}, {McGaugh}, {Di Cintio},
  {Brook}, \& {Schombert}}]{Katz2017}
{Katz}, H., {Lelli}, F., {McGaugh}, S.~S., {et~al.} 2017, \mnras, 466, 1648

\bibitem[{{Kauffmann} {et~al.}(1997){Kauffmann}, {Nusser}, \&
  {Steinmetz}}]{Kauffmann1997}
{Kauffmann}, G., {Nusser}, A., \& {Steinmetz}, M. 1997, \mnras, 286, 795

\bibitem[{{Kawata} {et~al.}(2014){Kawata}, {Gibson}, {Barnes}, {Grand}, \&
  {Rahimi}}]{Kawata2014}
{Kawata}, D., {Gibson}, B.~K., {Barnes}, D.~J., {Grand}, R.~J.~J., \& {Rahimi},
  A. 2014, \mnras, 438, 1208

\bibitem[{{Kilbinger} {et~al.}(2011){Kilbinger}, {Benabed}, {Cappe}, {Cardoso},
  {Coupon}, {Fort}, {McCracken}, {Prunet}, {Robert}, \&
  {Wraith}}]{Kilbinger2011}
{Kilbinger}, M., {Benabed}, K., {Cappe}, O., {et~al.} 2011, arXiv1101.0950
  [\eprint[arXiv]{1101.0950}]

\bibitem[{{Landy} \& {Szalay}(1993)}]{Landy1993}
{Landy}, S.~D. \& {Szalay}, A.~S. 1993, \apj, 412, 64

\bibitem[{{Le F{\`e}vre} {et~al.}(2005){Le F{\`e}vre}, {Guzzo}, {Meneux},
  {Pollo}, {Cappi}, {Colombi}, {Iovino}, {Marinoni}, {McCracken}, {Scaramella},
  {Bottini}, {Garilli}, {Le Brun}, {Maccagni}, {Picat}, {Scodeggio}, {Tresse},
  {Vettolani}, {Zanichelli}, {Adami}, {Arnaboldi}, {Arnouts}, {Bardelli},
  {Blaizot}, {Bolzonella}, {Charlot}, {Ciliegi}, {Contini}, {Foucaud},
  {Franzetti}, {Gavignaud}, {Ilbert}, {Marano}, {Mathez}, {Mazure}, {Merighi},
  {Paltani}, {Pell{\`o}}, {Pozzetti}, {Radovich}, {Zamorani}, {Zucca}, {Bondi},
  {Bongiorno}, {Busarello}, {Lamareille}, {Mellier}, {Merluzzi}, {Ripepi}, \&
  {Rizzo}}]{OLF2005}
{Le F{\`e}vre}, O., {Guzzo}, L., {Meneux}, B., {et~al.} 2005, \aap, 439, 877

\bibitem[{{Le F{\`e}vre} {et~al.}(2017){Le F{\`e}vre}, {Lemaux}, {Nakajima},
  {Schaerer}, {Talia}, {Zamorani}, {Cassata}, {Garilli}, {Maccagni},
  {Pentericci}, {Tasca}, {Zucca}, {Amorin}, {Bardelli}, {Cimatti},
  {Giavalisco}, {Guaita}, {Hathi}, {Marchi}, {Vanzella}, {Vergani}, \&
  {Dunlop}}]{OLF2017}
{Le F{\`e}vre}, O., {Lemaux}, B.~C., {Nakajima}, K., {et~al.} 2017, ArXiv
  e-prints [\eprint[arXiv]{1710.10715}]

\bibitem[{{Le F{\`e}vre} {et~al.}(2015){Le F{\`e}vre}, {Tasca}, {Cassata},
  {Garilli}, {Le Brun}, {Maccagni}, {Pentericci}, {Thomas}, {Vanzella},
  {Zamorani}, {Zucca}, {Amorin}, {Bardelli}, {Capak}, {Cassar{\`a}},
  {Castellano}, {Cimatti}, {Cuby}, {Cucciati}, {de la Torre}, {Durkalec},
  {Fontana}, {Giavalisco}, {Grazian}, {Hathi}, {Ilbert}, {Lemaux}, {Moreau},
  {Paltani}, {Ribeiro}, {Salvato}, {Schaerer}, {Scodeggio}, {Sommariva},
  {Talia}, {Taniguchi}, {Tresse}, {Vergani}, {Wang}, {Charlot}, {Contini},
  {Fotopoulou}, {L{\'o}pez-Sanjuan}, {Mellier}, \& {Scoville}}]{OLF2015}
{Le F{\`e}vre}, O., {Tasca}, L.~A.~M., {Cassata}, P., {et~al.} 2015, \aap, 576,
  A79

\bibitem[{{Leauthaud} {et~al.}(2012){Leauthaud}, {Tinker}, {Bundy}, {Behroozi},
  {Massey}, {Rhodes}, {George}, {Kneib}, {Benson}, {Wechsler}, {Busha},
  {Capak}, {Cort{\^e}s}, {Ilbert}, {Koekemoer}, {Le F{\`e}vre}, {Lilly},
  {McCracken}, {Salvato}, {Schrabback}, {Scoville}, {Smith}, \&
  {Taylor}}]{Leauthaud2012}
{Leauthaud}, A., {Tinker}, J., {Bundy}, K., {et~al.} 2012, \apj, 744, 159

\bibitem[{{Lee} {et~al.}(2006){Lee}, {Giavalisco}, {Gnedin}, {Somerville},
  {Ferguson}, {Dickinson}, \& {Ouchi}}]{Lee2006}
{Lee}, K.-S., {Giavalisco}, M., {Gnedin}, O.~Y., {et~al.} 2006, \apj, 642, 63

\bibitem[{{Lilly} {et~al.}(1996){Lilly}, {Le Fevre}, {Hammer}, \&
  {Crampton}}]{Lilly1996}
{Lilly}, S.~J., {Le Fevre}, O., {Hammer}, F., \& {Crampton}, D. 1996, \apjl,
  460, L1

\bibitem[{{Lin} {et~al.}(2012){Lin}, {Dickinson}, {Jian}, {Merson}, {Baugh},
  {Scott}, {Foucaud}, {Wang}, {Yan}, {Yan}, {Cheng}, {Guo}, {Helly}, {Kirsten},
  {Koo}, {Lagos}, {Meger}, {Messias}, {Pope}, {Simard}, {Grogin}, \&
  {Wang}}]{Lin2012}
{Lin}, L., {Dickinson}, M., {Jian}, H.-Y., {et~al.} 2012, \apj, 756, 71

\bibitem[{{L{\'o}pez-Sanjuan} {et~al.}(2011){L{\'o}pez-Sanjuan}, {Le
  F{\`e}vre}, {de Ravel}, {Cucciati}, {Ilbert}, {Tresse}, {Bardelli},
  {Bolzonella}, {Contini}, {Garilli}, {Guzzo}, {Maccagni}, {McCracken},
  {Mellier}, {Pollo}, {Vergani}, \& {Zucca}}]{LopezSanjuan2011}
{L{\'o}pez-Sanjuan}, C., {Le F{\`e}vre}, O., {de Ravel}, L., {et~al.} 2011,
  \aap, 530, A20

\bibitem[{{L{\'o}pez-Sanjuan} {et~al.}(2013){L{\'o}pez-Sanjuan}, {Le
  F{\`e}vre}, {Tasca}, {Epinat}, {Amram}, {Contini}, {Garilli},
  {Kissler-Patig}, {Moultaka}, {Paioro}, {Perret}, {Queyrel}, {Tresse},
  {Vergani}, \& {Divoy}}]{LopezSanjuan2013}
{L{\'o}pez-Sanjuan}, C., {Le F{\`e}vre}, O., {Tasca}, L.~A.~M., {et~al.} 2013,
  \aap, 553, A78

\bibitem[{{Magliocchetti} \& {Porciani}(2003)}]{Magliocchetti2003}
{Magliocchetti}, M. \& {Porciani}, C. 2003, \mnras, 346, 186

\bibitem[{{Marulli} {et~al.}(2013){Marulli}, {Bolzonella}, {Branchini},
  {Davidzon}, {de la Torre}, {Granett}, {Guzzo}, {Iovino}, {Moscardini},
  {Pollo}, {Abbas}, {Adami}, {Arnouts}, {Bel}, {Bottini}, {Cappi}, {Coupon},
  {Cucciati}, {De Lucia}, {Fritz}, {Franzetti}, {Fumana}, {Garilli}, {Ilbert},
  {Krywult}, {Le Brun}, {Le F{\`e}vre}, {Maccagni}, {Ma{\l}ek}, {McCracken},
  {Paioro}, {Polletta}, {Schlagenhaufer}, {Scodeggio}, {Tasca}, {Tojeiro},
  {Vergani}, {Zanichelli}, {Burden}, {Di Porto}, {Marchetti}, {Marinoni},
  {Mellier}, {Nichol}, {Peacock}, {Percival}, {Phleps}, {Wolk}, \&
  {Zamorani}}]{Marulli2013}
{Marulli}, F., {Bolzonella}, M., {Branchini}, E., {et~al.} 2013, \aap, 557, A17

\bibitem[{{Mashchenko} {et~al.}(2008){Mashchenko}, {Wadsley}, \&
  {Couchman}}]{Mashchenko2008}
{Mashchenko}, S., {Wadsley}, J., \& {Couchman}, H.~M.~P. 2008, Science, 319,
  174

\bibitem[{{Mason} {et~al.}(2015){Mason}, {Trenti}, \& {Treu}}]{Mason2015}
{Mason}, C.~A., {Trenti}, M., \& {Treu}, T. 2015, \apj, 813, 21

\bibitem[{{Massey} {et~al.}(2007){Massey}, {Rhodes}, {Ellis}, {Scoville},
  {Leauthaud}, {Finoguenov}, {Capak}, {Bacon}, {Aussel}, {Kneib}, {Koekemoer},
  {McCracken}, {Mobasher}, {Pires}, {Refregier}, {Sasaki}, {Starck},
  {Taniguchi}, {Taylor}, \& {Taylor}}]{Massey2007}
{Massey}, R., {Rhodes}, J., {Ellis}, R., {et~al.} 2007, \nat, 445, 286

\bibitem[{{McCracken} {et~al.}(2015){McCracken}, {Wolk}, {Colombi},
  {Kilbinger}, {Ilbert}, {Peirani}, {Coupon}, {Dunlop}, {Milvang-Jensen},
  {Caputi}, {Aussel}, {B{\'e}thermin}, \& {Le F{\`e}vre}}]{McCracken2015}
{McCracken}, H.~J., {Wolk}, M., {Colombi}, S., {et~al.} 2015, \mnras, 449, 901

\bibitem[{{McLure} {et~al.}(2013){McLure}, {Dunlop}, {Bowler}, {Curtis-Lake},
  {Schenker}, {Ellis}, {Robertson}, {Koekemoer}, {Rogers}, {Ono}, {Ouchi},
  {Charlot}, {Wild}, {Stark}, {Furlanetto}, {Cirasuolo}, \&
  {Targett}}]{McLure2013}
{McLure}, R.~J., {Dunlop}, J.~S., {Bowler}, R.~A.~A., {et~al.} 2013, \mnras,
  432, 2696

\bibitem[{{Meneux} {et~al.}(2009){Meneux}, {Guzzo}, {de la Torre}, {Porciani},
  {Zamorani}, {Abbas}, {Bolzonella}, {Garilli}, {Iovino}, {Pozzetti}, {Zucca},
  {Lilly}, {Le F{\`e}vre}, {Kneib}, {Carollo}, {Contini}, {Mainieri},
  {Renzini}, {Scodeggio}, {Bardelli}, {Bongiorno}, {Caputi}, {Coppa},
  {Cucciati}, {de Ravel}, {Franzetti}, {Kampczyk}, {Knobel}, {Kova{\v c}},
  {Lamareille}, {Le Borgne}, {Le Brun}, {Maier}, {Pell{\`o}}, {Peng}, {Perez
  Montero}, {Ricciardelli}, {Silverman}, {Tanaka}, {Tasca}, {Tresse},
  {Vergani}, {Bottini}, {Cappi}, {Cimatti}, {Cassata}, {Fumana}, {Koekemoer},
  {Leauthaud}, {Maccagni}, {Marinoni}, {McCracken}, {Memeo}, {Oesch}, \&
  {Scaramella}}]{Meneux2009}
{Meneux}, B., {Guzzo}, L., {de la Torre}, S., {et~al.} 2009, \aap, 505, 463

\bibitem[{{Meneux} {et~al.}(2008){Meneux}, {Guzzo}, {Garilli}, {Le F{\`e}vre},
  {Pollo}, {Blaizot}, {De Lucia}, {Bolzonella}, {Lamareille}, {Pozzetti},
  {Cappi}, {Iovino}, {Marinoni}, {McCracken}, {de la Torre}, {Bottini}, {Le
  Brun}, {Maccagni}, {Picat}, {Scaramella}, {Scodeggio}, {Tresse}, {Vettolani},
  {Zanichelli}, {Abbas}, {Adami}, {Arnouts}, {Bardelli}, {Bongiorno},
  {Charlot}, {Ciliegi}, {Contini}, {Cucciati}, {Foucaud}, {Franzetti},
  {Gavignaud}, {Ilbert}, {Marano}, {Mazure}, {Merighi}, {Paltani}, {Pell{\`o}},
  {Radovich}, {Vergani}, {Zamorani}, \& {Zucca}}]{Meneux2008}
{Meneux}, B., {Guzzo}, L., {Garilli}, B., {et~al.} 2008, \aap, 478, 299

\bibitem[{{Meneux} {et~al.}(2006){Meneux}, {Le F{\`e}vre}, {Guzzo}, {Pollo},
  {Cappi}, {Ilbert}, {Iovino}, {Marinoni}, {McCracken}, {Bottini}, {Garilli},
  {Le Brun}, {Maccagni}, {Picat}, {Scaramella}, {Scodeggio}, {Tresse},
  {Vettolani}, {Zanichelli}, {Adami}, {Arnouts}, {Arnaboldi}, {Bardelli},
  {Bolzonella}, {Charlot}, {Ciliegi}, {Contini}, {Foucaud}, {Franzetti},
  {Gavignaud}, {Marano}, {Mazure}, {Merighi}, {Paltani}, {Pell{\`o}},
  {Pozzetti}, {Radovich}, {Zamorani}, {Zucca}, {Bondi}, {Bongiorno},
  {Busarello}, {Cucciati}, {Gregorini}, {Lamareille}, {Mathez}, {Mellier},
  {Merluzzi}, {Ripepi}, \& {Rizzo}}]{Meneux2006}
{Meneux}, B., {Le F{\`e}vre}, O., {Guzzo}, L., {et~al.} 2006, \aap, 452, 387

\bibitem[{{Metcalf} \& {Madau}(2001)}]{Metcalf2001}
{Metcalf}, R.~B. \& {Madau}, P. 2001, \apj, 563, 9

\bibitem[{{Meylan} {et~al.}(2006){Meylan}, {Jetzer}, {North}, {Schneider},
  {Kochanek}, \& {Wambsganss}}]{Meylan2006}
{Meylan}, G., {Jetzer}, P., {North}, P., {et~al.}, eds. 2006, {Gravitational
  Lensing: Strong, Weak and Micro}

\bibitem[{{Miller} {et~al.}(2014){Miller}, {Ellis}, {Newman}, \&
  {Benson}}]{Miller2014}
{Miller}, S.~H., {Ellis}, R.~S., {Newman}, A.~B., \& {Benson}, A. 2014, \apj,
  782, 115

\bibitem[{{Mo} \& {White}(1996)}]{Mo1996}
{Mo}, H.~J. \& {White}, S.~D.~M. 1996, \mnras, 282, 347

\bibitem[{{Mostek} {et~al.}(2013){Mostek}, {Coil}, {Cooper}, {Davis}, {Newman},
  \& {Weiner}}]{Mostek2013}
{Mostek}, N., {Coil}, A.~L., {Cooper}, M., {et~al.} 2013, \apj, 767, 89

\bibitem[{{Moster} {et~al.}(2013){Moster}, {Naab}, \& {White}}]{Moster2013}
{Moster}, B.~P., {Naab}, T., \& {White}, S.~D.~M. 2013, \mnras, 428, 3121

\bibitem[{{Moustakas} \& {Metcalf}(2003)}]{Moustakas2003}
{Moustakas}, L.~A. \& {Metcalf}, R.~B. 2003, \mnras, 339, 607

\bibitem[{{Newman} {et~al.}(2012){Newman}, {Genzel}, {F{\"o}rster-Schreiber},
  {Shapiro Griffin}, {Mancini}, {Lilly}, {Renzini}, {Bouch{\'e}}, {Burkert},
  {Buschkamp}, {Carollo}, {Cresci}, {Davies}, {Eisenhauer}, {Genel}, {Hicks},
  {Kurk}, {Lutz}, {Naab}, {Peng}, {Sternberg}, {Tacconi}, {Vergani}, {Wuyts},
  \& {Zamorani}}]{Newman2012}
{Newman}, S.~F., {Genzel}, R., {F{\"o}rster-Schreiber}, N.~M., {et~al.} 2012,
  \apj, 761, 43

\bibitem[{{Norberg} {et~al.}(2002){Norberg}, {Baugh}, {Hawkins}, {Maddox},
  {Madgwick}, {Lahav}, {Cole}, {Frenk}, {Baldry}, {Bland-Hawthorn}, {Bridges},
  {Cannon}, {Colless}, {Collins}, {Couch}, {Dalton}, {De Propris}, {Driver},
  {Efstathiou}, {Ellis}, {Glazebrook}, {Jackson}, {Lewis}, {Lumsden},
  {Peacock}, {Peterson}, {Sutherland}, \& {Taylor}}]{Norberg2002}
{Norberg}, P., {Baugh}, C.~M., {Hawkins}, E., {et~al.} 2002, \mnras, 332, 827

\bibitem[{{O{\~n}orbe} {et~al.}(2015){O{\~n}orbe}, {Boylan-Kolchin}, {Bullock},
  {Hopkins}, {Kere{\v s}}, {Faucher-Gigu{\`e}re}, {Quataert}, \&
  {Murray}}]{Onorbe2015}
{O{\~n}orbe}, J., {Boylan-Kolchin}, M., {Bullock}, J.~S., {et~al.} 2015,
  \mnras, 454, 2092

\bibitem[{{Ogiya} \& {Mori}(2014)}]{Ogiya2014}
{Ogiya}, G. \& {Mori}, M. 2014, \apj, 793, 46

\bibitem[{{Ouchi} {et~al.}(2005){Ouchi}, {Hamana}, {Shimasaku}, {Yamada},
  {Akiyama}, {Kashikawa}, {Yoshida}, {Aoki}, {Iye}, {Saito}, {Sasaki},
  {Simpson}, \& {Yoshida}}]{Ouchi2005}
{Ouchi}, M., {Hamana}, T., {Shimasaku}, K., {et~al.} 2005, \apjl, 635, L117

\bibitem[{{Papastergis} \& {Shankar}(2016)}]{Papastergis2016}
{Papastergis}, E. \& {Shankar}, F. 2016, \aap, 591, A58

\bibitem[{{Parsa} {et~al.}(2016){Parsa}, {Dunlop}, {McLure}, \&
  {Mortlock}}]{Parsa2016}
{Parsa}, S., {Dunlop}, J.~S., {McLure}, R.~J., \& {Mortlock}, A. 2016, \mnras,
  456, 3194

\bibitem[{{Peacock} \& {Smith}(2000)}]{Peacock2000}
{Peacock}, J.~A. \& {Smith}, R.~E. 2000, \mnras, 318, 1144

\bibitem[{{Peebles}(1980)}]{Peebles1980}
{Peebles}, P.~J.~E. 1980, {The large-scale structure of the universe}

\bibitem[{{P{\'e}rez-Gonz{\'a}lez} {et~al.}(2008){P{\'e}rez-Gonz{\'a}lez},
  {Rieke}, {Villar}, {Barro}, {Blaylock}, {Egami}, {Gallego}, {Gil de Paz},
  {Pascual}, {Zamorano}, \& {Donley}}]{Perez2008}
{P{\'e}rez-Gonz{\'a}lez}, P.~G., {Rieke}, G.~H., {Villar}, V., {et~al.} 2008,
  \apj, 675, 234

\bibitem[{{Pettini} {et~al.}(2001){Pettini}, {Shapley}, {Steidel}, {Cuby},
  {Dickinson}, {Moorwood}, {Adelberger}, \& {Giavalisco}}]{Pettini2001}
{Pettini}, M., {Shapley}, A.~E., {Steidel}, C.~C., {et~al.} 2001, \apj, 554,
  981

\bibitem[{{Planck Collaboration} {et~al.}(2014){Planck Collaboration}, {Ade},
  {Aghanim}, {Armitage-Caplan}, {Arnaud}, {Ashdown}, {Atrio-Barandela},
  {Aumont}, {Baccigalupi}, {Banday}, \& et~al.}]{Planck2014}
{Planck Collaboration}, {Ade}, P.~A.~R., {Aghanim}, N., {et~al.} 2014, \aap,
  571, A16

\bibitem[{{Pollo} {et~al.}(2006){Pollo}, {Guzzo}, {Le F{\`e}vre}, {Meneux},
  {Cappi}, {Franzetti}, {Iovino}, {McCracken}, {Marinoni}, {Zamorani},
  {Bottini}, {Garilli}, {Le Brun}, {Maccagni}, {Picat}, {Scaramella},
  {Scodeggio}, {Tresse}, {Vettolani}, {Zanichelli}, {Adami}, {Arnouts},
  {Bardelli}, {Bolzonella}, {Charlot}, {Ciliegi}, {Contini}, {Foucaud},
  {Gavignaud}, {Ilbert}, {Marano}, {Mazure}, {Merighi}, {Paltani}, {Pell{\`o}},
  {Pozzetti}, {Radovich}, {Zucca}, {Bondi}, {Bongiorno}, {Busarello},
  {Cucciati}, {Gregorini}, {Lamareille}, {Mathez}, {Mellier}, {Merluzzi},
  {Ripepi}, \& {Rizzo}}]{Pollo2006}
{Pollo}, A., {Guzzo}, L., {Le F{\`e}vre}, O., {et~al.} 2006, \aap, 451, 409

\bibitem[{{Pollo} {et~al.}(2005){Pollo}, {Meneux}, {Guzzo}, {Le F{\`e}vre},
  {Blaizot}, {Cappi}, {Iovino}, {Marinoni}, {McCracken}, {Bottini}, {Garilli},
  {Le Brun}, {Maccagni}, {Picat}, {Scaramella}, {Scodeggio}, {Tresse},
  {Vettolani}, {Zanichelli}, {Adami}, {Arnaboldi}, {Arnouts}, {Bardelli},
  {Bolzonella}, {Charlot}, {Ciliegi}, {Contini}, {Foucaud}, {Franzetti},
  {Gavignaud}, {Ilbert}, {Marano}, {Mathez}, {Mazure}, {Merighi}, {Paltani},
  {Pell{\`o}}, {Pozzetti}, {Radovich}, {Zamorani}, {Zucca}, {Bondi},
  {Bongiorno}, {Busarello}, {Gregorini}, {Lamareille}, {Mellier}, {Merluzzi},
  {Ripepi}, \& {Rizzo}}]{Pollo2005}
{Pollo}, A., {Meneux}, B., {Guzzo}, L., {et~al.} 2005, \aap, 439, 887

\bibitem[{{Pontzen} \& {Governato}(2012)}]{Pontzen2012}
{Pontzen}, A. \& {Governato}, F. 2012, \mnras, 421, 3464

\bibitem[{{Press} \& {Schechter}(1974)}]{Press1974}
{Press}, W.~H. \& {Schechter}, P. 1974, \apj, 187, 425

\bibitem[{{Read} {et~al.}(2017){Read}, {Iorio}, {Agertz}, \&
  {Fraternali}}]{Read2017}
{Read}, J.~I., {Iorio}, G., {Agertz}, O., \& {Fraternali}, F. 2017, \mnras,
  467, 2019

\bibitem[{{Reddy} \& {Steidel}(2009)}]{Reddy2009}
{Reddy}, N.~A. \& {Steidel}, C.~C. 2009, \apj, 692, 778

\bibitem[{{Rines} {et~al.}(2013){Rines}, {Geller}, {Diaferio}, \&
  {Kurtz}}]{Rines2013}
{Rines}, K., {Geller}, M.~J., {Diaferio}, A., \& {Kurtz}, M.~J. 2013, \apj,
  767, 15

\bibitem[{{Robertson}(2010)}]{Robertson2010}
{Robertson}, B.~E. 2010, \apj, 713, 1266

\bibitem[{{Rubin} {et~al.}(1978){Rubin}, {Thonnard}, \& {Ford}}]{Rubin1978}
{Rubin}, V.~C., {Thonnard}, N., \& {Ford}, Jr., W.~K. 1978, \apjl, 225, L107

\bibitem[{{Sawala} {et~al.}(2013){Sawala}, {Frenk}, {Crain}, {Jenkins},
  {Schaye}, {Theuns}, \& {Zavala}}]{Sawala2013}
{Sawala}, T., {Frenk}, C.~S., {Crain}, R.~A., {et~al.} 2013, \mnras, 431, 1366

\bibitem[{{Sawala} {et~al.}(2015){Sawala}, {Frenk}, {Fattahi}, {Navarro},
  {Bower}, {Crain}, {Dalla Vecchia}, {Furlong}, {Jenkins}, {McCarthy}, {Qu},
  {Schaller}, {Schaye}, \& {Theuns}}]{Sawala2015}
{Sawala}, T., {Frenk}, C.~S., {Fattahi}, A., {et~al.} 2015, \mnras, 448, 2941

\bibitem[{{Sawala} {et~al.}(2011){Sawala}, {Guo}, {Scannapieco}, {Jenkins}, \&
  {White}}]{Sawala2011}
{Sawala}, T., {Guo}, Q., {Scannapieco}, C., {Jenkins}, A., \& {White}, S. 2011,
  \mnras, 413, 659

\bibitem[{{Sawicki} \& {Thompson}(2006)}]{Sawicki2006}
{Sawicki}, M. \& {Thompson}, D. 2006, \apj, 648, 299

\bibitem[{{Schechter}(1976)}]{Schechter1976}
{Schechter}, P. 1976, \apj, 203, 297

\bibitem[{{Seljak}(2000)}]{Seljak2000}
{Seljak}, U. 2000, \mnras, 318, 203

\bibitem[{{Shapley} {et~al.}(2003){Shapley}, {Steidel}, {Pettini}, \&
  {Adelberger}}]{Shapley2003}
{Shapley}, A.~E., {Steidel}, C.~C., {Pettini}, M., \& {Adelberger}, K.~L. 2003,
  \apj, 588, 65

\bibitem[{{Sheth} {et~al.}(2001){Sheth}, {Mo}, \& {Tormen}}]{Sheth2001}
{Sheth}, R.~K., {Mo}, H.~J., \& {Tormen}, G. 2001, \mnras, 323, 1

\bibitem[{{Skibba} {et~al.}(2015){Skibba}, {Coil}, {Mendez}, {Blanton}, {Bray},
  {Cool}, {Eisenstein}, {Guo}, {Miyaji}, {Moustakas}, \& {Zhu}}]{Skibba2015}
{Skibba}, R.~A., {Coil}, A.~L., {Mendez}, A.~J., {et~al.} 2015, \apj, 807, 152

\bibitem[{{Springel} {et~al.}(2005){Springel}, {White}, {Jenkins}, {Frenk},
  {Yoshida}, {Gao}, {Navarro}, {Thacker}, {Croton}, {Helly}, {Peacock}, {Cole},
  {Thomas}, {Couchman}, {Evrard}, {Colberg}, \& {Pearce}}]{Springel2005}
{Springel}, V., {White}, S.~D.~M., {Jenkins}, A., {et~al.} 2005, \nat, 435, 629

\bibitem[{{Steidel} {et~al.}(2010){Steidel}, {Erb}, {Shapley}, {Pettini},
  {Reddy}, {Bogosavljevi{\'c}}, {Rudie}, \& {Rakic}}]{Steidel2010}
{Steidel}, C.~C., {Erb}, D.~K., {Shapley}, A.~E., {et~al.} 2010, \apj, 717, 289

\bibitem[{{Talia} {et~al.}(2017){Talia}, {Brusa}, {Cimatti}, {Lemaux},
  {Amorin}, {Bardelli}, {Cassar{\`a}}, {Cucciati}, {Garilli}, {Grazian},
  {Guaita}, {Hathi}, {Koekemoer}, {Le F{\`e}vre}, {Maccagni}, {Nakajima},
  {Pentericci}, {Pforr}, {Schaerer}, {Vanzella}, {Vergani}, {Zamorani}, \&
  {Zucca}}]{Talia2017}
{Talia}, M., {Brusa}, M., {Cimatti}, A., {et~al.} 2017, \mnras, 471, 4527

\bibitem[{{Tasca} {et~al.}(2015){Tasca}, {Le F{\`e}vre}, {Hathi}, {Schaerer},
  {Ilbert}, {Zamorani}, {Lemaux}, {Cassata}, {Garilli}, {Le Brun}, {Maccagni},
  {Pentericci}, {Thomas}, {Vanzella}, {Zucca}, {Amorin}, {Bardelli},
  {Cassar{\`a}}, {Castellano}, {Cimatti}, {Cucciati}, {Durkalec}, {Fontana},
  {Giavalisco}, {Grazian}, {Paltani}, {Ribeiro}, {Scodeggio}, {Sommariva},
  {Talia}, {Tresse}, {Vergani}, {Capak}, {Charlot}, {Contini}, {de la Torre},
  {Dunlop}, {Fotopoulou}, {Koekemoer}, {L{\'o}pez-Sanjuan}, {Mellier}, {Pforr},
  {Salvato}, {Scoville}, {Taniguchi}, \& {Wang}}]{Tasca2015}
{Tasca}, L.~A.~M., {Le F{\`e}vre}, O., {Hathi}, N.~P., {et~al.} 2015, \aap,
  581, A54

\bibitem[{{Tasca} {et~al.}(2014){Tasca}, {Le F{\`e}vre}, {L{\'o}pez-Sanjuan},
  {Wang}, {Cassata}, {Garilli}, {Ilbert}, {Le Brun}, {Lemaux}, {Maccagni},
  {Tresse}, {Bardelli}, {Contini}, {Charlot}, {Cucciati}, {Fontana},
  {Giavalisco}, {Kneib}, {Salvato}, {Taniguchi}, {Vergani}, {Zamorani}, \&
  {Zucca}}]{Tasca2014}
{Tasca}, L.~A.~M., {Le F{\`e}vre}, O., {L{\'o}pez-Sanjuan}, C., {et~al.} 2014,
  \aap, 565, A10

\bibitem[{{Tegmark} {et~al.}(2004){Tegmark}, {Blanton}, {Strauss}, {Hoyle},
  {Schlegel}, {Scoccimarro}, {Vogeley}, {Weinberg}, {Zehavi}, {Berlind},
  {Budavari}, {Connolly}, {Eisenstein}, {Finkbeiner}, {Frieman}, {Gunn},
  {Hamilton}, {Hui}, {Jain}, {Johnston}, {Kent}, {Lin}, {Nakajima}, {Nichol},
  {Ostriker}, {Pope}, {Scranton}, {Seljak}, {Sheth}, {Stebbins}, {Szalay},
  {Szapudi}, {Verde}, {Xu}, {Annis}, {Bahcall}, {Brinkmann}, {Burles},
  {Castander}, {Csabai}, {Loveday}, {Doi}, {Fukugita}, {Gott}, {Hennessy},
  {Hogg}, {Ivezi{\'c}}, {Knapp}, {Lamb}, {Lee}, {Lupton}, {McKay}, {Kunszt},
  {Munn}, {O'Connell}, {Peoples}, {Pier}, {Richmond}, {Rockosi}, {Schneider},
  {Stoughton}, {Tucker}, {Vanden Berk}, {Yanny}, {York}, \& {SDSS
  Collaboration}}]{Tegmark2004}
{Tegmark}, M., {Blanton}, M.~R., {Strauss}, M.~A., {et~al.} 2004, \apj, 606,
  702

\bibitem[{{Tegmark} \& {Peebles}(1998)}]{Tegmark1998}
{Tegmark}, M. \& {Peebles}, P.~J.~E. 1998, \apjl, 500, L79

\bibitem[{{Thomas} {et~al.}(2016){Thomas}, {Le F{\`e}vre}, {Scodeggio},
  {Cassata}, {Garilli}, {Le Brun}, {Lemaux}, {Maccagni}, {Pforr}, {Tasca},
  {Zamorani}, {Bardelli}, {Hathi}, {Tresse}, \& {Zucca}}]{Thomas2016}
{Thomas}, R., {Le F{\`e}vre}, O., {Scodeggio}, M., {et~al.} 2016, ArXiv
  1602.01841 [\eprint[arXiv]{1602.01841}]

\bibitem[{{Van Waerbeke} {et~al.}(2000){Van Waerbeke}, {Mellier}, {Erben},
  {Cuillandre}, {Bernardeau}, {Maoli}, {Bertin}, {McCracken}, {Le F{\`e}vre},
  {Fort}, {Dantel-Fort}, {Jain}, \& {Schneider}}]{VanWaerbeke2000}
{Van Waerbeke}, L., {Mellier}, Y., {Erben}, T., {et~al.} 2000, \aap, 358, 30

\bibitem[{{Wake} {et~al.}(2011){Wake}, {Whitaker}, {Labb{\'e}}, {van Dokkum},
  {Franx}, {Quadri}, {Brammer}, {Kriek}, {Lundgren}, {Marchesini}, \&
  {Muzzin}}]{Wake2011}
{Wake}, D.~A., {Whitaker}, K.~E., {Labb{\'e}}, I., {et~al.} 2011, \apj, 728, 46

\bibitem[{{Wang} {et~al.}(2007){Wang}, {Li}, {Kauffmann}, \& {De
  Lucia}}]{Wang2007}
{Wang}, L., {Li}, C., {Kauffmann}, G., \& {De Lucia}, G. 2007, \mnras, 377,
  1419

\bibitem[{{Weiner} {et~al.}(2009){Weiner}, {Coil}, {Prochaska}, {Newman},
  {Cooper}, {Bundy}, {Conselice}, {Dutton}, {Faber}, {Koo}, {Lotz}, {Rieke}, \&
  {Rubin}}]{Weiner2009}
{Weiner}, B.~J., {Coil}, A.~L., {Prochaska}, J.~X., {et~al.} 2009, \apj, 692,
  187

\bibitem[{{Weinmann} {et~al.}(2012){Weinmann}, {Pasquali}, {Oppenheimer},
  {Finlator}, {Mendel}, {Crain}, \& {Macci{\`o}}}]{Weinmann2012}
{Weinmann}, S.~M., {Pasquali}, A., {Oppenheimer}, B.~D., {et~al.} 2012, \mnras,
  426, 2797

\bibitem[{{Wetzel} {et~al.}(2009){Wetzel}, {Cohn}, \& {White}}]{Wetzel2009}
{Wetzel}, A.~R., {Cohn}, J.~D., \& {White}, M. 2009, \mnras, 395, 1376

\bibitem[{{White}(1976)}]{White1976}
{White}, S.~D.~M. 1976, \mnras, 177, 717

\bibitem[{{White} {et~al.}(1987){White}, {Davis}, {Efstathiou}, \&
  {Frenk}}]{White1987}
{White}, S.~D.~M., {Davis}, M., {Efstathiou}, G., \& {Frenk}, C.~S. 1987, \nat,
  330, 451

\bibitem[{{White} \& {Rees}(1978)}]{White1978}
{White}, S.~D.~M. \& {Rees}, M.~J. 1978, \mnras, 183, 341

\bibitem[{{Wraith} {et~al.}(2009){Wraith}, {Kilbinger}, {Benabed}, {Capp{\'e}},
  {Cardoso}, {Fort}, {Prunet}, \& {Robert}}]{Wraith2009}
{Wraith}, D., {Kilbinger}, M., {Benabed}, K., {et~al.} 2009, \prd, 80, 023507

\bibitem[{{Yang} {et~al.}(2012){Yang}, {Mo}, {van den Bosch}, {Zhang}, \&
  {Han}}]{Yang2012}
{Yang}, X., {Mo}, H.~J., {van den Bosch}, F.~C., {Zhang}, Y., \& {Han}, J.
  2012, \apj, 752, 41

\bibitem[{{Zehavi} {et~al.}(2004){Zehavi}, {Weinberg}, {Zheng}, {Berlind},
  {Frieman}, {Scoccimarro}, {Sheth}, {Blanton}, {Tegmark}, {Mo}, {Bahcall},
  {Brinkmann}, {Burles}, {Csabai}, {Fukugita}, {Gunn}, {Lamb}, {Loveday},
  {Lupton}, {Meiksin}, {Munn}, {Nichol}, {Schlegel}, {Schneider}, {SubbaRao},
  {Szalay}, {Uomoto}, {York}, \& {SDSS Collaboration}}]{Zehavi2004}
{Zehavi}, I., {Weinberg}, D.~H., {Zheng}, Z., {et~al.} 2004, \apj, 608, 16

\bibitem[{{Zehavi} {et~al.}(2011){Zehavi}, {Zheng}, {Weinberg}, {Blanton},
  {Bahcall}, {Berlind}, {Brinkmann}, {Frieman}, {Gunn}, {Lupton}, {Nichol},
  {Percival}, {Schneider}, {Skibba}, {Strauss}, {Tegmark}, \&
  {York}}]{Zehavi2011}
{Zehavi}, I., {Zheng}, Z., {Weinberg}, D.~H., {et~al.} 2011, \apj, 736, 59

\bibitem[{{Zheng} {et~al.}(2005){Zheng}, {Berlind}, {Weinberg}, {Benson},
  {Baugh}, {Cole}, {Dav{\'e}}, {Frenk}, {Katz}, \& {Lacey}}]{Zheng2005}
{Zheng}, Z., {Berlind}, A.~A., {Weinberg}, D.~H., {et~al.} 2005, \apj, 633, 791

\bibitem[{{Zheng} {et~al.}(2007){Zheng}, {Coil}, \& {Zehavi}}]{Zheng2007}
{Zheng}, Z., {Coil}, A.~L., \& {Zehavi}, I. 2007, \apj, 667, 760

\bibitem[{{Zwicky}(1937)}]{Zwicky1937}
{Zwicky}, F. 1937, \apj, 86, 217

\end{thebibliography}

\appendix
\section{Correction for the luminosity and stellar mass function evolution}
\label{app:ev_correction}
\begin{figure}
 \centering
  \subfloat{\includegraphics[angle=270]{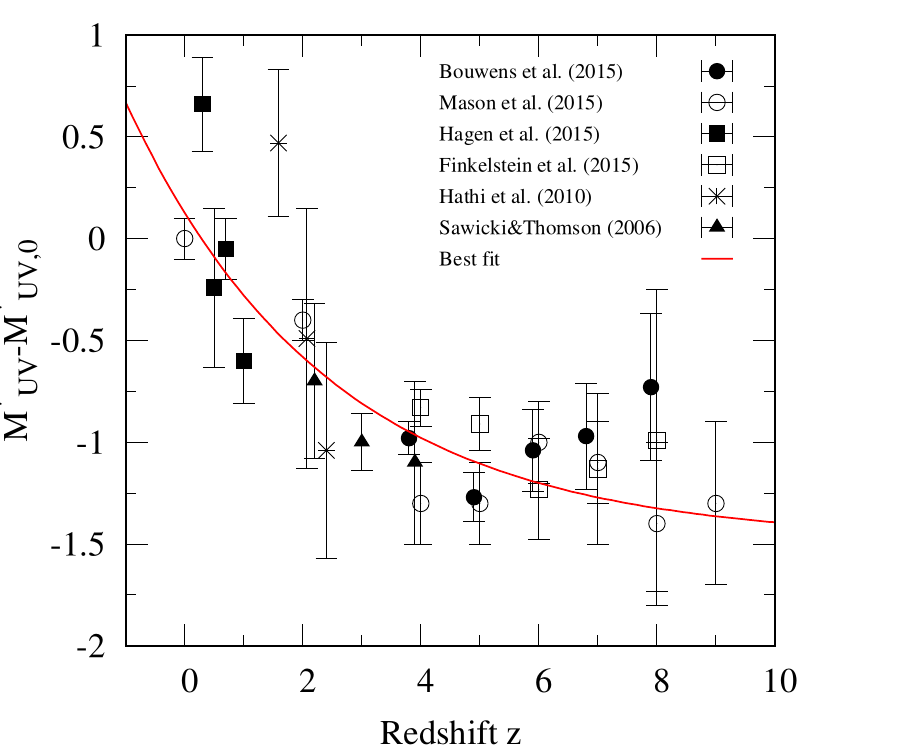}} \\ [-1ex]
  \subfloat{\includegraphics[angle=270]{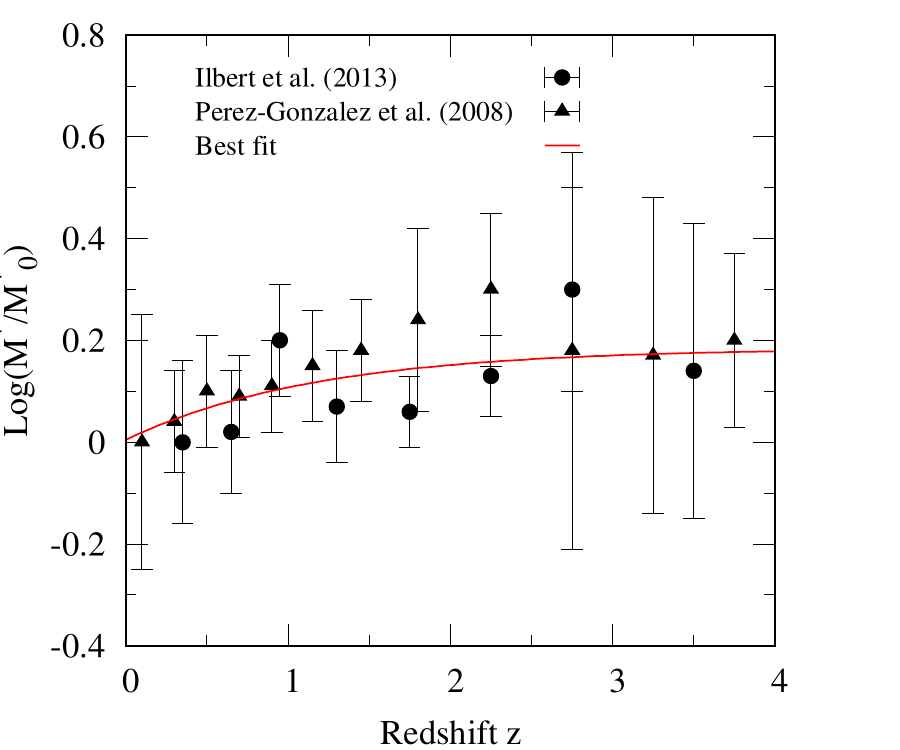}}
 \caption{A compilation of the values of Schechter characteristic UV galaxy luminosity $M^*_{UV}-M^*_{UV,0}$ (upper panel) and  Schechter characteristic stellar mass $\log\left(M^*(z)/M^*_0\right)$ (lower panel).
	  The symbols represent the measurements taken from various works \citep[]{Bouwens2015, Mason2015, Hagen2015, Finkelstein2015, Ilbert2013, Hathi2010, Perez2008, Sawicki2006} as described in the legend.
	  In each plot the solid red line shows the best-fitting exponential function given by Eq. \ref{eq:lum_corr} and Eq. \ref{eq:mass_corr} for the upper and lower panel, respectively.}
 \label{fig:lum_char_params}
\end{figure}
The mass, shape, number density of stars in the galaxies are constantly evolving with time. 
Consequently we observe the overall changes in luminosity and stellar mass of the galaxy populations at different epochs.
The influence of these changes on the absolute magnitude and stellar mass of the galaxy population are reflected in the evolution of the luminosity and stellar mass functions, respectively.
Particularly in the evolution of the $M^*$ parameter, from the best-fitted Schechter function \citep{Schechter1976}, which describes the characteristic absolute magnitude (or stellar mass) of the galaxy population at given epoch.

The luminosity and stellar mass functions have been extensively studied in the literature, even at extremely high redshift ranges \citep[e.g., ][]{Lilly1996,Bouwens2007, Reddy2009, Robertson2010, McLure2013} and all the evidence to date suggests a brightening of the galaxy population when moving back in time.
In the redshift range $2<z<4$, one of the most recent studies of the galaxy UV luminosity function from \cite{Parsa2016} (based on the combination of data from the Hubble Ultra Deep Field (HUDF), CANDELS/GOODS-South, and UltraVISTA/COSMOS surveys), shows a brightening in the UV characteristic luminosity from $M^*_{UV}=-19.61\pm0.07$ at $z\sim1.7$ to $M^*_{UV}=-20.71\pm0.1$ at $z\sim3.8$.
At even higher redshift ranges ($4<z<8$) \cite{Bouwens2015} finds that the characteristic UV galaxy luminosity does not change its value significantly and  at $z\sim3.8$ is $M^*_{UV}=-20.88\pm0.08$, while at at $z\sim8$ $M^*_{UV}=-20.63\pm0.36$.

In our study we focus on the luminosity and stellar mass dependencies of the galaxy clustering.
In order to draw conclusions and compare our results with data from different epochs, we need to address the evolutionary brightening of galaxies. 
Hence, we normalised the absolute magnitudes and stellar masses, at each redshift, to the corresponding value of the characteristic luminosity $M^*_{UV}$ or characteristic stellar mass $\log M^*$

Using measurements of the UV characteristic absolute magnitudes from \cite{Bouwens2015, Mason2015, Hagen2015, Finkelstein2015, Hathi2010} and \cite{Sawicki2006}, we construct the $M^*_{UV}(z)-M^*_{UV,0}$ function, as presented in the upper panel of Fig. \ref{fig:lum_char_params}, where the $M^*_{UV,0}$ is the characteristic luminosity for galaxies at $z=0$. 
Then the best-fitting exponential function in form,
\begin{equation}
 M^*_{UV}(z)-M^*_{UV,0} = -1.32 + 1.44\exp\left(-z/2.93\right),
 \label{eq:lum_corr}
\end{equation}
has been used to normalise the absolute magnitudes of galaxies used in this paper.
For each galaxy we take  $M_{UV} = M'_{UV} - (M^*_{UV} - M_{UV,0}^*)$, where $M'_{UV}$ is the original (not corrected) absolute magnitude.

We proceeded similarly to normalise the galaxy stellar masses.
We took the characteristic stellar masses measured by \cite{Ilbert2013} and \cite{Perez2008} in the redshift range $0<z<4$ and fitted it with a simple exponential function, as presented in the lower panel of Fig. \ref{fig:lum_char_params}.
As before, the resulting best-fitting function,
\begin{equation}
 \log\left(\frac{M^*(z)}{M^*_0}\right) = -0.18\exp\left(-z/1.18\right) + 0.18,
 \label{eq:mass_corr}
\end{equation}
has been used to normalise all stellar masses of the galaxies used in this study.

\section{Tests of sample variation - a large structure in the COSMOS field at $z\sim3$}
\label{app:cosmic_varaince}
\begin{figure}
 \centering
 \includegraphics[angle=270]{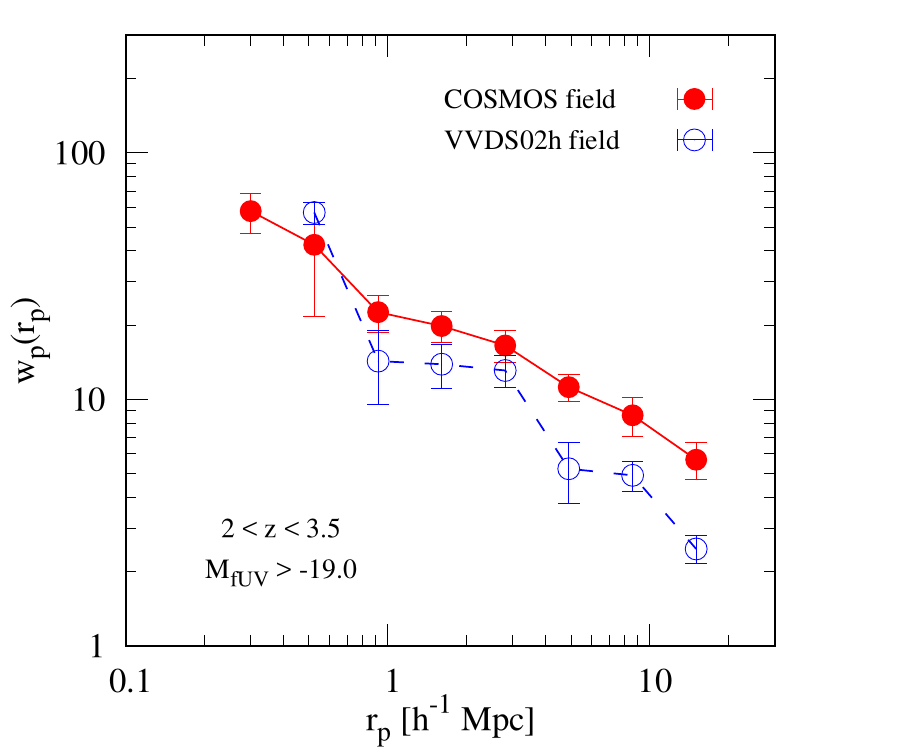}
 \caption{Projected two-point correlation function $w_p(r_p)$ measured independently for the $M_{UV} > -19.0$ galaxies in two VUDS fields. 
	  Red filled circles represent the correlation function measurements for the COSMOS field galaxy sample, while open blue circles show similar measurements for the VVDS-02h field sample.}
 \label{fig:variance}
\end{figure}
The correlation function measurements presented in this work are obtained from three independent VUDS fields (COSMOS, VVDS-02h and ECDFS). 
The differences between these fields, like their angular size and number of galaxies, are accounted for by using an appropriate weighting scheme (see Sec. \ref{sec:method}).
This weighting scheme favours the biggest and the most populated fields in order to retrieve the best correlation function signal for all separations $r_p$.
At the same time, the differences between the correlation functions measured for the different fields yield information about the cosmic variance.

As a representative example in Fig. \ref{fig:variance} we show a comparison of the independent correlation function measurements for the $M_{UV} > -19.0$ galaxy sample from two VUDS fields: COSMOS (red filled circles) and VVDS-02h (open blue circles).
Please note that in the further discussion we neglect the ECDFS field.
Because its small size ($S_{eff} = 0.11$ deg$^2$), the measurement of the correlation function in this field does not contribute to the final $w_p(r_p)$ measurement at scales $r_p > 5$ $h^{-1}$ Mpc on which the discussion below is focused.

The most significant cosmic variance effect appears at large separations $r_p >5$ $h^{-1}$ Mpc. 
At these large scales we observe a significant difference between the two correlation function measurements, as presented in Fig. \ref{fig:variance}.
The values of $w_p(r_p)$ measured at $r_p >5$ $h^{-1}$ Mpc for the COSMOS field are approximately two times higher than the correlation function signal obtained for the VVDS-02h field.
Naturally, this difference has an impact on the overall composite correlation function measurements presented in this work. 
The COSMOS field contains of the largest number of galaxies spread across the biggest effective surface (see Tab. \ref{tab:sample_numbers}). 
Therefore, the clustering results obtained for this field have the biggest impact on the final composite correlation function measurements, and this results in the higher vales of the correlation function with respect to the best HOD models seen in Fig. \ref{fig:cf_results_hod}.
For all UV absolute magnitude and stellar mass selected sub-samples, the correlation function measurement, at scales $r_p > 5$ $h^{-1}$ Mpc, is higher on average by a factor of 1.7 with respect to the HOD model.  

The flattening of the correlation function measured in the COSMOS field at large scales is likely related to the existence of an extremely large structure of galaxies, possibly a proto-supercluster or a massive filament, at $z\sim2.5$, which spans a size comparable to to the entire filed covered by VUDS-COSMOS (Cucciati et al. 2017, in prep).
This would be the first observation of such a structure at high redshift.
This hypothesis requires further investigation and will be addressed in the dedicated follow up research.

Ideally, to get the most robust measurements of the correlation function, one would exclude members of this structure from the measurements, however (1) the members of this structure have not been fully identified yet and (2) this would significantly lower the sample statistic and probably make it impossible to perform correlation function measurements for the luminosity and stellar mass selected galaxy samples.   

\end{document}